\documentclass[10pt,twocolumn,twoside]{IEEEtran}

\usepackage{amsmath,graphicx,url,times,amssymb}
\usepackage{amsopn}
\usepackage{bm} % Math packages

\usepackage{subfigure}
\usepackage{acronym}
\usepackage{balance}
\usepackage{caption}
\usepackage{color}
\usepackage[usenames,dvipsnames]{xcolor}
\usepackage{amsthm}
\usepackage{tikz}

\usepackage{multicol}

\usepackage{dblfloatfix}
\usetikzlibrary{shapes,arrows,calc,backgrounds,external}
\usepackage{pgfplots}
\usepackage{pgfbaselayers}
\usepackage{pgfshape_mic}
\usepackage[]{units}
\usetikzlibrary{fit,positioning}

\usepackage{cite}
\usepackage{multirow}

\usepackage{algorithm}
\usepackage{algorithmic}
\usepackage{hhline}
\usepackage{booktabs}
\usepackage{array}
\usepackage{url}
\usepackage{setspace}

\usepackage{soul,xcolor} % use this (many fancier options)
\setstcolor{red}
\usepackage[normalem]{ulem}
\newcommand\redout{\bgroup\markoverwith{\textcolor{red}{\rule[.5ex]{2pt}{0.4pt}}}\ULon}

\definecolor{grey30}{RGB}{84,84,84}
\definecolor{ored}{RGB}{255,36,0}
\definecolor{metal}{RGB}{35,107,142}
\definecolor{dsblue}{RGB}{0,104,139}

\newcommand{\hermitian}{{\textrm{H}}}

\newcommand{\argmax}{\operatornamewithlimits{argmax}}
\newcommand{\vect}[1]{\mathbf{#1}}
\newcommand{\Tr}{\textrm{Tr}}

\acrodef{VEM}{variational Expectation-Maximization}
\acrodef{MMSE}{minimum mean square error}
\acrodef{MCWF}{multichannel Wiener filter}
\acrodef{MVDR}{minimum variance distortionless response}
\acrodef{STFT}{short-time Fourier transform}
\acrodef{TF}{time-frequency}
\acrodef{VAD}{voice activity detector}
\acrodef{RIR}{room impulse response}
\acrodef{RTF}{relative transfer function}
\acrodef{SPP}{speech presence probability}
\acrodef{EVD}{eigenvalue decomposition}
\acrodef{GEVD}{generalized eigenvalue decomposition}
\acrodef{ML}{maximum likelihood}
\acrodef{MAP}{maximum a posteriori}
\acrodef{EM}{Expectation-Maximization}
\acrodef{PDF}{probability density function}
\acrodef{LDS}{linear dynamical system}
\acrodef{KL}{Kullback-Leibler}
\acrodef{VMVDR}{variational MVDR}
\acrodef{PF}{postfilter}
\acrodef{VPF}{variational postfilter}
\acrodef{LS}{least-squares}
\acrodef{PM}{posterior mean}
\acrodef{MCMC}{Monte-Carlo Markov Chain}
\acrodef{RLS}{recursive least squares}
\acrodef{WLS}{weighted least-squares}

\acrodef{SNR}{signal-to-noise ratio}
\acrodef{RSNR}{reverberant signal-to-noise ratio}
\acrodef{SRR}{signal-to-reverberation ratio}
\acrodef{SRNR}{signal-to-reverberation-plus-noise ratio}

\acrodef{ULA}{uniform linear array}

\acrodef{PSD}{power spectral density}
\acrodef{AIR}{acoustic impulse response}
\acrodef{ATF}{acoustic transfer function}
\acrodef{PESQ}{perceptual evaluation of speech quality}
\acrodef{LSD}{log-spectral distance}
\acrodef{LLR}{log-likelihood ratio}
\acrodef{fwSNRseg}{frequency-weighted segmental SNR}
\acrodef{NPM}{normalized projection misalignment}
\acrodef{TDOA}{time difference of arrival}
\acrodef{DOA}{direction of arrival}
\acrodef{BIU}{Bar-Ilan University}
\acrodef{FIM}{Fisher information matrix}
\acrodef{MVU}{minimum variance unbiased}
\acrodef{CRB}{Cram\'{e}r-Rao Bound}
\acrodef{PSD}{power spectral density}
\acrodef{MMSE}{minimum mean square error}
\acrodef{MSE}{mean square error}
\acrodef{nMSE}{normalized mean square error}
\acrodef{ML}{maximum likelihood}
\acrodef{MLE}{maximum likelihood estimator}
\acrodef{BM}{blocking matrix}
\acrodef{RETF}{relative early transfer function}
\acrodef{ETF}{early transfer function}
\acrodef{RDTF}{relative direct-path transfer function}
\acrodef{DTF}{direct-path transfer function}
\acrodef{LCMV}{linearly constrained minimum variance}
\acrodef{INR}{interference-to-white noise ratio}
\acrodef{ASR}{automatic speech recognition}

\begin{document}
\sloppy
\parskip 0pt

\title{ML Estimation and CRBs for Reverberation, Speech and Noise PSDs 
in Rank-Deficient Noise Field
}
\author{Yaron~Laufer,~\IEEEmembership{Student Member,~IEEE}, 
Bracha~Laufer-Goldshtein,~\IEEEmembership{Student Member,~IEEE}, 
and Sharon~Gannot,~\IEEEmembership{Senior Member,~IEEE}
\thanks{The authors are with the Faculty of Engineering, Bar-Ilan University,
Ramat-Gan, 5290002, Israel (e-mail: Yaron.Laufer@biu.ac.il, Bracha.Laufer@biu.ac.il, Sharon.Gannot@biu.ac.il).
}}

%\markboth{ IEEE/ACM TRANSACTIONS ON AUDIO, SPEECH, AND LANGUAGE PROCESSING}
%{Laufer \textit{\MakeLowercase{et al}.}: ML Estimation and CRB\MakeLowercase{s} for Reverberation, Speech and Noise 
%PSD\MakeLowercase{s} in Rank-Deficient Noise Field}

\maketitle

\begin{abstract}
Speech communication systems are prone to performance degradation
in reverberant and noisy acoustic environments.
Dereverberation and noise reduction algorithms typically require 
several model parameters, e.g.~the speech, reverberation and noise power spectral densities (PSDs).\acused{PSD}
A commonly used assumption is that 
the noise \ac{PSD} matrix is known.
However, in practical acoustic scenarios, the noise \ac{PSD} matrix
is unknown and should be estimated 
along with the speech and reverberation \acp{PSD}.
%%%%
In this paper, we consider the case of rank-deficient noise \ac{PSD} matrix, 
which arises when the noise signal consists of multiple directional 
noise sources, whose number is less than the number of microphones.
%%%
We derive two closed-form \acp{MLE}.
The first is a non-blocking-based estimator which jointly estimates
the speech, reverberation and noise \acp{PSD},
and the second is a blocking-based estimator, which first blocks the speech signal
and then jointly estimates the reverberation and noise \acp{PSD}.
Both estimators are analytically compared and analyzed, and \acp{MSE} expressions are derived.
Furthermore, \acp{CRB} on the estimated \acp{PSD} are derived.
%%%%%%
The proposed estimators are examined using both
simulation and real reverberant and noisy signals,
demonstrating the advantage of the proposed method compared to competing estimators.
\end{abstract}


\vspace{-2mm}
%%%%%%%%%%%%%%%%%%%%%%%%%%%%%%%%%%%%%%%%%%%%%%%%%%%%%%%
\section{Introduction}
\label{sec:intro}

In many hands-free scenarios,
the measured microphone signals are corrupted by additive background noise,
which may originate from both environmental sources and from microphone responses.
%%%%
Apart from noise, if the recording takes place in an enclosed space,
the recorded signals may also contain multiple sound reflections from walls and other objects in the room,
resulting in reverberation. 
%%%%
As the level of noise and reverberation increases, the perceived quality and intelligibility of 
the speech signal deteriorate, 
which in turn affect the performance of speech communication systems, 
as well as \ac{ASR} systems.

%%%%%%%%%%%%%%%%%%%%%%%%%%%%%%%%%%%%%%%%%%%%%%%%%%%%%%%%%%%%%%%
In order to reduce the effects of reverberation and noise, speech enhancement algorithms
are required, which aim at recovering the clean speech source from the recorded microphone signals.
%%%%
Speech dereverberation and noise reduction algorithms often require 
the power spectral densities (PSDs) of the speech, reverberation and noise components.
%%%%%%%%%%%%%%%%%%%%%
In the multichannel framework, a commonly used assumption 
(see e.g.~in \cite{kjems2012maximum,braun2013dereverberation,kuklasinski2014maximum,braun2015multichannel,schwartz2015maximum,schwartz2016joint_ICASSP,
kuklasinski2016maximum_ICASSP,schwartz2016joint_EUSIPCO,kuklasinski2016maximum,braun2018evaluation})
is that the late reverberant signal is modelled as a spatially homogeneous sound field,
with a time-varying \ac{PSD} multiplied by a spherical diffuse time-invariant spatial coherence matrix.
As the spatial coherence matrix depends only on the microphone geometry, it can be calculated in advance.
However, the reverberation \ac{PSD} is an unknown parameter that should be estimated.
%%%%
Numerous methods exist for estimating the reverberation \ac{PSD}.
They are broadly divided into two classes, 
namely non-blocking-based estimators and blocking-based estimators.
The non-blocking-based approach jointly estimates the \acp{PSD} of the late reverberation and speech.
The estimation is carried out using the \acf{ML} criterion \cite{kuklasinski2014maximum,schwartz2016joint_ICASSP}
or in the \ac{LS} sense, by minimizing the Frobenius norm of an error \ac{PSD} matrix \cite{schwartz2016joint_EUSIPCO}.
In the blocking-based method, the desired speech signal is first blocked using a \ac{BM},
and then the reverberation \ac{PSD} is estimated.
Estimators in this class are also based on the \ac{ML} approach \cite{kjems2012maximum,schwartz2015maximum,kuklasinski2016maximum_ICASSP,kuklasinski2016maximum}
or the \ac{LS} criteria \cite{braun2013dereverberation,braun2015multichannel}.

%%%%%%%%%%%%%%%%%%%%%%%%%%%%%%
All previously mentioned methods 
do not include an estimator for the noise \ac{PSD}.
In \cite{kjems2012maximum,kuklasinski2014maximum}, a noiseless scenario is assumed.
In \cite{braun2013dereverberation,braun2015multichannel,schwartz2015maximum,schwartz2016joint_ICASSP,
schwartz2016joint_EUSIPCO,kuklasinski2016maximum_ICASSP,kuklasinski2016maximum,braun2018evaluation},
the noise \ac{PSD} matrix is assumed to be known in advance, or that an estimate is available.
Typically, the noise \ac{PSD} matrix is assumed to be time-invariant,
and therefore can be estimated during speech-absent periods using a \acf{VAD}.
However, in practical acoustic scenarios the spectral characteristics of the noise might be time-varying,
e.g. when the noise environment includes a background radio or TV, and thus a \ac{VAD}-based algorithm may fail.
Therefore, the noise \ac{PSD} matrix has to be included in the estimation procedure.

%%%%%%%%%%%%%%%%%%%%%%
Some papers in the field deal with performance analysis of the proposed estimators.
We give a brief review of the commonly used tools to assess the quality of an estimator.
Theoretical analysis of estimators typically consists of calculating the bias and the \acf{MSE},
which coincides with the variance for unbiased estimators.
%%%%%
The \acf{CRB} is an important tool to evaluate the quality of any unbiased estimator,
since it gives a lower bound on the \ac{MSE}.
An estimator that is unbiased and attains the \ac{CRB}, is called \textit{efficient}.
%%%
The \acf{MLE} is \textit{asymptotically} efficient \cite{Kay1993},
namely attains the \ac{CRB} when the amount of samples is large.

%%%%%%%%%%
Theoretical analysis of \ac{PSD} estimators in the noise-free scenario
was addressed in \cite{jensen2015analysis,kuklasinski2015multi}.
%%%%%%%%
In~\cite{jensen2015analysis}, \acp{CRB} were derived for the reverberation
and the speech \acp{MLE} proposed in \cite{kuklasinski2014maximum}.
These \acp{MLE} are efficient, i.e.~attain the \ac{CRB} for any number of samples.
In addition, it was pointed out that the non-blocking-based reverberation \ac{MLE} derived in \cite{kuklasinski2014maximum}
is identical to the blocking-based \ac{MLE} proposed in \cite{kjems2012maximum}.
%%%
In \cite{kuklasinski2015multi}, it was shown that the non-blocking-based reverberation \ac{MLE} 
of \cite{kuklasinski2014maximum} obtains a lower \ac{MSE} compared to a noiseless version of
the blocking-based \ac{LS} estimator derived in \cite{braun2013dereverberation}.

%%%%%
In the noisy case, quality assessment was discussed in \cite{kuklasinski2016maximum}
and \cite{schwartz2017cramer}.
%%%%%
In~\cite{kuklasinski2016maximum}, it was numerically demonstrated that an iterative blocking-based \ac{MLE}
yields lower \ac{MSE} than the blocking-based \ac{LS} estimator proposed in \cite{braun2013dereverberation}. 
%%%%%
In~\cite{schwartz2017cramer}, closed-form \acp{CRB} were derived
for the two previously proposed \acp{MLE} of the reverberation \ac{PSD}, 
namely the blocking-based estimator in \cite{schwartz2015maximum}
and the non-blocking-based estimator in \cite{schwartz2016joint_ICASSP}.
The \ac{CRB} for the non-blocking-based reverberation estimator was shown to be lower than
the \ac{CRB} for the blocking-based estimator.
However, it was shown that in the noiseless case, both reverberation \acp{MLE} are identical 
and both \acp{CRB} coincide.

%%%%
As opposed to previous works, the assumption of known noise \ac{PSD} matrix is not made in \cite{kodrasi18joint}.
The noise is modelled as a spatially homogeneous sound field, 
with a time-varying \ac{PSD} multiplied by a time-invariant spatial coherence matrix.
It is assumed that the spatial coherence matrix of the noise
is known in advance, while the time-varying \ac{PSD} is unknown.
Two different estimators were developed, based on the \ac{LS} method. 
In the first one, a joint estimator for the speech, noise and late reverberation \acp{PSD} was developed.
As an alternative, a blocking-based estimator was proposed, in which the speech signal is first blocked by a \ac{BM},
and then the noise and reverberation \acp{PSD} are jointly estimated.
%%%%%%%%%%%%%%%
This method was further extended in \cite{tammen2019joint} to jointly estimate 
the various \acp{PSD} along with the \ac{RETF} vector, using an alternating least-squares method.
%%%%%
However, the model used in \cite{kodrasi18joint} and \cite{tammen2019joint}
only fits spatially homogeneous noise fields that are characterized by a full-rank covariance matrix.
Moreover, in \cite{kuklasinski2016maximum} it was claimed that the \ac{ML} approach 
is preferable over the \ac{LS} estimation procedure.

%%%
Recently, a confirmatory factor analysis was used in \cite{koutrouvelis2019robust}
to jointly estimate the \acp{PSD} and the \ac{RETF}.
The noise is modelled as a microphone-self noise, with a time-invariant diagonal \ac{PSD} matrix.
A closed-form solution is not available for this case, thus requiring iterative optimization techniques.

%%%%%%%%%%%%%%%%%
In this paper, we treat the noise \ac{PSD} matrix as an unknown parameter. 
We assume that the noise PSD matrix is a \textit{rank-deficient} matrix,
as opposed to the spatially homogeneous assumption considered in \cite{kodrasi18joint}.
%%%%%%%%%%%%%%%%%%%%%%%%%%%%%%%%%
This scenario arises when the noise signal consists of a set of directional 
noise sources, whose number is smaller than the number of microphones.
%%%%%%%%%%%%%%%%%%%%%%%%%%%%%%%%
We assume that the positions of the noise sources are fixed, 
while their spectral \ac{PSD} matrix is time-varying,
e.g.~when the acoustic environment includes radio or TV.
%%%%%
It should be emphasized that, in contrast to \cite{kodrasi18joint}
which estimates only a scalar \ac{PSD} of the noise,
in our model the entire spectral \ac{PSD} matrix of the noise is estimated,
and thus the case of multiple non-stationary noise sources, can be handled.
%%%%%%%%%%%%%%%%%%%%%%%%%%%%%%%%
We derive closed-form \acp{MLE} of the various \acp{PSD},
for both the non-blocking-based and the blocking-based methods.
The proposed estimators are analytically studied and compared, and the corresponding \acp{MSE} expressions are derived.
Furthermore, \acp{CRB} for estimating the various \acp{PSD} are derived.

%%%%%%%%%%%%%%%%%%%
An important benefit of considering the rank-deficient noise as a separated problem,
is due to the form of the solution.
%%%%%
In the \ac{ML} framework, a closed-form solution exists for
the noiseless case \cite{kjems2012maximum,kuklasinski2014maximum}
but not for the full-rank noise scenario, thus requiring 
iterative optimization techniques \cite{schwartz2015maximum,schwartz2016joint_ICASSP,kuklasinski2016maximum}
(as opposed to \ac{LS} method that has closed-form solutions in both cases).
%%%%%
However, we show here that when the noise \ac{PSD} matrix is a rank-deficient matrix,
closed-form \ac{MLE} exists, which yields simpler and faster  estimation procedure
with low computational complexity, and is not sensitive to local maxima.

%%%%%%%%%%%%%%%%%%%%%%%%%%%
The remainder of the paper is organized as follows.
Section~\ref{sec:problem} presents the problem formulation, and describes the probabilistic model.
Section~\ref{sec:ML_estimators} derives the \acp{MLE} for both the non-blocking-based and the blocking-based methods,
and Section~\ref{sec:CRB} presents the \ac{CRB} derivation.
Section~\ref{sec:Experimental_Study} demonstrates the performance of the proposed estimators by an experimental study
based on both simulated data and recorded \acp{RIR}.
The paper is concluded in Section~\ref{sec:conclusions}.

\vspace{-2mm}
%%%%%%%%%%%%%%%%%%%%%%%%%%%%%%%%%%%%%%%%%%%%%%%%%%%%%%%%%%%%%%%%%%%%%%%%%%%
\section{Problem Formulation}
\label{sec:problem}

In this section, we formulate the dereverberation and noise reduction problem.
Scalars are denoted with regular lowercase letters,
vectors are denoted with bold lowercase letters and
matrices are denoted  with bold uppercase letters.
A list of notations used in our derivations is given in Table~\ref{tab:Definitions}.

\subsection{Signal Model}
Consider a speech signal received by $N$ microphones, in a noisy and reverberant acoustic environment.
We work with the \ac{STFT} representation of the measured signals. 
Let $k\in[1,K]$ denote the frequency bin index, and $m\in[1,M]$ denote the time frame index.
The $N$-channel observation signal $\mathbf{y}(m,k)\in \mathbb{C}^{N}$ writes
\begin{equation} \label{eq:sig_model}
\mathbf{y}(m,k) = \mathbf{x}_{\text{e}}(m,k)+\mathbf{r}(m,k) + \mathbf{u}(m,k),
\end{equation}
where $\mathbf{x}_{\text{e}}(m,k)\in \mathbb{C}^{N}$ denotes the direct and early reverberation speech component,
$\mathbf{r}(m,k)\in \mathbb{C}^{N}$ denotes the late reverberation speech component
and $\mathbf{u}(m,k) \in \mathbb{C}^{N}$ denotes the noise.
The direct and early reverberation speech component is given by
$\mathbf{x}_{\text{e}}(m,k)=\mathbf{g}_{d}(k)s(m,k)$, 
where $s(m,k)\in \mathbb{C}$ is the direct and early speech component
as received by the first microphone (designated as a reference microphone),
and
$\mathbf{g}_{d}(k)=[1,g_{d,2}(k),\cdots, g_{d,N}(k)]^{\top}\in \mathbb{C}^{N}$
is the time-invariant \acf{RETF} vector between the reference microphone and all microphones.
%%%%%
In this paper, we follow previous works in the field, e.g.~\cite{braun2013dereverberation,kuklasinski2015multi,schwartz2017cramer,kodrasi18joint},
and neglect the early reflections. 
Thus, the target signal $s(m,k)$ is approximated as the direct component
at the reference microphone,
and $\mathbf{g}_{d}(k)$ reduces to the \ac{RDTF} vector.
%%%%%%%%%%%%%%%%%%%%%%%%%%%%%%%%%%%%%%%%%%%%%%%%%%%%%%
It is assumed that the noise signal consists of $T$ noise sources, i.e.
\begin{equation}\label{eq:directional_noises}
\mathbf{u}(m,k)=\vect{A}_{\mathbf{u}}(k)\mathbf{s}_{\mathbf{u}}(m,k),
\end{equation}
where $\mathbf{s}_{\mathbf{u}}(m,k) \in \mathbb{C}^{T}$ denotes the vector of noise sources
and $\vect{A}_{\mathbf{u}}(k) \in \mathbb{C}^{N \times T}$
is the noise \ac{ATF} matrix, assumed to be time-invariant.
%%%%
It is further assumed that $T \leq N-2$.

\begin{table}[t!]
\caption{Nomenclature}
\label{tab:Definitions}
\centering
\begin{tabular}{c c l}
\toprule
\toprule
Symbol && Meaning  \\
\midrule
$(\cdot)^\top$              &&  transpose \\
$(\cdot)^\hermitian$        && conjugate transpose \\ 
$\left(\cdot \right)^{*}$   && complex conjugate \\ 
$\vert\cdot\vert$           &&  determinant of a matrix \\ 
$\text{Tr}[\cdot]$          &&  trace of a matrix\\ 
$\otimes$                   &&  Kronecker product \\ 
$\text{vec}(\cdot)$         &&  stacking the columns of a matrix on top of \\
                            &&  one another \\
\midrule
$N$ && No.~of microphones, $n\in\{1,\ldots,N\}$\\
$M$ && No.~of time frames, $m\in\{1,\ldots,M\}$ \\
$K$ && No.~of frequency bins, $k\in\{1,\ldots,K\}$\\
$T$ && No.~of noise sources, $t\in\{1,\ldots,T\}$\\
\midrule
$\mathbf{y}$ && Observation signal\\
$s$ && Direct speech component \\
$\mathbf{g}_{d}$ && Relative direct-path transfer function \\
$\mathbf{r}$ && Late reverberation component \\
$\mathbf{u}$ && Noise signal \\
$\vect{A}_{\mathbf{u}}$ && Noise ATF matrix \\
$\mathbf{s}_{\mathbf{u}}$ && Vector of noise sources \\
$\phi_{S}$ && Speech PSD \\
$\phi_{R}$ && Reverberation PSD \\
$\bm{\Gamma}_R$ && Reverberation coherence matrix \\
$\bm{\Psi}_{\mathbf{u}}$ && Noise PSD matrix \\
\bottomrule
\bottomrule
\end{tabular}
\end{table}

\vspace{-2mm}
%%%%%%%%%%%%%%%%%%%%%%%%%%%%%%%%%%%%%%%%%%%%%%%%%%%%%%%
\subsection{Probabilistic Model}
The speech \ac{STFT} coefficients are assumed to follow a zero-mean complex Gaussian 
distribution\footnote{
The multivariate complex Gaussian \ac{PDF} is given by 
$\mathcal{N}_c(\mathbf{a};\bm{\mu}_a,\bm{\Phi}_{a})=
\frac{1}{\vert\pi\bm{\Phi}_{a}\vert}
\exp\big(-(\mathbf{a}-\bm{\mu}_a)^\hermitian\bm{\Phi}_{a}^{-1}(\mathbf{a}-\bm{\mu}_a)\big),
$
where $\bm{\mu}_a$ is the mean vector and $\bm{\Phi}_{a}$ is an Hermitian positive definite complex covariance matrix 
\cite{wooding1956multivariate}.
}
with a time-varying \ac{PSD} $\phi_{S}(m,k)$. 
Hence, the \ac{PDF} of the speech writes:
\begin{equation}
p\big(s(m,k);\phi_{S}(m,k)\big)=\mathcal{N}_c\big(s(m,k);0,\phi_{S}(m,k)\big).
\end{equation}
%%%%
The late reverberation signal is modelled by a zero-mean 
complex multivariate Gaussian distribution: 
\begin{equation}
p\big(\mathbf{r}(m,k); \bm{\Phi}_{\mathbf{r}}(m,k)\big)
= \mathcal{N}_c\big(\mathbf{r}(m,k);\bm{0},\bm{\Phi}_{\mathbf{r}}(m,k)\big).
\end{equation}
The reverberation \ac{PSD} matrix is modelled as a spatially homogeneous and
isotropic sound field, with a time-varying \ac{PSD}, 
$\bm{\Phi}_{\mathbf{r}}(m,k)=\phi_{R}(m,k)\bm{\Gamma}_R(k)$.
It is assumed that the time-invariant coherence matrix $\bm{\Gamma}_R(k)$
can be modelled by a spherically diffuse sound field \cite{cron1962spatial}:
\begin{equation}\label{eq:diffuse}
\Gamma_{R,ij}(k)=\text{sinc}\left(\frac{2\pi f_{s}k}{K}\frac{d_{ij}}{c}\right),
\end{equation}
where $\text{sinc}(x) = \sin(x)/x$,
$d_{ij}$ is the inter-distance between microphones $i$ and $j$,
$f_s$ denotes the sampling frequency
and $c$ is the sound velocity.

%%%%%%%%%%%%%%%%%%%%%%%%%%%%%%%%
The noise sources vector is modelled by a zero-mean complex multivariate Gaussian distribution
with a time-varying \ac{PSD} matrix $\bm{\Psi}_{\mathbf{u}}(m,k)\in \mathbb{C}^{T \times T}$:
\begin{align}\label{eq:su}
\hspace{-0.025cm}
p\big(\mathbf{s}_{\mathbf{u}}(m,k); \bm{\Psi}_{\mathbf{u}}(m,k)\big)
= \mathcal{N}_c\big(\mathbf{s}_{\mathbf{u}}(m,k);\bm{0},\bm{\Psi}_{\mathbf{u}}(m,k)\big).
\end{align}
%%%%%%%%%%%%%%%%%%%%%%%%%%%%%%%%
Using \eqref{eq:directional_noises} and \eqref{eq:su},
it follows that $\mathbf{u}(m,k)$ has a zero-mean complex multivariate Gaussian distribution
with a \ac{PSD} matrix
$\bm{\Phi}_{\mathbf{u}}(m,k) \in \mathbb{C}^{N \times N}$,
given by
\begin{equation}\label{eq:Phi_v_0}
\bm{\Phi}_{\mathbf{u}}(m,k)=\vect{A}_{\mathbf{u}}(k)\bm{\Psi}_{\mathbf{u}}(m,k) \vect{A}_{\mathbf{u}}^{\hermitian}(k).
\end{equation}
Note that
$\text{rank}(\bm{\Phi}_{\mathbf{u}})=\text{rank}(\bm{\Psi}_{\mathbf{u}})=T\leq N-2$,
i.e.~$\bm{\Phi}_{\mathbf{u}}$ is a rank-deficient matrix.
%%%%%%%%%%%%%%%%%%%%%%%%%%%%%%%%%%%%%%%%%%%%%%%%%%%%%
The \ac{PDF} of $\mathbf{y}(m,k)$ writes
\begin{equation}
p\big(\mathbf{y}(m,k); \bm{\Phi}_{\mathbf{y}}(m,k)\big)
= \mathcal{N}_c\big(\mathbf{y}(m,k);\bm{0},\bm{\Phi}_{\mathbf{y}}(m,k)\big),
\end{equation}
where $\bm{\Phi}_{\mathbf{y}}$ is the \ac{PSD} matrix of the input signals.
Assuming that the components in \eqref{eq:sig_model} are independent,
$\bm{\Phi}_{\mathbf{y}}$ is given by
\begin{align}\label{eq:Phi_y_rank_deff}
  \bm{\Phi}_{\mathbf{y}}(m,k)  
  &= \phi_{S}(m,k) \mathbf{g}_{d}(k) \mathbf{g}_{d}^\hermitian(k) 
  + \phi_{R}(m,k) \, \bm{\Gamma}_{R}(k)
  \nonumber \\
  & \quad 
%+\vect{V}(k)\bm{\Lambda}(m,k)\vect{V}^{\hermitian}(k).
+\vect{A}_{\mathbf{u}}(k)\bm{\Psi}_{\mathbf{u}}(m,k) \vect{A}_{\mathbf{u}}^{\hermitian}(k).
\end{align}

%%%%%%%%%%%%%%%%%%%%%%%%%%%%%%%%
A commonly used dereverberation and noise reduction technique
is to estimate the speech signal using the multichannel \ac{MMSE} estimator,
which yields the \ac{MCWF}, given by \cite{van2002optimum}:
\begin{equation}\label{eq:MCWF}
\hat{s}_{\text{MCWF}}(m,k) 
=\frac{\mathbf{g}_{d}^\hermitian(k)\bm{\Phi}_{i}^{-1}(m,k) }
{\mathbf{g}_{d}^\hermitian(k)\bm{\Phi}_{i}^{-1}(m,k) \mathbf{g}_{d}(k)
+\phi_{S}^{-1}(m,k) }
\mathbf{y}(m,k),
\end{equation}
where
\begin{equation}\label{eq:psi}
  \bm{\Phi}_{i}(m,k)  \triangleq \phi_{R}(m,k)  \, \vect{\Gamma}_R(k) 
+\vect{A}_{\mathbf{u}}(k)\bm{\Psi}_{\mathbf{u}}(m,k) \vect{A}_{\mathbf{u}}^{\hermitian}(k)
\end{equation}
denotes the total interference \ac{PSD} matrix.
%%%%
For implementing \eqref{eq:MCWF}, we assume that the \ac{RDTF} vector $\mathbf{g}_{d}$
and the spatial coherence matrix $\bm{\Gamma}_R$ are known in advance.
%%%
The \ac{RDTF} depends only on the \ac{DOA} of the speaker and the geometry of the microphone array,
and thus it can be constructed based on a \ac{DOA} estimate.
The spatial coherence matrix is calculated using \eqref{eq:diffuse},
based on the spherical diffuseness assumption.
%%%%%%%%%%%%%%%%%%%%%%%
It follows that estimators of the late reverberation, speech and noise \acp{PSD} 
are required for evaluating the \ac{MCWF}.

%%%%
It should be noted that estimating directly the complete $N \times N$ time-varying noise \ac{PSD} matrix $\vect{\Phi}_{\mathbf{u}}(m,k)$,
along with the speech and reverberation \acp{PSD}, is a complex problem, and 
a closed-form \ac{ML} solution is not available.
%%%
In this paper, we rely on the decomposition of the noise \ac{PSD} matrix
in \eqref{eq:Phi_v_0}, along with the rank-deficiency assumption.
%%%
By utilizing the time-invariant \ac{ATF} matrix $\vect{A}_\mathbf{u}(k)$,
we can construct a projection matrix onto the subspace orthogonal to the noise subspace, 
which generates nulls towards the noise sources. 
In the following section, we show that this step enables the derivation of closed-form \acp{MLE} for the various \acp{PSD}.

%%%%%%%%% 
The noise ATF matrix $\mathbf{A}_{\mathbf{u}}(k)$ is in general not available, 
since the estimation of the individual \acp{ATF} requires that each noise is active separately.
Instead, we only assume the availability of a speech-absent segment, denoted by $m_0$, in which all noise sources are concurrently active. 
Based on this segment, we compute a basis for the noise subspace that can be used instead of the unknown \ac{ATF} matrix. 
To this end, we compute the noise \ac{PSD} matrix $\vect{\Phi}_{\mathbf{u}}(m_0,k)$, and apply the \ac{EVD} to the resulting matrix.
Recall that the rank of $\vect{\Phi}_{\mathbf{u}}$ is $T$. Accordingly, 
a $T$-rank representation of the noise \ac{PSD} matrix is formed by the computed eigenvalues and eigenvectors:
%%%%%%%%%%%%%%%
\begin{equation}\label{eq:EVD1}
\bm{\Phi}_{\mathbf{u}}(m_0,k)
%=\sum_{t=1}^{T}\lambda_t(m_0,k)\mathbf{v}_t(k)\mathbf{v}_t^\hermitian(k)
=\vect{V}(k)\bm{\Lambda}(m_0,k)\vect{V}^{\hermitian}(k),
\end{equation}
where 
$\bm{\Lambda}(m_0,k)\in \mathbb{C}^{T \times T}$ is the eigenvalues matrix
(comprised of the non-zero eigenvalues)
and 
$
\vect{V}(k)=\left[\mathbf{v}_1(k),\cdots, \mathbf{v}_T(k) \right]
\in \mathbb{C}^{N \times T}
$
is the corresponding eigenvectors matrix.
$\vect{V}(k)$ is a basis that spans the noise \acp{ATF} subspace,
and thus \cite{markovich2009multichannel}
\begin{align}\label{eq:span}
\vect{A}_{\mathbf{u}}(k)=\vect{V}(k)\vect{G}(k),
\end{align}
where $\vect{G}(k) \in \mathbb{C}^{T \times T}$ 
consists of projection coefficients of the original \acp{ATF} on the basis vectors.
%%%%%%%%%%
Substituting \eqref{eq:span} into \eqref{eq:directional_noises} and then into \eqref{eq:sig_model}, 
yields
\begin{equation} \label{eq:sig_model_V}
\mathbf{y}(m,k) = \mathbf{g}_{d}(k)s(m,k)+\mathbf{r}(m,k) + \vect{V}(k)\mathbf{s}_{\mathbf{v}}(m,k),
\end{equation}
where 
$\mathbf{s}_{\mathbf{v}}(m,k)\triangleq\vect{G}(k)\mathbf{s}_{\mathbf{u}}(m,k)$.
%%%
It follows that the noise \ac{PSD} matrix in \eqref{eq:Phi_v_0} can be recast as
\begin{align}\label{eq:phi_u_V}
\bm{\Phi}_{\mathbf{u}}(m,k)
=\vect{V}(k)
\bm{\Psi}_{\mathbf{v}}(m,k)
\vect{V}^{\hermitian}(k),
\end{align}
where 
$\bm{\Psi}_{\mathbf{v}}(m,k)=\vect{G}(k)\bm{\Psi}_{\mathbf{u}}(m,k) \vect{G}^{\hermitian}(k)$.
%%%%%%%%%%%%%%%
Using this basis change, the \ac{MCWF} in \eqref{eq:MCWF} is now computed with
\begin{equation}\label{eq:psi_2}
\bm{\Phi}_{i}(m,k)  
=
\phi_{R}(m,k)  \, \vect{\Gamma}_R(k) 
+
%\vect{A}_{\mathbf{u}}(k)\bm{\Psi}_{\mathbf{u}}(m,k) \vect{A}_{\mathbf{u}}^{\hermitian}(k)
\vect{V}(k)\bm{\Psi}_{\mathbf{v}}(m,k)\vect{V}^{\hermitian}(k).
\end{equation}
%%%%%%%%%%%%%%%
As a result, rather than requiring the knowledge of the exact noise \ac{ATF} matrix,
we use $\vect{V}$ that is learned from a speech-absent segment.
Due to this basis change, we will need to estimate $\bm{\Psi}_{\mathbf{v}}$
instead of $\bm{\Psi}_{\mathbf{u}}$.

%%%%%%%%%%%%%%%%%%%%%%%%%%%%%%%%%
To summarize, estimators of $\phi_R$, $\phi_S$ and  $\bm{\Psi}_{\mathbf{v}}$ are required.
%%%%
For the sake of brevity, the frame index $m$ and the frequency bin index $k$ 
are henceforth omitted whenever possible.

\vspace{-2mm}
%%%%%%%%%%%%%%%%%%%%%%%%%%%%%%%%%%%%%%%%%%%%%%%%%%%%%%%%%%%%%%%%%%%%%%%%%%%%%%
\section{\ac{ML} Estimators}
\label{sec:ML_estimators}

We propose two \ac{ML}-based methods: 
(i) Non-blocking-based estimation: Simultaneous \ac{ML} estimation of the reverberation, speech and noise \acp{PSD};
and 
(ii) Blocking-based estimation: Elimination of the speech \ac{PSD} using a \acf{BM}, and then joint \ac{ML} estimation of the reverberation and noise \acp{PSD}.
Both methods are then compared and analyzed.

\vspace{-2mm}
\subsection{Non-Blocking-Based Estimation}  \label{sec:NON_BLOCKING_ML}
We start with the joint \ac{ML} estimation of the reverberation, speech and noise \acp{PSD}.
Based on the short-time stationarity assumption \cite{jensen2015analysis,kuklasinski2016maximum},
it is assumed that the \acp{PSD} are approximately constant
across small number of consecutive time frames, denoted by $L$.
%%%%
We therefore denote 
$\bar{\mathbf{y}}(m)\in \mathbb{C}^{LN}$ 
as the concatenation of $L$ successive observations of $\mathbf{y}(m)$:
\begin{equation}\label{eq:y_bar}
  \bar{\mathbf{y}}(m)\triangleq 
  \left[\mathbf{y}^\top(m-L+1), \cdots , \mathbf{y}^\top(m) \right]^\top.
\end{equation}
The set of unknown parameters is denoted by
$\bm{\phi}(m)= \left[\phi_{R}(m),\phi_{S}(m),\bm{\psi}_{V}^\top(m)\right]^\top$,
where 
$\bm{\psi}_{V}$ is a vector containing all elements of $\bm{\Psi}_{\mathbf{v}}$, 
i.e.~$\bm{\psi}_{V}=\text{vec}\left(\bm{\Psi}_{\mathbf{v}}\right)$.
%%%%%%%%%%%
Assuming that the $L$ consecutive signals in $\bar{\mathbf{y}}$ are i.i.d.,
the \ac{PDF} of $\bar{\mathbf{y}}$ writes (see e.g.~\cite{schwartz2017cramer}):
\begin{flalign}\label{eq:likelihood_y}
& p\big(\bar{\mathbf{y}}(m);\bm{\phi}(m) \big)
 \nonumber &&\\
&\quad =  \left( \frac{1}{\pi^{N} |\bm{\Phi}_{\mathbf{y}}(m)|}
 \exp \Big(- \Tr \left[ \bm{\Phi}^{-1}_{\mathbf{y}}(m)  \vect{R}_{\mathbf{y}}(m) \right]   \Big)\right)^L,&&
 \raisetag{1.3\normalbaselineskip}
 \end{flalign}
where $\vect{R}_{\mathbf{y}}$ is the sample covariance matrix, given by
\begin{equation}\label{eq:smooth0}
    \vect{R}_{\mathbf{y}}(m) = \frac{1}{L} \sum^{m}_{\ell=m-L+1} \mathbf{y}(\ell) \mathbf{y}^\hermitian(\ell).
\end{equation}
%%%%%%%%%%%%%%%%%%%%%%
The \ac{MLE} of the set $\bm{\phi}(m)$ is therefore given by
\begin{equation} \label{eq:ML_observed}
 \bm{\phi}^{\textrm{ML},\bar{\mathbf{y}}}(m) = 
 \argmax_{\bm{\phi}(m)} 
 \log p\big(\bar{\mathbf{y}}(m);\bm{\phi}(m) \big).
\end{equation}
%%%%%
To the best of our knowledge, for the general noisy scenario
this problem is considered as having no closed-form solution.
However, we will show that when the noise \ac{PSD} matrix $\bm{\Phi}_{\mathbf{u}}$ is rank-deficient,
with $T =\text{rank}(\bm{\Phi}_{\mathbf{u}})\leq N-2$, a closed-form solution exists.
%%%
In the following, we present the proposed estimators.
The detailed derivations appear in the Appendices.

%%%%%%%%%%%%%%%%%%%%%%%%%%
In Appendix~\ref{app:Phi_r_No_BM}, 
it is shown that the \ac{MLE} of $\phi_{R}(m)$ is given by:
\begin{align}\label{eq:phi_R_closed-form}
\phi^{\mathrm{ML},\bar{\mathbf{y}}}_{R}(m)
=\frac{1}{N-(T+1)}
\Tr \big[\vect{Q} \vect{R}_{\mathbf{y}}(m)\vect{\Gamma}_{R}^{-1}  \big],
\end{align}
where $\vect{Q} \in \mathbb{C}^{N  \times N}$ is given by 
\begin{align}\label{eq:Q1}
\vect{Q}=\vect{I}_{N}-\vect{A}\left(\vect{A}^\hermitian\bm{\Gamma}_R^{-1}\vect{A} \right)^{-1}\vect{A}^\hermitian\bm{\Gamma}_R^{-1},
\end{align}
and $\vect{A} \in \mathbb{C}^{N  \times (T+1)}$
is the \textit{speech-plus-noise} subspace
\begin{align}\label{eq:A_def}
\vect{A}=\left[\mathbf{g}_{d},\mathbf{v}_1,\cdots, \mathbf{v}_T \right].
\end{align}
The matrix $\vect{Q}$ is a projection matrix onto the subspace orthogonal
to the speech-plus-noise subspace.
The role of $\vect{Q}$ is to block the directions of the desired speech and noise signals, in order to estimate the reverberation level.

%%%%%%%%%%%%%%%%%%%%%%%%%%%%%%%%%%%%%%%
Once we obtain the \ac{MLE} for the late reverberation \ac{PSD},
the \acp{MLE} for the speech and noise \acp{PSD} can be computed.
In Appendix~\ref{app:Phi_S_No_BM}, it is shown that the \ac{MLE} for the speech \ac{PSD} writes
\begin{align}\label{eq:phi_S_closed-form}
\phi^{\mathrm{ML},\bar{\mathbf{y}}}_{S}(m)
=\mathbf{w}_{s}^\hermitian
\left(\vect{R}_{\mathbf{y}}(m)-\phi^{\mathrm{ML},\bar{\mathbf{y}}}_{R}(m)\bm{\Gamma}_R  \right)
\mathbf{w}_{s},
\end{align}
where $\mathbf{w}_{s} \in \mathbb{C}^{N}$ is a \ac{MVDR} beamformer that extracts
the speech signal while eliminating the noise, given by  
\begin{equation}\label{eq:MVDR_S}
\mathbf{w}_{s}^\hermitian
=\frac{\mathbf{g}_{d}^\hermitian\bm{P}_{v}^{\perp}\bm{\Gamma}_R^{-1}}
{\mathbf{g}_{d}^\hermitian\bm{P}_{v}^{\perp}\bm{\Gamma}_R^{-1}\mathbf{g}_{d}},
\end{equation}
and $\bm{P}_{v}^{\perp}\in \mathbb{C}^{N  \times N}$ is a projection matrix 
onto the subspace orthogonal to the noise subspace, given by
\begin{equation}
\bm{P}_{v}^{\perp}=\vect{I}_{N}-\bm{\Gamma}_R^{-1}\vect{V}\left(\vect{V}^\hermitian\bm{\Gamma}_R^{-1}\vect{V} \right) ^{-1}\vect{V}^\hermitian.
\end{equation}
Note that
\begin{align}\label{eq:w_mvdr_dot_gd_V}
\mathbf{w}_{s}^\hermitian \mathbf{g}_{d}= 1
\hspace{0.1cm},\hspace{0.1cm}
\mathbf{w}_{s}^\hermitian \vect{V} = \vect{0}.
\end{align}
The estimator in \eqref{eq:phi_S_closed-form} can be interpreted as the variance of the noisy observations
minus the estimated variance of the reverberation, at the output of the 
\ac{MVDR} beamformer \cite{jensen2015analysis}.

%%%%%%%
In Appendix~\ref{app:Phi_v_No_BM}, it is shown that the \ac{MLE} of the noise \ac{PSD} can be computed with
\begin{align}\label{eq:phi_v1}
\bm{\Psi}_{\mathbf{v}}^{\mathrm{ML},\bar{\mathbf{y}}}(m)
=
\vect{W}_{\mathbf{u}}^\hermitian
\left(\vect{R}_{\mathbf{y}}(m)-\phi^{\mathrm{ML},\bar{\mathbf{y}}}_{R}(m)\bm{\Gamma}_R  \right)
\vect{W}_{\mathbf{u}},
\end{align}
where $\vect{W}_{\mathbf{u}} \in \mathbb{C}^{N  \times T}$ is a multi-source \ac{LCMV} beamformer
that extracts the noise signals while eliminating the speech signal:
\begin{equation}\label{eq:LCMV_V}
\vect{W}_{\mathbf{u}}^\hermitian= 
\left(\vect{V}^\hermitian\bm{P}_{g}^{\perp}\bm{\Gamma}_R^{-1}\vect{V} \right)^{-1}
\vect{V}^\hermitian\bm{P}_{g}^{\perp}\bm{\Gamma}_R^{-1},
\end{equation}
and $\bm{P}_{g}^{\perp}\in \mathbb{C}^{N  \times N}$ is a projection matrix 
onto the subspace orthogonal to the speech subspace, given by
\begin{equation}\label{eq:Pg+}
\bm{P}_{g}^{\perp}=\vect{I}_{N}-\frac{\bm{\Gamma}_R^{-1}\mathbf{g}_{d}\mathbf{g}_{d}^\hermitian}
{\mathbf{g}_{d}^\hermitian\bm{\Gamma}_R^{-1}\mathbf{g}_{d}}.
\end{equation}
Note that
\begin{align}\label{eq:w_lcmv_dot_gd,V}
\vect{W}_{\mathbf{u}}^\hermitian \mathbf{g}_{d}= \bm{0}
\hspace{0.1cm},\hspace{0.1cm}
\vect{W}_{\mathbf{u}}^\hermitian \vect{V}= \vect{I}_T.
\end{align}
%%%%%%%%%%%%%%%%%%%%%%%
Interestingly, the projection matrix $\vect{Q}$ can be recast as a linear combination of the above beamformers
(see Appendix~\ref{app:Q_simplified}):
\begin{align}\label{eq:Q3}
\vect{Q}=\vect{I}_{N}
-\mathbf{g}_{d} \mathbf{w}_{s}^\hermitian
-\vect{V}\vect{W}_{\mathbf{u}}^\hermitian .
\end{align}
%%%%%%%%%%%%%
Using \eqref{eq:w_mvdr_dot_gd_V}, \eqref{eq:w_lcmv_dot_gd,V} and \eqref{eq:Q3},
it can also be noted that $\vect{Q}$ is orthogonal to both beamformers
\begin{align}\label{eq:w_mvdr_s_Q_w_lcmv_v_Q}
\mathbf{w}_{s}^\hermitian\vect{Q}=\mathbf{0}
\hspace{0.1cm},\hspace{0.1cm}
\vect{W}_{\mathbf{u}}^\hermitian\vect{Q}
=\mathbf{0}.
\end{align}
%%%%%%
In the noiseless case, i.e.~when $T=0$, $\vect{Q}$ reduces to
\begin{align}\label{eq:Q_noiseless}
\vect{Q}
=\vect{I}_{N}- 
\mathbf{g}_{d}
\mathbf{w}_{s_0}^\hermitian
=\left(\bm{P}_{g}^{\perp}\right)^\hermitian,
\end{align}
where $\mathbf{w}_{s_0}^\hermitian=\frac{\mathbf{g}_{d}^\hermitian\bm{\Gamma}_R^{-1}}
{\mathbf{g}_{d}^\hermitian\bm{\Gamma}_R^{-1}\mathbf{g}_{d}}$,
leading to the same closed-form estimators as in \cite[Eq. (7)]{kuklasinski2014maximum}:
\begin{align}\label{eq:phi_R_closed-form_noiseless}
\phi^{\mathrm{ML},\bar{\mathbf{y}}}_{R}(m)
&=\frac{1}{N-1}
\Tr 
\left[
\left(\vect{I}_{N}- 
\mathbf{g}_{d}
\mathbf{w}_{s_0}^\hermitian
\right) 
\vect{R}_{\mathbf{y}}(m) \vect{\Gamma}_{R}^{-1} 
\right],
\\
\phi^{\mathrm{ML},\bar{\mathbf{y}}}_{S}(m)
&=\mathbf{w}_{s_0}^\hermitian
\left(\vect{R}_{\mathbf{y}}(m)-\phi^{\mathrm{ML},\bar{\mathbf{y}}}_{R}(m)\bm{\Gamma}_R  \right)
\mathbf{w}_{s_0}.
\end{align}

\vspace{-2mm}
%%%%%%%%%%%%%%%%%%%%%%%%%%%%%%%%%%%%%%%%%%%%%%%%%%%%%%%%%%%%%%%%%%%%%%%%%%
\subsection{Blocking-Based Estimation} \label{sec:Blocking_ML}
As a second approach, we first block the speech component using a \ac{BM}, 
and then jointly estimate the \acp{PSD} of the reverberation and noise.
Let $\vect{B} \in \mathbb{C}^{N \times (N-1)}$ denote the \ac{BM}, which satisfies 
$\vect{B}^\hermitian \mathbf{g}_{d}=\mathbf{0}$. 
The output of the \ac{BM} is given by
\begin{align}\label{}
 \mathbf{z}(m) \triangleq \vect{B}^\hermitian\, \mathbf{y}(m)
 	= \vect{B}^\hermitian \big( \vect{r}(m) + \vect{u}(m) \big).
\end{align}
The \ac{PDF} of $\mathbf{z}(m)\in \mathbb{C}^{N-1}$ therefore writes:
\begin{equation}\label{eq:u_pdf}
p\big(\mathbf{z}(m); \bm{\Phi}_{\mathbf{z}}(m)\big)
= \mathcal{N}_c\big(\mathbf{z}(m);\bm{0},\bm{\Phi}_{\mathbf{z}}(m)\big),
\end{equation}
with the \ac{PSD} matrix $\bm{\Phi}_{\mathbf{z}}(m)\in \mathbb{C}^{(N-1) \times (N-1)}$
given by
\begin{equation}\label{eq:Phi_u}
  \bm{\Phi}_{\mathbf{z}}(m) = 
  \vect{B}^\hermitian 
%  \left(\phi_{R} \, \vect{\Gamma}_R+   \phi_{V} \, \vect{\Gamma}_V     \right) 
\bm{\Phi}_{i}(m)
  \vect{B},
\end{equation}
where $\bm{\Phi}_{i}$ is the total interference matrix, defined in
%\eqref{eq:psi}
\eqref{eq:psi_2}.
%%%%%
Multiplying \eqref{eq:phi_u_V} from left by $\vect{B}^\hermitian$ and from right by $\vect{B}$,
the noise \ac{PSD} matrix at the output of the \ac{BM} writes
\begin{equation}\label{eq:EVD2}
%\vect{B}^\hermitian\bm{\Gamma}_{V}\vect{B}
\vect{B}^\hermitian\bm{\Phi}_{\mathbf{u}}(m)\vect{B}
%=\sum_{t=1}^{T}\lambda_t(m)\tilde{\mathbf{v}}_t\tilde{\mathbf{v}}_t^\hermitian
=\widetilde{\vect{V}}\bm{\Psi}_{\mathbf{v}}(m)\widetilde{\vect{V}}^{\hermitian},
\end{equation}
where 
$\widetilde{\vect{V}} \in \mathbb{C}^{(N-1)  \times T}$
is the \textit{reduced} noise subspace:
\begin{align}\label{eq:V_til_BV}
\widetilde{\vect{V}}
\triangleq\vect{B}^\hermitian \vect{V}
=\left[\tilde{\mathbf{v}}_1,\cdots, \tilde{\mathbf{v}}_T \right].
\end{align}
%%%
Substituting \eqref{eq:psi_2} into \eqref{eq:Phi_u} and using \eqref{eq:V_til_BV},
yields
\begin{equation}\label{eq:Phi_u2}
  \bm{\Phi}_{\mathbf{z}}(m) 
=   \phi_{R}(m) \vect{B}^\hermitian \vect{\Gamma}_R \vect{B}
% + \vect{B}^\hermitian\bm{\Phi}_{\mathbf{u}}(m)\vect{B}
+\widetilde{\vect{V}}\bm{\Psi}_{\mathbf{v}}(m)\widetilde{\vect{V}}^{\hermitian}.
\end{equation}
%%%%%%%%%%%%%%%%%%%%%
Under this model, the parameter set of interest is 
$\tilde{\bm{\phi}}(m)= \left[\phi_{R}(m),\bm{\psi}_{V}^\top(m)\right]^\top$. 
%%%
Let $\bar{\mathbf{z}}\in \mathbb{C}^{L(N-1)}$ be defined similarly to $\bar{\mathbf{y}}$ in \eqref{eq:y_bar},
as a concatenation of $L$ i.i.d.~consecutive frames.
Similarly to $\bm{\Phi}_{\mathbf{y}}(m)$, it is assumed that $\bm{\Phi}_{\mathbf{z}}(m)$ is fixed during the entire segment.
The \ac{PDF} of $\bar{\mathbf{z}}$ therefore writes
\begin{flalign}\label{eq:blocked_pdf}
& p\big(\bar{\mathbf{z}}(m);\tilde{\bm{\phi}}(m)\big)
 \nonumber &&\\
&\hspace{0.0cm} =  \left( \frac{1}{\pi^{N-1} |\bm{\Phi}_{\mathbf{z}}(m)|}
 \exp \Big(- \Tr\left[ \bm{\Phi}^{-1}_{\mathbf{z}}(m)  \vect{R}_{\mathbf{z}}(m) \right] \Big)\right)^L,&&
 \raisetag{1.3\normalbaselineskip}
 \end{flalign}
where 
$\vect{R}_{\mathbf{z}}(m)$
is given by
 \begin{equation}\label{eq:smooth_u}
      \vect{R}_{\mathbf{z}}(m) = \frac{1}{L}\sum^{m}_{\ell=m-L+1}  \mathbf{z} (\ell)\mathbf{z}^\hermitian(\ell)= \vect{B}^\hermitian \vect{R}_{\mathbf{y}}(m) \vect{B}.
 \end{equation}
The \ac{MLE} of $\tilde{\bm{\phi}}(m)$ is obtained by solving:
\begin{equation} \label{eq:ML_blocked}
 \tilde{\bm{\phi}}^{\textrm{ML},\bar{\mathbf{z}}}(m)   
 = \argmax_{\tilde{\bm{\phi}}(m)} \log p\big(\bar{\mathbf{z}}(m);\, \tilde{\bm{\phi}}(m) \big).
\end{equation}
%%%%%%%%%%%%%%%%%%%%%%%%%%%%%%%%%%%%%%%%%%%%%%%%%%%%%%%%
To the best of our knowledge, this problem is also considered as having no closed-form solution.
Again, we argue that if the noise \ac{PSD} matrix satisfies
$T =\text{rank}(\bm{\Phi}_{\mathbf{u}})\leq N-2$,
then we can obtain a closed-form solution.
%%%%%%%%%%%%%%%%%
In Appendix~\ref{app:Phi_r_BM}, the following \ac{MLE} is obtained:
\begin{align}\label{eq:phi_R_BM_closed-form}
\phi^{\mathrm{ML},\bar{\mathbf{z}}}_{R}(m)
=\frac{1}{N-1-T}
\Tr \left[\widetilde{\vect{Q}} \vect{R}_{\mathbf{z}}(m)\left(\vect{B}^\hermitian\bm{\Gamma}_{R}\vect{B} \right)^{-1}  \right],
\end{align}
where $\widetilde{\vect{Q}} \in \mathbb{C}^{(N-1)  \times (N-1)}$ is given by 
\begin{align}\label{eq:Q_BM_1}
\widetilde{\vect{Q}}=\vect{I}_{N-1}-\widetilde{\vect{V}}\left(\widetilde{\vect{V}}^\hermitian\left(\vect{B}^\hermitian\bm{\Gamma}_{R}\vect{B} \right)^{-1}\widetilde{\vect{V}} \right)^{-1}\widetilde{\vect{V}}^\hermitian\left(\vect{B}^\hermitian\bm{\Gamma}_{R}\vect{B} \right)^{-1}.
\end{align}
After the \ac{BM} was applied, the remaining role of $\widetilde{\vect{Q}}$ 
is to block the noise signals, in order to estimate the reverberation level.
Note that 
$\Tr \left[\vect{Q}\right ]=\Tr \left[\widetilde{\vect{Q}} \right ]= N-1-T$.

%%%%%%%%%%%%%%%%%%%%%
Given $\phi^{\mathrm{ML},\bar{\mathbf{z}}}_{R}$,
it is shown in Appendix~\ref{app:Phi_v_BM} that the \ac{MLE} for the noise \ac{PSD} writes
\begin{align}\label{eq:phi_v_BM1}
\bm{\Psi}_{\mathbf{v}}^{\mathrm{ML},\bar{\mathbf{z}}}(m)
%\hspace{0.2cm}
=\widetilde{\vect{W}}_{\mathbf{u}}^\hermitian
\left(\vect{R}_{\mathbf{z}}(m)-\phi^{\mathrm{ML},\bar{\mathbf{z}}}_{R}(m)
\left(\vect{B}^\hermitian\bm{\Gamma}_R \vect{B}\right) \right)
\widetilde{\vect{W}}_{\mathbf{u}},
\end{align}
where $\widetilde{\vect{W}}_{\mathbf{u}} \in \mathbb{C}^{(N-1)  \times T}$ is
a multi-source \ac{LCMV} beamformer, directed towards the noise signals after the \ac{BM},
given by  
\begin{equation}\label{eq:LCMV_BM}
\widetilde{\vect{W}}_{\mathbf{u}}^\hermitian= 
\left(\widetilde{\vect{V}}^\hermitian
\left(\vect{B}^\hermitian\bm{\Gamma}_R \vect{B}\right)^{-1}\widetilde{\vect{V}} \right)^{-1}
\widetilde{\vect{V}}^\hermitian
\left(\vect{B}^\hermitian\bm{\Gamma}_R \vect{B}\right)^{-1}.
\end{equation}
%%%%%%%%%%%%%%%%%%%%
Note that with this notation, $\widetilde{\vect{Q}}$ in \eqref{eq:Q_BM_1}
can be recast as
\begin{align}\label{eq:Q_BM2}
\widetilde{\vect{Q}}=\vect{I}_{N-1}
-\widetilde{\vect{V}}\widetilde{\vect{W}}_{\mathbf{u}}^\hermitian.
\end{align}
%%%%
Since
$\widetilde{\vect{W}}_{\mathbf{u}}^\hermitian \widetilde{\vect{V}} = \vect{I}_T$,
it also follows that
\begin{align}\label{eq:w_tilde_lcmv_v_Q_tilde}
\widetilde{\vect{W}}_{\mathbf{u}}^\hermitian\widetilde{\vect{Q}}
=\mathbf{0}.
\end{align}
%%%%
Also, in Appendix~\ref{app:w_B_relation} it is shown that
\begin{equation}\label{eq:w_relation}
\vect{W}_{\mathbf{u}}=\vect{B}\widetilde{\vect{W}}_{\mathbf{u}},
\end{equation}
namely the \ac{LCMV} of \eqref{eq:LCMV_V}, used in the non-blocking-based approach,
can be factorized into two stages:
The first is a \ac{BM} that blocks the speech signal, 
followed by a modified \ac{LCMV}, which recovers the noise signals at the output of the \ac{BM}.

\vspace{-2mm}
%%%%%%%%%%%%%%%%%%%%%%%%%%%%%%%%%%%%%%%%%%%%%%%%%%%%%%%%
\subsection{Comparing the \acp{MLE}}
\label{sec:comparison}
In this section, the obtained blocking-based and non-blocking-based \acp{MLE} are compared. 
%%%%%%%%%
We will use the following identity, that is proved in \cite[Appendix A]{schwartz2017cramer}:
\begin{equation}   \label{eq:Identity_Phi_R}
 \vect{B}   \left(  \vect{B}^\hermitian \vect{\Gamma}_R \vect{B}  \right)^{-1}  \vect{B}^\hermitian 
= 
 \vect{\Gamma}_R^{-1} - \frac{\vect{\Gamma}_R^{-1} \mathbf{g}_{d} \mathbf{g}_{d}^\hermitian\vect{\Gamma}_R^{-1} }{ \mathbf{g}_{d}^\hermitian\vect{\Gamma}_R^{-1} \mathbf{g}_{d} }.
\end{equation}
Substituting \eqref{eq:Pg+} into \eqref{eq:Identity_Phi_R} yields
\begin{equation}   \label{eq:Identity_Phi_R2}
 \vect{B}   \left(  \vect{B}^\hermitian \vect{\Gamma}_R \vect{B}  \right)^{-1}  \vect{B}^\hermitian 
= \bm{P}_{g}^{\perp}\vect{\Gamma}_R^{-1}.
\end{equation}

%%%%%%%%%%%%%%%%%%%%%%%%
\subsubsection{Comparing the reverberation \ac{PSD} estimators}
First, we compare the reverberation \ac{PSD} estimators in 
\eqref{eq:phi_R_closed-form} and \eqref{eq:phi_R_BM_closed-form}.
%%%%%%%%%
Substituting \eqref{eq:Q_BM2} into \eqref{eq:phi_R_BM_closed-form}
and then using \eqref{eq:V_til_BV}, \eqref{eq:smooth_u}, \eqref{eq:w_relation}
and \eqref{eq:Identity_Phi_R}, yields the following equation:
\begin{align}\label{eq:phi_R_BM_closed-form2}
&\phi^{\mathrm{ML},\bar{\mathbf{z}}}_{R}
=\frac{1}{N-1-T}
\nonumber \\
&\times
\Tr \left[
\left(\vect{I}_{N}-\frac{\mathbf{g}_{d}\mathbf{g}_{d}^\hermitian \bm{\Gamma}_R^{-1}}
{\mathbf{g}_{d}^\hermitian \bm{\Gamma}_R^{-1}\mathbf{g}_{d}} \right)
\left(\vect{I}_{N}-\vect{V}\vect{W}_{\mathbf{u}}^\hermitian \right)
 \vect{R}_{\mathbf{y}}(m)\bm{\Gamma}_{R}^{-1}  
 \right].
\end{align}
Using \eqref{eq:Q3} and noting that
$\mathbf{g}_{d}^\hermitian\bm{\Gamma}_R^{-1}\vect{Q}=\bm{0}$,
yields \eqref{eq:phi_R_closed-form}.
It follows that both estimators are identical:
\begin{align}\label{eq:phi_R_identical}
\phi^{\mathrm{ML},\bar{\mathbf{y}}}_{R}=
\phi^{\mathrm{ML},\bar{\mathbf{z}}}_{R}.
\end{align}
%%%%%%%%%%%%%%%%
It should be noted that in \cite{jensen2015analysis,schwartz2017cramer} 
the two \acp{MLE} of the reverberation \ac{PSD} were shown to be identical in the noiseless case.
Here we extend this result to the noisy case,
when the noise \ac{PSD} matrix is a rank-deficient matrix.

\subsubsection{Comparing the noise \ac{PSD} estimators}
The noise \ac{PSD} estimators in \eqref{eq:phi_v1}
and \eqref{eq:phi_v_BM1} are now compared.
Substituting \eqref{eq:smooth_u} into \eqref{eq:phi_v_BM1}
and then using \eqref{eq:w_relation} and \eqref{eq:phi_R_identical},
yields the same expression as in \eqref{eq:phi_v1}, and therefore
\begin{align}
\bm{\Psi}_{\mathbf{v}}^{\mathrm{ML},\bar{\mathbf{y}}}=
\bm{\Psi}_{\mathbf{v}}^{\mathrm{ML},\bar{\mathbf{z}}}.
\end{align}

\vspace{-2mm}
%%%%%%%%%%%%%%%%%%%%%%%%%%%
\subsection{\ac{MSE} Calculation}

In the sequel, the theoretical performance of the proposed \ac{PSD} estimators is analyzed. 
Since the non-blocking-based and the blocking-based \acp{MLE} 
were proved in section~\ref{sec:comparison} to be identical
for both reverberation and noise \acp{PSD},
it suffices to analyze the non-blocking-based \acp{MLE}.

\subsubsection{Theoretical performance of the reverberation \ac{PSD} estimators}

It is well known that for an unbiased estimator, the \ac{MSE} is identical to the variance.
We therefore start by showing that the non-blocking-based \ac{MLE} in \eqref{eq:phi_R_closed-form} is unbiased.
Using \eqref{eq:smooth0}, the expectation of \eqref{eq:phi_R_closed-form} writes 
\begin{align}\label{eq:phi_R_Expectation}
\mathbb{E}\left(\phi^{\mathrm{ML},\bar{\mathbf{y}}}_{R}(m)\right)
=\frac{1}{N-1-T}
\Tr \left[\bm{\Phi}_{\mathbf{y}}(m)\vect{\Gamma}_{R}^{-1}\vect{Q}  
\right].
\end{align}
Then, we use the following property (see \eqref{eq:PhiQ_prop}):
\begin{equation}\label{eq:Gamma_r_Q_identity}
\vect{\Gamma}_{R}^{-1}\vect{Q}=
\phi_{R}(m)\bm{\Phi}_{\mathbf{y}}^{-1}(m)\vect{Q},
\end{equation}
to obtain
\begin{align}\label{eq:phi_R_Expectation2}
\mathbb{E}\left(\phi^{\mathrm{ML},\bar{\mathbf{y}}}_{R}(m)\right)
=\frac{\phi_{R}(m)}{N-1-T}
\Tr \left[\vect{Q}  \right]
=\phi_{R}(m).
\end{align}
%%%%%%%%%%%
It follows that the reverberation \ac{MLE} is unbiased, and thus the \ac{MSE} is identical to the variance.
%%%%
Using the i.i.d. assumption, the variance of the non-blocking-based \ac{MLE} in \eqref{eq:phi_R_closed-form} 
is given by
\begin{flalign}\label{eq:phi_R_variance}
\text{var}\left(\phi^{\mathrm{ML},\bar{\mathbf{y}}}_{R}(m)\right)
=\frac{1}{L}\frac{1}{\left(N-1-T\right)^2}
\text{var}\big( \mathbf{y}^\hermitian (m) 
\vect{\Gamma}_{R}^{-1}\vect{Q}
\mathbf{y}(m) \big).&&
\end{flalign}
In order to simplify \eqref{eq:phi_R_variance}, the following identity is used.
For a positive definite Hermitian form 
$\mathbf{x}^\hermitian\mathbf{Z}\mathbf{x}$,
where $\mathbf{x} \sim \mathcal{N}_c(\bm{0},\bm{\Phi}_{\mathbf{x}})$
and $\mathbf{Z}$ a Hermitian matrix, the variance is given by
\cite[p.~513, Eq. (15.30)]{Kay1993}:
\begin{align}
\text{var}\left(\mathbf{x}^\hermitian\mathbf{Z}\mathbf{x}\right)
&=\Tr \left[\bm{\Phi}_{\mathbf{x}}\mathbf{Z}\bm{\Phi}_{\mathbf{x}}\mathbf{Z} \right]
\label{eq:Kay_identity_15.30}.
\end{align}   
%%%%%%
Since $\mathbf{y}(m)\sim \mathcal{N}_c\big(\bm{0},\bm{\Phi}_{\mathbf{y}}(m)\big)$
and $\vect{\Gamma}_{R}^{-1}\vect{Q}$ is a Hermitian matrix
(note that $\vect{Q}^\hermitian=\vect{\Gamma}_{R}^{-1}\vect{Q}\vect{\Gamma}_{R}$),
we obtain
\begin{flalign}\label{eq:phi_R_variance2}
\text{var}\left(\phi^{\mathrm{ML},\bar{\mathbf{y}}}_{R}(m)\right)
&=\frac{1}{L}\frac{1}{\left(N-1-T\right)^2}
\nonumber &&\\
& \hspace{0.1cm}\times
\Tr
\left[\bm{\Phi}_{\mathbf{y}}(m) \vect{\Gamma}_{R}^{-1}\vect{Q}
      \bm{\Phi}_{\mathbf{y}}(m) \vect{\Gamma}_{R}^{-1}\vect{Q}\right].&&
\end{flalign}
Finally, using \eqref{eq:Gamma_r_Q_identity} and \eqref{eq:QQ_prop},
the variance writes
\begin{align}\label{eq:phi_R_variance3}
\text{var}\left(\phi^{\mathrm{ML},\bar{\mathbf{y}}}_{R}(m)\right)
=\frac{1}{L}\frac{\phi_{R}^2(m)}{N-1-T}.
\end{align}
Note that in the noiseless case, namely $T=0$, the variance reduces to the 
one derived in \cite{jensen2015analysis,kuklasinski2015multi}.

%%%%%%%%%%%%%%%
\subsubsection{Theoretical performance of the noise \ac{PSD} estimators}

Using \eqref{eq:smooth0}, \eqref{eq:Phi_y_rank_deff} and \eqref{eq:w_lcmv_dot_gd,V},
and based on the unbiasedness of $\phi^{\mathrm{ML},\bar{\mathbf{y}}}_{R}(m)$,
it can be shown that 
$\bm{\Psi}_{\mathbf{v}}^{\mathrm{ML},\bar{\mathbf{y}}}(m)$
in \eqref{eq:phi_v1} is an unbiased estimator of 
$\bm{\Psi}_{\mathbf{v}}(m)$.

Next, we calculate the variance of the diagonal terms of $\bm{\Psi}_{\mathbf{v}}^{\mathrm{ML},\bar{\mathbf{y}}}(m)$.
To this end, we write the $(i,j)$ entry of $\bm{\Psi}_{\mathbf{v}}^{\mathrm{ML},\bar{\mathbf{y}}}(m)$
in \eqref{eq:phi_v1} as 
\begin{align}\label{eq:lambda_v1}
\left[\bm{\Psi}_{\mathbf{v}}^{\mathrm{ML},\bar{\mathbf{y}}}(m)\right]_{ij}
&= \vect{w}_{i}^\hermitian
\left(\vect{R}_{\mathbf{y}}(m)-\phi^{\mathrm{ML},\bar{\mathbf{y}}}_{R}(m)\bm{\Gamma}_R  \right)
\vect{w}_{j},
%\hspace{0.15cm} ; \hspace{0.15cm} i,j=1,\ldots,T
\end{align}
for $i,j=1,\ldots,T$, where $\vect{w}_{i}$ is the $i^{\text{th}}$ column of 
the matrix $\vect{W}_{\mathbf{u}}$ in \eqref{eq:LCMV_V}.
Using a partitioned matrix to simplify $\vect{w}_{i}$, 
it can be shown that 
\begin{equation}\label{eq:wt}
\mathbf{w}_{i}^\hermitian
=\frac{\mathbf{v}_{i}^\hermitian\bm{P}_{\bar{\vect{V}}_{i}}^{\perp}\bm{P}_{g}^{\perp}\bm{\Gamma}_R^{-1}}
{\mathbf{v}_{i}^\hermitian\bm{P}_{\bar{\vect{V}}_{i}}^{\perp}\bm{P}_{g}^{\perp}\bm{\Gamma}_R^{-1}\mathbf{v}_{i}},
\end{equation}
where $\bar{\vect{V}}_{i}\in \mathbb{C}^{N \times T-1}$ is composed of all the vectors in $\vect{V}$ except $\mathbf{v}_{i}$, i.e.
$\bar{\vect{V}}_{i}=\left[\mathbf{v}_1,\cdots,\mathbf{v}_{i-1},\mathbf{v}_{i+1},\cdots,\mathbf{v}_T\right]$,
and $\bm{P}_{\bar{\vect{V}}_{i}}^{\perp}\in \mathbb{C}^{N  \times N}$ is the corresponding projection matrix 
onto the subspace orthogonal to $\bar{\vect{V}}_{i}$
\begin{equation}
\bm{P}_{\bar{\vect{V}}_{i}}^{\perp}=\vect{I}_{N}-\bm{P}_{g}^{\perp}\bm{\Gamma}_R^{-1}\bar{\vect{V}}_{i}\left(\bar{\vect{V}}_{i}^\hermitian\bm{P}_{g}^{\perp}\bm{\Gamma}_R^{-1}\bar{\vect{V}}_{i} \right)^{-1}\bar{\vect{V}}_{i}^\hermitian.
\end{equation}
It can be verified that 
$\vect{w}_{i}^\hermitian \mathbf{v}_{j}=\delta_{ij}$.
%%%%%%%%%%
Denote the diagonal terms of $\bm{\Psi}_{\mathbf{v}}$ by 
$\lbrace\psi_{i}\rbrace_{i=1}^{T}$.
In Appendix~\ref{app:lambda_v_variance}, it is shown that
\begin{align}\label{eq:lambda_V_variance}
&\text{var}\left(\psi_i^{\mathrm{ML},\bar{\mathbf{y}}}(m)\right)
\nonumber \\
&\quad =\frac{\psi_i^2(m)}{L}
\left[ 
\left(\frac{1+\xi_i(m)}{\xi_i(m)}\right)^2+\frac{1}{N-1-T}\frac{1}{\xi_i^2(m)}\right],
\end{align}
where $\xi_i(m)$ is defined as the \textit{noise-to-reverberation ratio} 
at the output of $\vect{w}_{i}$:
\begin{equation}\label{eq:xi}
\xi_i(m)
\triangleq\frac{\psi_i(m)}{\vect{w}_{i}^\hermitian
\phi_R(m)\bm{\Gamma}_R\vect{w}_{i}}
=\frac{\psi_i(m)}{\phi_R(m)}\mathbf{v}_{i}^\hermitian
\bm{P}_{\bar{\vect{V}}_{i}}^{\perp}
\bm{P}_{g}^{\perp}\bm{\Gamma}_R^{-1}\mathbf{v}_{i}.
\end{equation}
Note that
\begin{align}\label{eq:lambda_V_variance_@eps_goes2inf}
 \lim_{\xi_i(m)\rightarrow \infty}\text{var}\left(\psi_i^{\mathrm{ML},\bar{\mathbf{y}}}(m)\right)
= \frac{\psi_i^2(m)}{L}.
\end{align}

%%%%%%%%%%%%%
\subsubsection{Theoretical performance of the speech \ac{PSD} estimator}

Using \eqref{eq:smooth0}, \eqref{eq:Phi_y_rank_deff} and \eqref{eq:w_mvdr_dot_gd_V}
and based on the unbiasedness of $\phi^{\mathrm{ML},\bar{\mathbf{y}}}_{R}(m)$,
it can be shown that 
$\phi^{\mathrm{ML},\bar{\mathbf{y}}}_{S}(m)$
is an unbiased estimator of $\phi_{S}(m)$.
%%%%%
In a similar manner to \eqref{eq:lambda_V_variance},
the variance of \eqref{eq:phi_S_closed-form} can be shown to be
\begin{align}\label{eq:phi_S_variance}
&\text{var}\left(\phi^{\mathrm{ML},\bar{\mathbf{y}}}_{S}(m)\right)
\nonumber \\
&\quad =\frac{\phi_{S}^2(m)}{L}
\left[ 
\left(\frac{1+\varepsilon(m)}{\varepsilon(m)}\right)^2+\frac{1}{N-1-T}\frac{1}{\varepsilon^2(m)}\right],
\end{align}
where $\varepsilon(m)$ is defined as the \textit{signal-to-reverberation ratio} 
at the output of $\mathbf{w}_{s}$:
\begin{equation}
\varepsilon(m)\triangleq\frac{\phi_{S}(m)}{\mathbf{w}_{s}^\hermitian
\phi_R(m)\bm{\Gamma}_R\mathbf{w}_{s}}
=\frac{\phi_{S}(m)}{\phi_R(m)}\mathbf{g}_{d}^\hermitian\bm{P}_{v}^{\perp}\bm{\Gamma}_R^{-1}\mathbf{g}_{d}.
\end{equation}
Note that
\begin{align}\label{eq:phi_S_variance_@eps_goes2inf}
 \lim_{\varepsilon(m)\rightarrow \infty}\text{var}\left(\phi^{\mathrm{ML},\bar{\mathbf{y}}}_{S}(m)\right) 
= \frac{\phi_{S}^2(m)}{L}.
\end{align}   
%%%%
In the noiseless case, i.e.~$T=0$, \eqref{eq:phi_S_variance}
becomes identical to the variance derived in \cite{jensen2015analysis,kuklasinski2015multi}.

%%%%%%%%%%%%%%%%%%%%%%%%%%%%%%%%%%%%%%%%%%%%%%%%%%%%%%%%%%%%%%%%%%%%%%%%%%%
\section{\ac{CRB} Derivation} 
\label{sec:CRB}
In this section, we derive the \ac{CRB} on the variance of any unbiased estimator of 
the various \acp{PSD}.

\vspace{-2mm}
%%%%%%%%%%%%%%%%%%%%%%%%%%%%%%%
\subsection{\ac{CRB} for the Late Reverberation PSD}
\label{sec:CRB_phi_R}

In Appendix~\ref{app:CRB_phi_R}, it is shown that
the \ac{CRB} on the reverberation \ac{PSD} writes
\begin{align}\label{eq:CRB1_2}
\text{CRB}\left(\phi_R \right)
=\frac{1}{L} \frac{\phi_R^{2}}{N-(T+1)}.
\end{align}
The resulting \ac{CRB} is identical to the \ac{MSE} derived in
\eqref{eq:phi_R_variance3}, and thus the proposed
\ac{MLE} is an efficient estimator.

\vspace{-2mm}
%%%%%%%%%%%%%%%%%%%%%%%%%
\subsection{\ac{CRB} for the Speech and Noise PSDs}

The \ac{CRB} on the speech \ac{PSD} is
identical to the \ac{MSE} derived in \eqref{eq:phi_S_variance},
as outlined in Appendix~\ref{app:CRB_phi_S}.
The \ac{CRB} on the noise \ac{PSD} can be derived similarly.
We conclude that the proposed \ac{PSD} estimators are efficient.

\vspace{-2mm}
%%%%%%%%%%%%%%%%%%%%%%%%%%%%%%%%%%%%%%%%%%%%%%%%%%%%%%%%%%%%%%%%%%%%%%%%%%%%%%%%%%%%%%%%%%%%%%
\section{Experimental Study}
\label{sec:Experimental_Study}

In this section, the proposed \acp{MLE} are evaluated in a synthetic Monte-Carlo simulation 
as well as on measurements of a real room environment.
%%%%
In Section~\ref{sec:Monte}, a Monte-Carlo simulation is conducted
in which signals are generated synthetically based on the assumed statistical model. 
The sensitivity of the proposed \acp{MLE} is examined with respect to the various model parameters,
and the \acp{MSE} of the proposed \acp{MLE} are compared to the corresponding \acp{CRB}.
%%%%
In Section~\ref{sec:Real_Recorded}, the proposed estimators are examined
in a real room environment, by utilizing them
for the task of speech dereverberation and noise reduction using the \ac{MCWF}.

\vspace{-2mm}
%%%%%%%%%%%%%%%%%%%%%%%%%%%%%%%%%%%%%%%%%%%%%%%%%%%%%%%
\subsection{Monte-Carlo Simulation}
\label{sec:Monte}

\subsubsection{Simulation Setup}

In order to evaluate the accuracy of the proposed estimators,
synthetic data was generated according to the signal model 
in \eqref{eq:sig_model}, by simulating $L$ i.i.d.~snapshots of single-tone 
signals, having a frequency of $f=2000$~Hz.
The signals are captured by a \ac{ULA} with $N$ microphones,
and inter-distance $d$ between adjacent microphones.
The desired signal component $s$ was drawn according to a complex Gaussian distribution 
with a zero-mean and a \ac{PSD} $\phi_{S}$.
The \ac{RDTF} is given by
\begin{equation}
\mathbf{g}_{d} =\exp \left(-j2\pi f \bm{\tau} \right),
\end{equation}
where $\bm{\tau} \in \mathbb{R}^{N}$ is the \ac{TDOA}
w.r.t.~the reference microphone, given by 
$\bm{\tau}=\frac{d\sin(\theta)}{c} \times \left[0,\cdots ,N-1 \right]^\top$,
and $\theta$ is the \ac{DOA}, defined as the broadside angle
measured w.r.t.~the perpendicular to the array. 
%%%%%%%%%%%
The reverberation component $\mathbf{r}$ was drawn according to a complex Gaussian distribution 
with a zero-mean and a \ac{PSD} matrix $\bm{\Phi}_{\mathbf{r}}=\phi_{R}\bm{\Gamma}_{R}$,
where $\bm{\Gamma}_{R}$ is modelled as an ideal spherical diffuse sound field, given by
$\Gamma_{R,ij}=\text{sinc}\left(2\pi f\frac{d\vert i-j \vert}{c} \right), \, i,j\in{1,\ldots,N}$.
%%%%%
The noise component was constructed as 
$\mathbf{u}=\vect{A}_{\mathbf{u}}\mathbf{s}_{\mathbf{u}}$
where $\mathbf{s}_{\mathbf{u}}$ denotes the $T$ noise sources, drawn according to 
a zero-mean complex Gaussian distribution with a random \ac{PSD} matrix $\bm{\Psi}_{\mathbf{u}}$,
and $\vect{A}_{\mathbf{u}}$ is an $N\times T$ random \ac{ATF} matrix.
%%%
For the estimation procedure, $\vect{V}$ is extracted 
by applying the \ac{EVD} to a set of noisy training samples, generated with different $\bm{\Psi}_{\mathbf{u}}$.

%%%%%
In the sequel, we examine the proposed estimators and bounds
as a function of the model parameters.
Specifically, the influence of the following parameters is examined:
i)~number of snapshots $L$; 
ii)~reverberation \ac{PSD} value $\phi_R$;
iii)~speech \ac{PSD} value $\phi_S$; and
iv)~noise power $P_{u}$,
which is defined as the Frobenius norm of the noise \ac{PSD} matrix, 
i.e.~$P_{u}\triangleq \Vert \bm{\Phi}_{\mathbf{u}} \Vert_{\text{F}}
=\sqrt{\Tr\left[\bm{\Phi}_{\mathbf{u}}^\hermitian \bm{\Phi}_{\mathbf{u}}\right]}$.
%%%%%
In each experiment, we changed the value of one parameter, while keeping the rest fixed.
The nominal values of the parameters are presented in Table~\ref{tab:parameters}.

\begin{table}
 \caption{Nominal Parameters}
 \label{tab:parameters}
\[
\begin{array}{|*{3}{c|}}
\hline
\text{Definition}                  &\text{Symbol}& \text{Nominal value} \\
\hline      \hline
\text{Microphone number }          &  N      &  8 \\
\hline
\text{Inter-microphone distance}   & d       &  6~\text{cm} \\ 
\hline
\text{Direction of arrival }       & \theta  &  0^\circ \\ 
\hline
\text{Frequency }                  & f  &  2~\text{kHz} \\ 
\hline
\text{Noise-\ac{PSD} rank}         & T       &  2 \\ 
\hline
\text{Number of Snapshots}         &  L      &  100 \\
\hline 
\text{Speech PSD }                 & \phi_S  &  0.5 \\ 
\hline
\text{Rev. PSD }                   & \phi_R  &  0.5 \\ 
\hline
%\text{Noise PSD }                  & \bm{\Lambda}_V &  ? \\ 
\text{Noise Power }                 & P_{u}  &  0.5 \\ 
\hline
\end{array}
\]
\end{table}

\begin{figure*}[t]
\centering
\subfigure[Increasing $L$ (from left to right)]{
\includegraphics[scale=0.5]{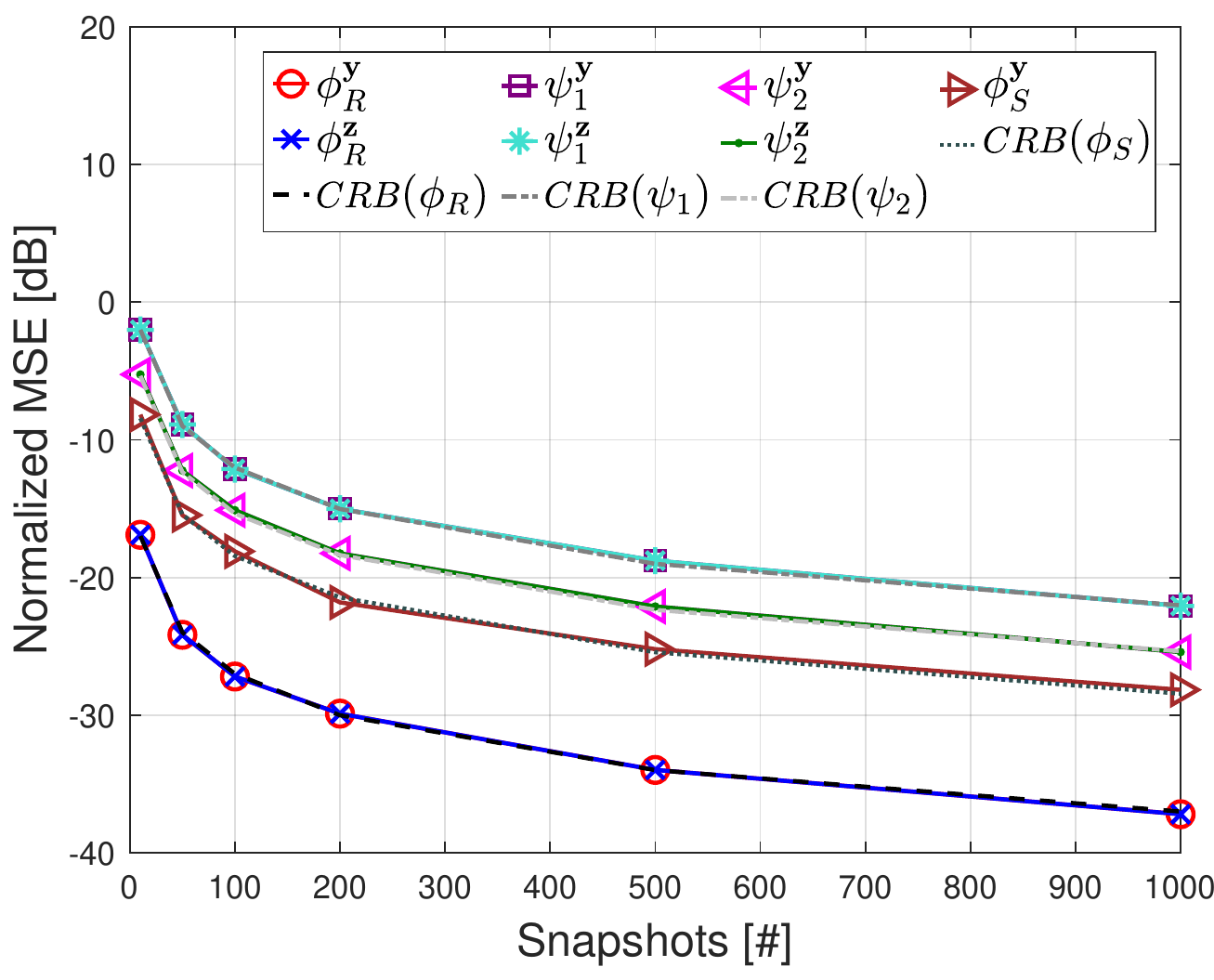}
\label{fig:MSE_phiR_vs_snapshots}
}
\subfigure[Decreasing $\phi_R$ (from left to right)]{
\includegraphics[scale=0.5]{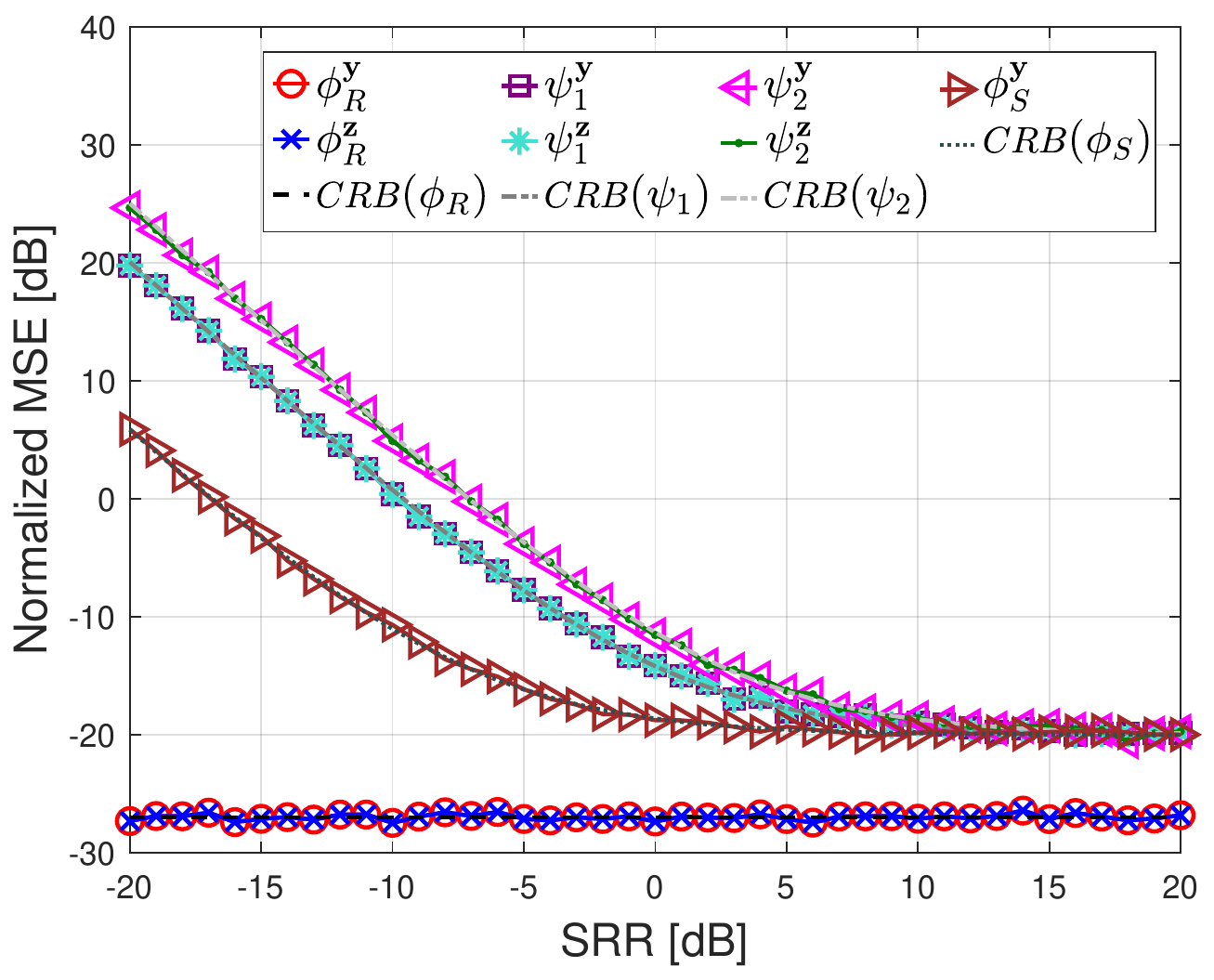}
\label{fig:MSE_phiR_vs_SRR}
}
\subfigure[Increasing $\phi_S$ (from left to right)]{
\includegraphics[scale=0.5]{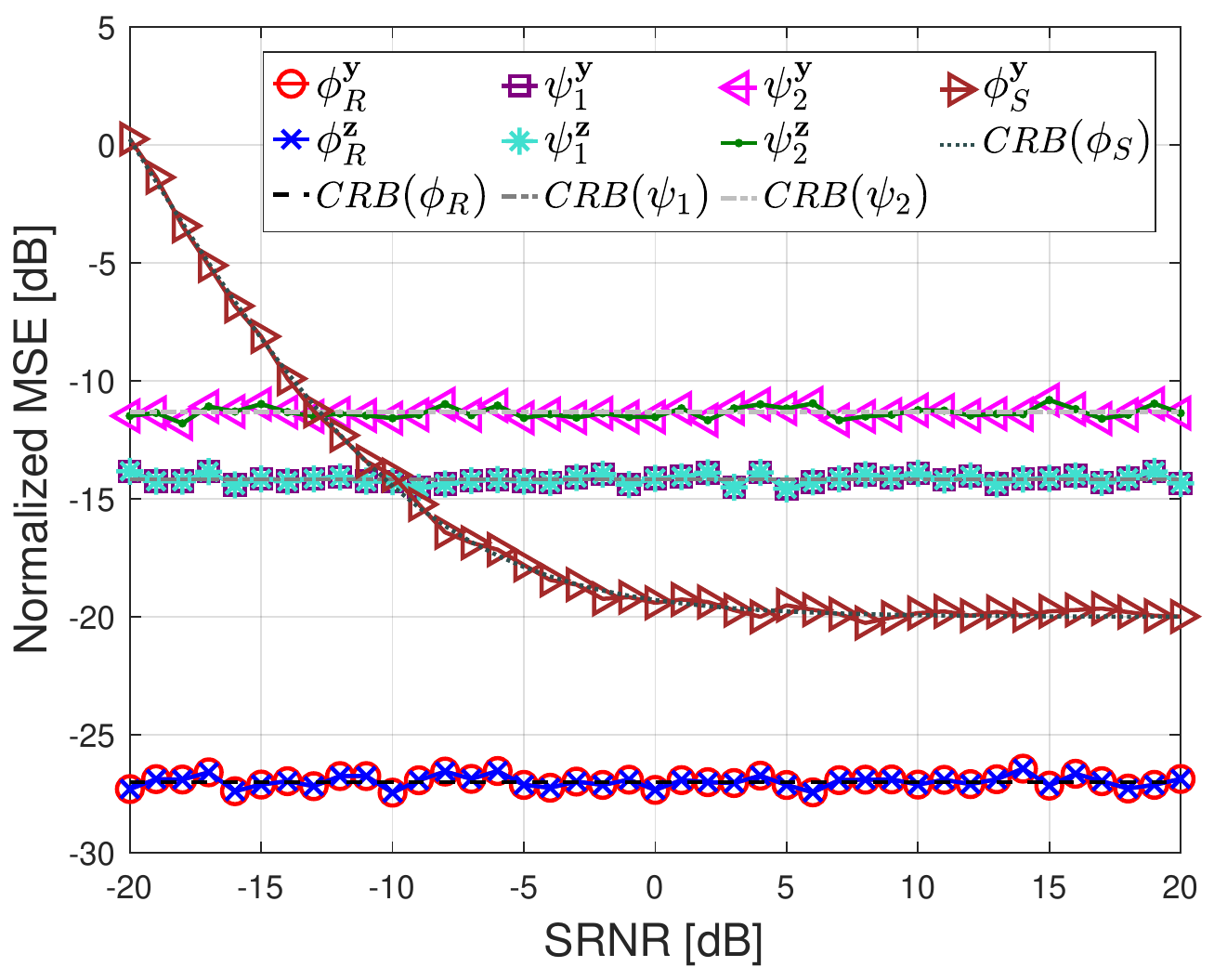}
\label{fig:MSE_phiR_vs_SRNR}
}
\subfigure[Decreasing $P_{u}$ (from left to right)]{
\includegraphics[scale=0.5]{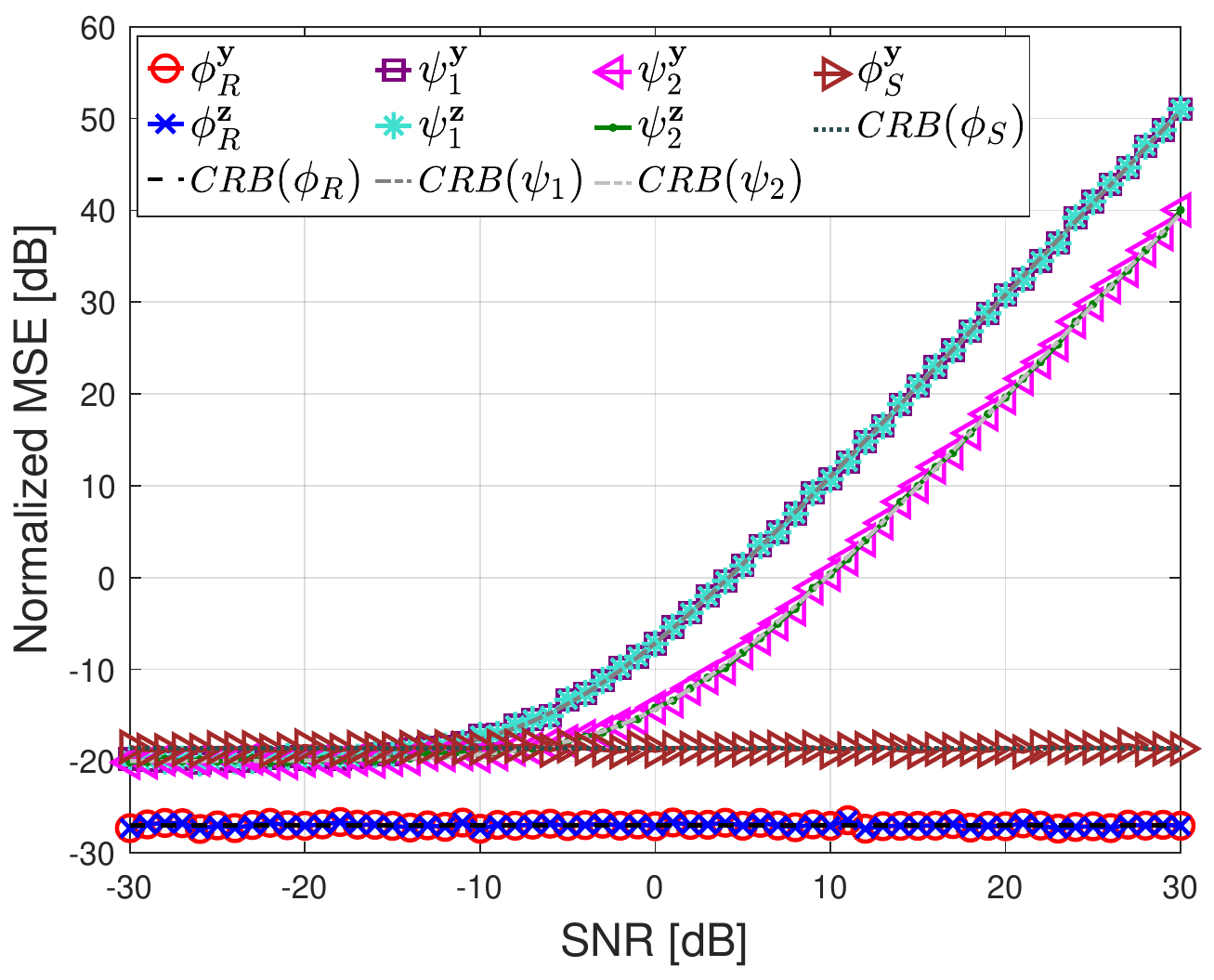}
\label{fig:MSE_phiR_vs_SNR}
}
\caption{Normalized \acp{MSE} and \acp{CRB} as a function of:
(a) Number of snapshots,
(b) \acs{SRR},
(c) \acs{SRNR}
and
(d) \acs{SNR}.
The parameters trends are depicted in the caption of each subfigure.}
\label{fig:simulation}
\end{figure*}

%%%%%%%%%%%%%%
For each scenario, we carried out $1000$ Monte-Carlo trials.
The reverberation \ac{PSD} $\phi_R$ was estimated in each trial
with  both \eqref{eq:phi_R_closed-form} and \eqref{eq:phi_R_BM_closed-form},
the noise \ac{PSD} was estimated with both \eqref{eq:phi_v1} and \eqref{eq:phi_v_BM1}
and the speech \ac{PSD} was estimated with \eqref{eq:phi_S_closed-form}.
The accuracy of the estimators was evaluated using the \ac{nMSE},
by averaging over the Monte-Carlo trials
and normalizing w.r.t.~the square of the corresponding \ac{PSD} value.
%%%%
For each quantity, the corresponding normalized \ac{CRB} was also computed,
in order to demonstrate the theoretical lower bound on the \ac{nMSE}.

%%%%%%%%%%%%%%%%%%
\subsubsection{Simulation Results}

In Fig.~\ref{fig:MSE_phiR_vs_snapshots}, the \acp{nMSE} 
are presented as a function of the number of snapshots, $L$.
Clearly, the \acp{nMSE} of all the estimators decrease as the number of snapshots increases.
%%%
As expected from the analytical study, it is evident that the non-blocking-based 
and the blocking-based \acp{MLE} yield the same \ac{nMSE},
for both the reverberation and noise \acp{PSD}.
Furthermore, for all quantities the \acp{nMSE} coincide with the corresponding \acp{CRB}.
%%%
It should be stressed that in practical applications of speech processing,
the \acp{PSD} are highly time-varying, and thus will not remain constant over a large number of snapshots.
This experiment only serves as a consistency check for the proposed estimators.

%%%%%%%%%%%%%%%%%%
We now study the effect of varying the reverberation level.
Let the \acf{SRR} be defined as $\text{SRR}\triangleq 10\log\left(\frac{\phi_S}{\phi_R} \right)$.
In this experiment, we change $\phi_R$ s.t.~the resulting \ac{SRR} ranges between  
$-20$~dB and $20$~dB.
In Fig.~\ref{fig:MSE_phiR_vs_SRR}, the \acp{nMSE} are presented as a function of \ac{SRR}. 
It is evident that the \acp{nMSE} of the reverberation \ac{PSD} estimators are independent of $\phi_R$,
and equal to $\frac{1}{L}\frac{1}{N-1-T}$, as manifested by \eqref{eq:phi_R_variance3}.
On the other hand, the \acp{nMSE} of the noise and speech \acp{MLE}
decrease as the reverberation level decreases,
until reaching the limiting value of $\frac{1}{L}$, which is inline with 
\eqref{eq:lambda_V_variance_@eps_goes2inf} and \eqref{eq:phi_S_variance_@eps_goes2inf},
respectively.

%%%%%%%%%%%%%%%%%%
We now examine the effect of changing the speech \ac{PSD} level.
Let the \acf{SRNR} be defined as $\text{SRNR}\triangleq 10\log\left(\frac{\phi_S}{\phi_R+P_{u}} \right)$.
We set $\phi_S$ to several values s.t.~the \ac{SRNR} ranges between $-20$~dB and $20$~dB.
In Fig.~\ref{fig:MSE_phiR_vs_SRNR}, the \acp{nMSE} are presented as a function of \ac{SRNR}. 
It is shown that the speech \ac{PSD} estimator is improved as the speech level increases.
For the reverberation and noise \ac{PSD} estimators, the performance is independent of $\phi_S$.
%%%%%%%%%%%%%
Obviously, the blocking-based estimators are not affected by the value of the speech \ac{PSD}.
It is further shown that the non-blocking-based reverberation and noise estimators also produce
\acp{nMSE} that are independent of $\phi_S$, as implied by \eqref{eq:phi_R_variance3} and \eqref{eq:lambda_V_variance}. 
%%%%
This behaviour can be explained as follows. Both the reverberation and the noise \ac{PSD} estimators apply a projection matrix onto the subspace orthogonal to the speech subspace, i.e.~eliminate the speech component by generating a null towards its direction, as manifested by \eqref{eq:Q1} and \eqref{eq:LCMV_V}, respectively.
As a result, they are not affected by any change in the speech level.

%%%%%%%%%%%%%%%%%%
The effect of increasing the noise level is now examined.
Let the \acf{SNR} be defined as $\text{SNR}\triangleq 10\log\left(\frac{\phi_S}{P_{u}} \right)$.
We change $P_{u}$ s.t.~the \ac{SNR} varies between $-30$~dB and $30$~dB.
In Fig.~\ref{fig:MSE_phiR_vs_SNR}, the \acp{nMSE} are presented as a function of \ac{SNR}. 
%%%%%%
It is evident that the performance of the noise \ac{PSD} estimators degrades as $P_u$ decreases.
In contrast, the \acp{nMSE} of the reverberation and speech \acp{MLE} are independent of the noise level, 
as manifested by \eqref{eq:phi_R_variance3} and \eqref{eq:phi_S_variance}, respectively. 
This is due to the fact that the reverberation and speech \acp{MLE}
eliminate the noise components by directing null towards the noise subspace, as implied by \eqref{eq:Q1} and \eqref{eq:MVDR_S}, respectively.

\vspace{-2mm}
%%%%%%%%%%%%%%%%%%%%%%%%%%%%%%%%%%%%%%%%%%%%%%%%%%%%%%%
\subsection{Experiments with Recorded Room Impulse Responses}
\label{sec:Real_Recorded}

The performance of the proposed \ac{PSD} estimators is now evaluated 
in a realistic acoustic scenario, for the task of speech dereverberation and noise reduction.
In our experiments, microphone signals were synthesized using real speech signals and measured \acp{RIR}.
The proposed \ac{PSD} estimators were used in order to calculate the \ac{MCWF}.

%%%%%%%%%%%%%%%%%%%%%%%%%%%%%%%%%%%%
\subsubsection{Competing Algorithms}
The proposed method is compared to \cite{kodrasi18joint},
in which the \ac{MCWF} is implemented using the blocking-based or the non-blocking-based \ac{LS} estimators.
Therein, a spatially homogeneous noise sound field is assumed, namely
$\bm{\Phi}_{\mathbf{u}}(m,k)=\phi_{U}(m,k)\bm{\Gamma}_U(k)$,
where $\bm{\Gamma}_U(k)$ is a known time-invariant spatial coherence matrix,
and $\phi_{U}(m,k)$ denotes the unknown time-varying \ac{PSD}, which has to be estimated,
along with the speech and reverberation \acp{PSD}.
%%%%%%%%
Although this method considers a different noise model, it is chosen as the baseline 
since this is the only work that estimates also the noise \ac{PSD}.

%%%%%%%%%%%%%%%%%%%%%%%%%%%%%%%%%%%%%%%%%%%
\subsubsection{Implementation of the \ac{MCWF}}

It is well-known that the \ac{MCWF} can be decomposed into an \ac{MVDR} beamformer followed by a single-channel Wiener
postfilter \cite{simmer2001post,balan2002microphone}:
\begin{equation}\label{eq:MCWF_decomp}
\hat{s}_{\text{MCWF}}(m) 
=
\underbrace{\frac{\hat{\gamma}(m)}{\hat{\gamma}(m)+1}}_{H_{W}(m)}
\underbrace{\frac{\mathbf{g}_{d}^\hermitian\hat{\bm{\Phi}}_{i}^{-1}(m) }
{\mathbf{g}_{d}^\hermitian\hat{\bm{\Phi}}_{i}^{-1}(m) \mathbf{g}_{d}}}_{\mathbf{w}^\hermitian_{\text{MVDR}}(m)}
\mathbf{y}(m),
\end{equation}
where 
\begin{equation}\label{eq:gamma_dir}
\hat{\gamma}(m)=\frac{\hat{\phi}_{S}(m)}{\hat{\phi}_{RE}(m)}
\end{equation}
denotes the \ac{SRNR} at the output of the \ac{MVDR}, and 
$\hat{\phi}_{RE}(m)\triangleq\left(\mathbf{g}_{d}^\hermitian\hat{\bm{\Phi}}_{i}^{-1}(m)\mathbf{g}_{d}\right)^{-1}$
is the residual interference at the \ac{MVDR} output.

%%%%%%%%%%%%%%%%%%%%%%%%
The implementation of \eqref{eq:gamma_dir} requires the estimate of the speech \ac{PSD},
which is missing in the blocking-based framework.
%%%%%%
By substituting the obtained blocking-based reverberation and noise estimates,
namely $\phi_{R}^{\mathrm{ML},\bar{\mathbf{z}}}$ and $\bm{\Psi}_{\mathbf{v}}^{\mathrm{ML},\bar{\mathbf{z}}}$,
into the general likelihood function in \eqref{eq:likelihood_y},
the maximization becomes a one-dimensional optimization problem,
and a closed-form solution is available \cite{ye1995maximum,kuklasinski2016maximum}:
\begin{align}\label{eq:phi_S_BM}
\phi^{\mathrm{ML},\bar{\mathbf{z}}}_S(m)
&=
\mathbf{w}^\hermitian_{\text{MVDR}}(m)
\Big(\vect{R}_{\mathbf{y}}(m)
-\phi^{\mathrm{ML},\bar{\mathbf{z}}}_R(m)\bm{\Gamma}_R
\nonumber \\
& \quad
-\vect{V}\bm{\Psi}_{\mathbf{v}}^{\mathrm{ML},\bar{\mathbf{z}}}(m)\vect{V}^\hermitian \Big)
\mathbf{w}_{\text{MVDR}}(m).
\end{align}
However, it was shown in \cite{kodrasi18joint} that rather than using \eqref{eq:gamma_dir},
better dereverberation performance is obtained by using the decision-directed approach \cite{ephraim1984speech}, 
where $\gamma(m)$ is estimated by 
\begin{equation}\label{eq:decision}
\hat{\gamma}_{\text{DD}}(m)=\beta\frac{\vert\hat{s}(m-1)\vert^2}{\hat{\phi}_{RE}(m-1)}+(1-\beta)
\frac{\hat{\phi}_{S_i}(m)}{\hat{\phi}_{RE}(m)},
\end{equation} 
where $\beta$ is a weighting factor, and 
$\hat{\phi}_{S_i}$ is an instantaneous estimate based on the \ac{MVDR} output \cite{schwartz2016joint_EUSIPCO}:
\begin{equation}\label{eq:phi_s_inst}
\hat{\phi}_{S_i}(m)=\text{max} \left(\vert\mathbf{w}^\hermitian_{\text{MVDR}}(m)\mathbf{y}(m) \vert^2-
\hat{\phi}_{RE}(m),0 \right).
\end{equation}
%%%
In our experiments, the \ac{MCWF} was implemented with the two variants of computing $\hat{\gamma}$:
i) The direct implementation in \eqref{eq:gamma_dir}, which will be referred to as Dir;
and
ii) the decision-directed implementation in \eqref{eq:decision} with the speech \ac{PSD} estimated as in \eqref{eq:phi_s_inst},
denoted henceforth as DD.

%%%%%%%%%%%%%%%%%%%%%%%%%%%%%%%%%%%%
\subsubsection{Performance Measures}
\label{sec:Performance_Measures}

The speech quality was evaluated in terms of three common objective measures, namely
\ac{PESQ} \cite{recommendation2001perceptual}, 
\ac{fwSNRseg} \cite{quackenbush1988objective}
and \ac{LLR} \cite{quackenbush1988objective}.
The measures were computed by comparing $\hat{s}_{\text{MCWF}}(m)$ with the direct speech signal as received
by the reference microphone, obtained by filtering the anechoic speech 
with the direct path.
%%%%%%
In the sequel, we present the improvement in these performance measures,
namely $\Delta\text{PESQ}$, $\Delta\text{fwSNRseg}$ and $\Delta\text{LLR}$,
computed as the measure difference between the output signal (i.e.~$\hat{s}_{\text{MCWF}}$)
and the noisy and reverberant signal at the reference microphone (i.e.~$y_1$).

%%%%%%%%%%%%%%%%%%%%%%%%%%%
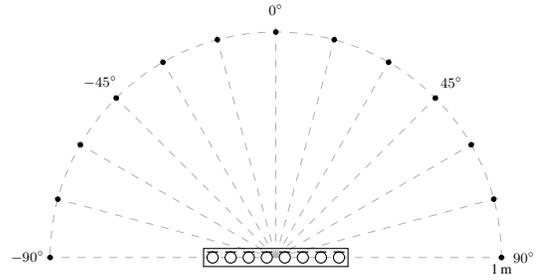
\begin{figure}[t]
  \centering
  \begin{tikzpicture}[scale=0.6,transform shape]
 \usetikzlibrary{calc}
 \tikzstyle{plain}=[]
 \tikzstyle{mic}=[microphone,draw,minimum width=2em, minimum height=2em,fill=white]

 % grid
 \draw [dashed,color=black!30] (5,0) arc(0:180:5);
 \foreach \a in {0, 15, ..., 180} {
   \draw [dashed,color=black!30,thin] (0,0) -- (\a:5.2);
   \foreach \r in {2} {
     \draw [fill=black] (\a :\r*2.5) circle (0.05);
   }
 }
 \foreach \aa in {90, 45, 0, -45, -90} {
   \node [plain] at (-\aa+90:5.5) {\small ${\aa}^\circ$};
 }
% \foreach \aa in {90,75,60,45,30,15,0,-15,-30,-45,-60,-75,-90} {
%   \node [plain] at (-\aa+90:5.5) {\small ${\aa}^\circ$};
% }
 
 % \node [plain] at (90:6) {\small \framebox{Speaker}};
 % \node [plain] at (45+3:5.9) {\small \framebox{Noise \#1}};
 % \node [plain] at (90+30-3:5.9) {\small \framebox{Noise \#2}};

 % mic array
 \foreach \x in {-14, -10, -6, -2, 2, 6, 10, 14} {
   \draw node [mic,scale=0.35] at (0.1*\x, 0) {};
 } 

 \node [rectangle, minimum width=32mm, minimum height=4mm, fill=none, draw] at (0,0){};

 \node [plain, yshift=-0.7em] at (0:5) {\small \unit[1]{m}};
  
\end{tikzpicture}
  \vspace{-0.1cm}
  \caption{Geometric setup.}
  \label{fig:illus}
\end{figure}
%%%%%%%%%%%%%%%%%%%%%%%%%

%ver 67 without SNR=20
\begin{table*}[!t]
\caption{Speech Quality Measures, $T_{60}=360$~msec}
\label{tab:table_T=2_TV2_360ms}
\centering
\small
\resizebox{0.99 \linewidth}{!}{%
\begin{tabular}{c c c c c c c c c c c c c c c}
\toprule
\toprule
& \multicolumn{4}{c}{$\Delta\text{PESQ}$} && \multicolumn{4}{c}{$\Delta\text{fwSNRseg}$~[dB]} && \multicolumn{4}{c}{$\Delta\text{LLR}$~[dB]} \\
\cmidrule{2-5} \cmidrule{7-10} \cmidrule{12-15}
Alg.\textbackslash RSNR
&  $0\text{dB}$ &  $5\text{dB}$ &  $10\text{dB}$ & $15\text{dB}$
&& $0\text{dB}$ &  $5\text{dB}$ &  $10\text{dB}$ & $15\text{dB}$
&& $0\text{dB}$ &  $5\text{dB}$ &  $10\text{dB}$ & $15\text{dB}$ \\ %[0.5ex]
%%%%%%%%%%%%%%%%%%%%%%%%%%%%%%%%%%%%%%%%%%%%%%%%%
\cmidrule{1-15}
Blocking LS \cite{kodrasi18joint} Dir    	       
& 0.56 & 0.70 & 0.77 & 0.76 && 11.85 & 11.02 & 9.80 & 8.67 && -0.55 & -0.52 & -0.45 & -0.37 \\

Blocking LS \cite{kodrasi18joint} DD   	           
& 0.60 & 0.75 & 0.85 & 0.88 && 13.05 & 11.67 & 10.23 & 8.88 && -0.50 & -0.45 & -0.36 & -0.27 \\

Blocking ML Dir 	    				           
& 0.74 & 0.88 & 0.93 & 0.90 && 12.56 & 11.77 & 10.61 & 9.39&& \textbf{-0.74}&\textbf{-0.65}&\textbf{-0.52}&\textbf{-0.40}\\

Blocking ML DD  	    		          			
& 0.79 & 0.93 & 0.99 & 0.99 && 15.57 & 13.16 & 10.75 & 8.66 &&-0.67 & -0.50 & -0.30 & -0.12 \\
\cmidrule{1-15}
Non-blocking LS \cite{kodrasi18joint} Dir   	  
& 0.70 & 0.81 & 0.85 & 0.81 && 13.52 & 11.69 & 9.93 & 8.57 && -0.54 & -0.49 & -0.41 & -0.33 \\

Non-blocking LS \cite{kodrasi18joint} DD    	   
& 0.67 & 0.79 & 0.89 & 0.89 && 13.56 & 11.82 & 10.03 & 8.57 && -0.55 & -0.48 & -0.38 & -0.28 \\

Non-blocking ML Dir 		    				   
&\textbf{0.98}&\textbf{1.12}&\textbf{1.15}&\textbf{1.12} && \textbf{15.59}&\textbf{13.69}&\textbf{11.67}&\textbf{9.77} &&\textbf{-0.74} & -0.61 & -0.46 & -0.32 \\

Non-blocking ML DD  		 	    			    
& 0.79 & 0.93 & 0.99 & 0.99 && 15.57 & 13.16 & 10.75 & 8.66 &&	-0.67 & -0.50 & -0.30 & -0.12 \\
\bottomrule
\bottomrule
\end{tabular}}
\end{table*}

%%%%%%%%%%%%%%%%%%%%%%%%%%%%%%%%%%%%%%%%%%%%%%%%%%%

%ver 67 without SNR=20
\begin{table*}[!t]
\caption{Speech Quality Measures, $T_{60}=610$~msec}
\label{tab:table_T=2_TV2_610ms}
\centering
\small
\resizebox{0.99 \linewidth}{!}{%
\begin{tabular}{c c c c c c c c c c c c c c c}
\toprule
\toprule
& \multicolumn{4}{c}{$\Delta\text{PESQ}$} && \multicolumn{4}{c}{$\Delta\text{fwSNRseg}$~[dB]} && \multicolumn{4}{c}{$\Delta\text{LLR}$~[dB]} \\
\cmidrule{2-5} \cmidrule{7-10} \cmidrule{12-15}
Alg.\textbackslash RSNR
&  $0\text{dB}$ &  $5\text{dB}$ &  $10\text{dB}$ & $15\text{dB}$
&& $0\text{dB}$ &  $5\text{dB}$ &  $10\text{dB}$ & $15\text{dB}$ 
&& $0\text{dB}$ &  $5\text{dB}$ &  $10\text{dB}$ & $15\text{dB}$ \\ %[0.5ex]
%%%%%%%%%%%%%%%%%%%%%%%%%%%%%%%%%%%%%%%%%%%%%%%%%
\cmidrule{1-15}
Blocking LS \cite{kodrasi18joint} Dir    	       
& 0.42 & 0.50 & 0.54 & 0.56 && 10.78 & 10.13 & 9.20 & 8.38 && -0.51 & -0.49 & -0.44& -0.37 \\

Blocking LS \cite{kodrasi18joint} DD   	           
& 0.46 & 0.54 & 0.62 & 0.67 && 12.49 & 11.54 & 10.46 & 9.47&& -0.45 & -0.41 & -0.36 & -0.30  \\

Blocking ML Dir 	    				          
& 0.51 & 0.59 & 0.63 & 0.63 && 10.10 & 9.57 & 8.68 & 7.95 && -0.58 &\textbf{-0.53}&\textbf{-0.46}&\textbf{-0.38} \\

Blocking ML DD  	    		          			
& 0.61 & 0.71 & 0.78&0.80 &&\textbf{14.93}&\textbf{13.10}&\textbf{11.30} &\textbf{9.67} 
&& -0.57 & -0.44 & -0.28 & -0.13 \\

\cmidrule{1-15}
Non-blocking LS \cite{kodrasi18joint} Dir   	
& 0.56 & 0.60 & 0.63 & 0.62 && 13.22 & 11.49 & 9.93 & 8.70 && -0.51 & -0.48 & -0.42 & -0.36 \\

Non-blocking LS \cite{kodrasi18joint} DD    	  
& 0.51 & 0.59 & 0.67 & 0.71 && 13.37 & 12.07 & 10.66 & 8.73 && -0.51 & -0.47 & -0.40 & -0.32 \\

Non-blocking ML Dir 		    				   
&\textbf{0.69}&\textbf{0.78}&\textbf{0.80} &\textbf{0.81}&& 13.88 & 12.39 & 10.81 & 9.47
&&\textbf{-0.62}&-0.52& -0.41 & -0.30 \\

Non-blocking ML DD  		 	    			    
& 0.61 & 0.71 & 0.78 & 0.80 &&\textbf{14.93}&\textbf{13.10}&\textbf{11.30} &\textbf{9.67}
&& -0.57 & -0.44 & -0.28 & -0.13 \\
\bottomrule
\bottomrule
\end{tabular}}
\end{table*}

%%%%%%%%%%%%%%%%%%%%%%%%%%
\subsubsection{Experimental Setup}
We used \acp{RIR} from the database presented in \cite{hadad2014multichannel}.
The impulse responses were recorded in the $6 \times 6 \times 2.4 $~m acoustic lab
of the Engineering Faculty at \ac{BIU}.
The reverberation time was set to $T_{60}\in \lbrace 360,610 \rbrace$~msec.
The RIRs were recorded by a \ac{ULA} of $N=8$ microphones with inter-distances of $8$~cm. 
A loudspeaker was positioned at a distance of $1$~m from the array, at angles $\lbrace-90^\circ,-75^\circ,-60^\circ,\ldots,90^\circ\rbrace$,
as illustrated in Fig.~\ref{fig:illus}.
The performance is evaluated on $10$ experiments with different speaker angles, which were randomly selected from the given set of angles.
%%%%%%%%%%%%%%%%%%%%%%%%%%%
The $10$ chosen \acp{RIR} were convolved with 3-5 sec long utterances of $5$ male and $5$ female
speakers that were drawn from the TIMIT database \cite{garofolo1993darpa}.

%%%%%%%%%%%%%%%%%%%
For the additive noise, we used \textit{non-stationary} directional noise signals.
Two non-stationary noise sources (with time-varying speech-like spectrum)
were positioned at a distance of $1$~m from the array.
In each experiment, the angles of the noise sources were randomly selected,
avoiding the angle of the speaker.
The microphone signals were synthesized by adding the noises to the reverberant
speech signals with several \ac{RSNR} levels.
%%%%%%
Finally, a $1$~sec noise segment was preceded to the reverberant speech signal.
Based on this segment, a sample noise covariance matrix is computed,
and then the \ac{EVD} is applied in order to extract the spatial basis $\vect{V}$
that replaces the noise \ac{ATF} matrix $\vect{A}_{\mathbf{u}}$
(see \eqref{eq:sig_model_V}).

%%%%%%%%%%%%%%%%%%
The following values of the parameters were used.
The sampling rate was set to $16$~KHz, the \ac{STFT} was computed with windows of 
$32$~msec and 75\% overlap between adjacent time frames.
%%%%%%%%%%%%%%%%%%%%%%%%%
As the experiment consists of real-life non-stationary speech signals,
the sample covariance matrix $\vect{R}_{\mathbf{y}}(m)$ 
was estimated using recursive averaging \cite{kuklasinski2016maximum}
with a smoothing factor of $\alpha=0.7$, rather than the moving-window averaging of \eqref{eq:smooth0}.
The same applies for $\vect{R}_{\mathbf{z}}(m)$.
%%%%%%%
The smoothing parameter for the decision-directed in \eqref{eq:decision}
was set to $\beta=0.9$.
The gain of the single channel postfilter was lower bounded by $-15$~dB.

%ver 75
\begin{figure*}[!t]
\centering
\includegraphics[scale=0.51]{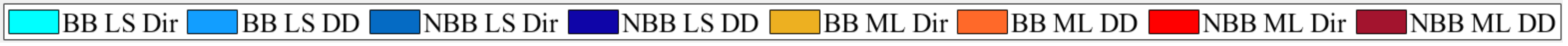}\\

\centering
\subfigure[\hspace{-0.6cm}]{
\includegraphics[scale=0.4]{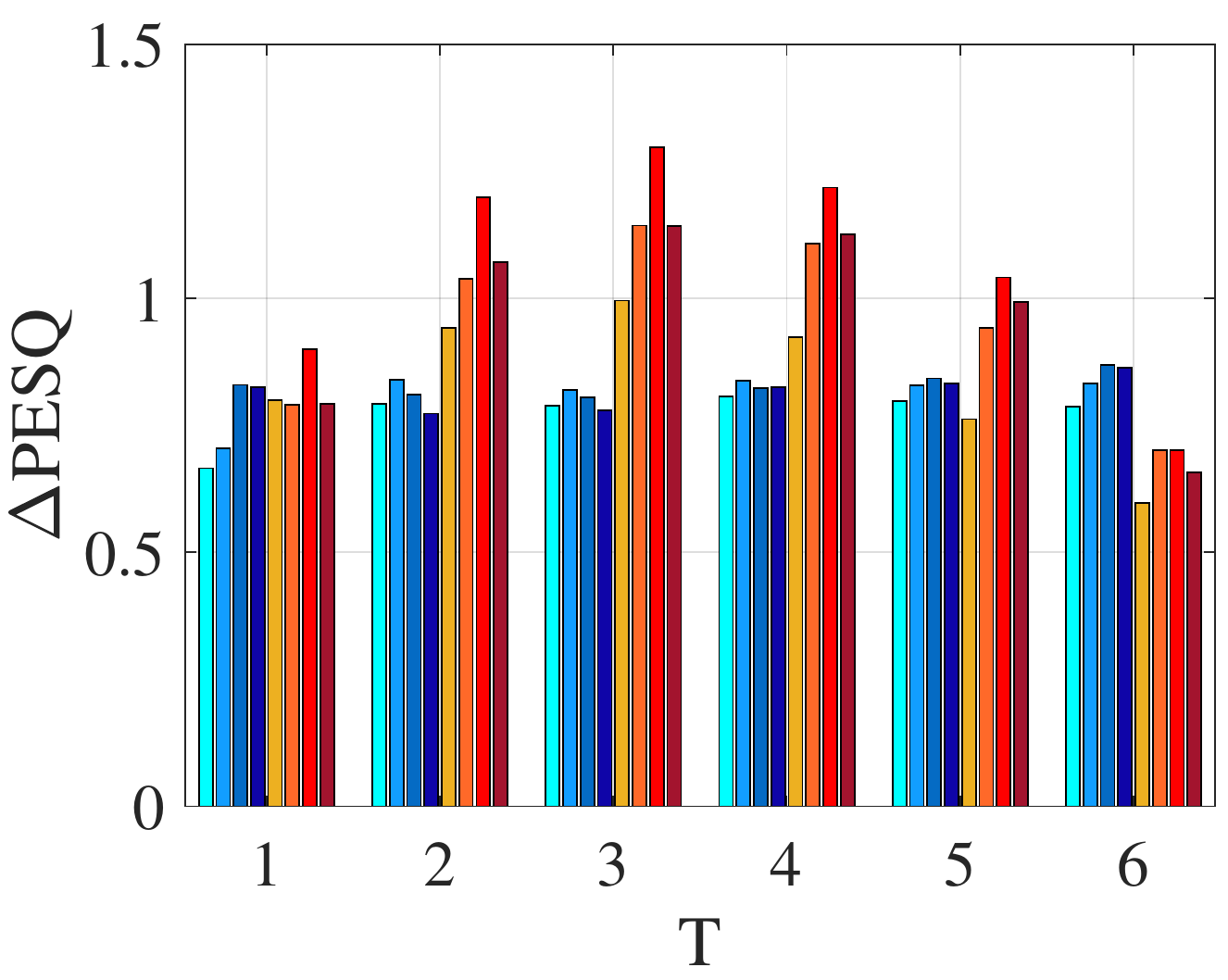}
\label{fig:DeltaPESQ}
}
\subfigure[\hspace{-0.6cm}]{
\includegraphics[scale=0.4]{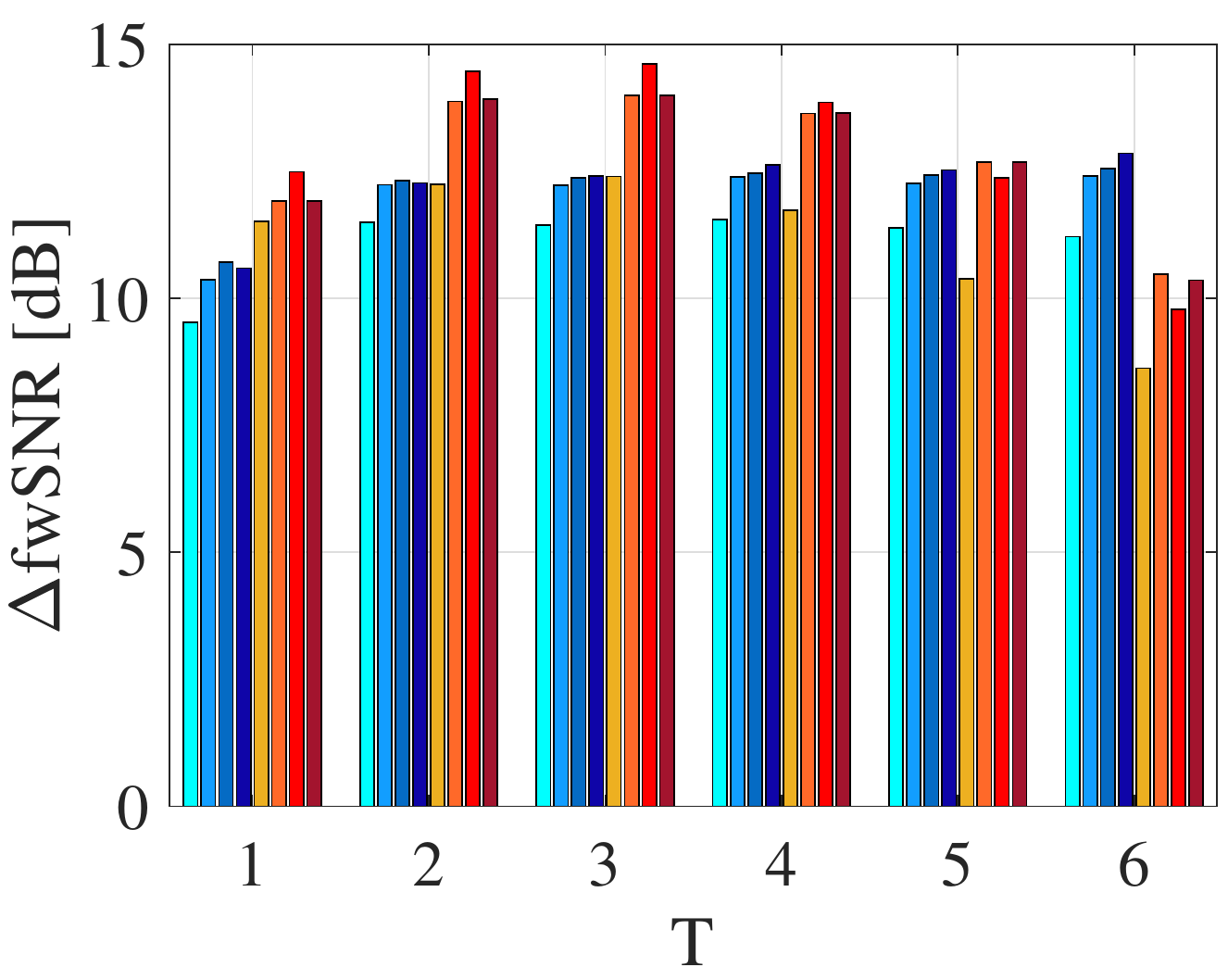}
\label{fig:DeltafwSNR}
}
\subfigure[\hspace{-0.6cm}]{
\includegraphics[scale=0.4]{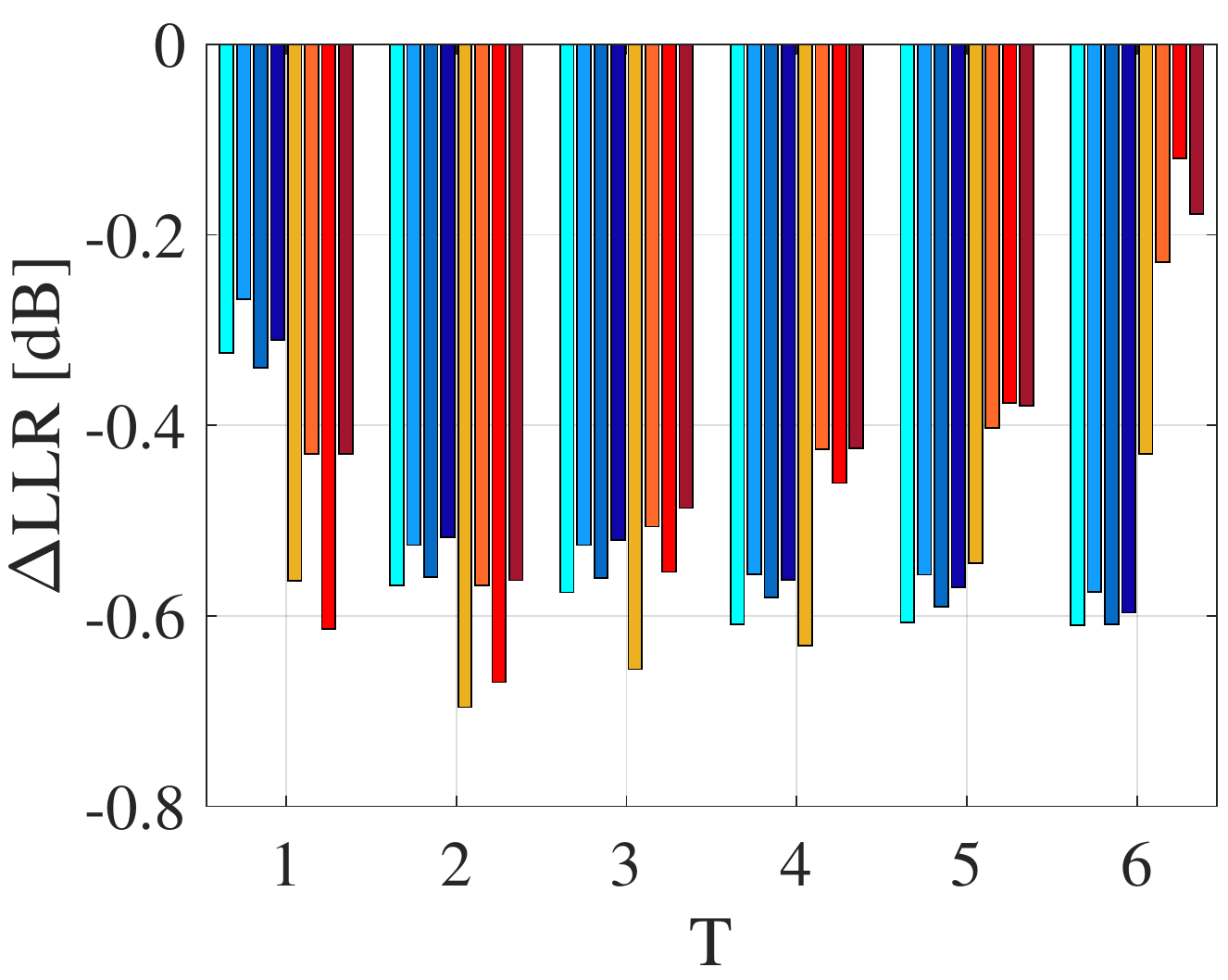}
\label{fig:DeltaLLR}
}
\caption{Results versus $T$, for $T_{60}=360$~msec, $N=8$, $\text{RSNR}=5\text{dB}$:
(a) $\Delta\text{PESQ}$,
(b) $\Delta\text{fwSNR}$,
and
(c) $\Delta\text{LLR}$.
}
\label{fig:Perf_vs_T_360ms}
\end{figure*}

%%%%%%
%ver 75
\begin{figure*}[!t]
\centering
\includegraphics[scale=0.51]{Perf_vs_T_T60=610_SNR=5_legend_crop}\\

\centering
\subfigure[\hspace{-0.6cm}]{
\includegraphics[scale=0.4]{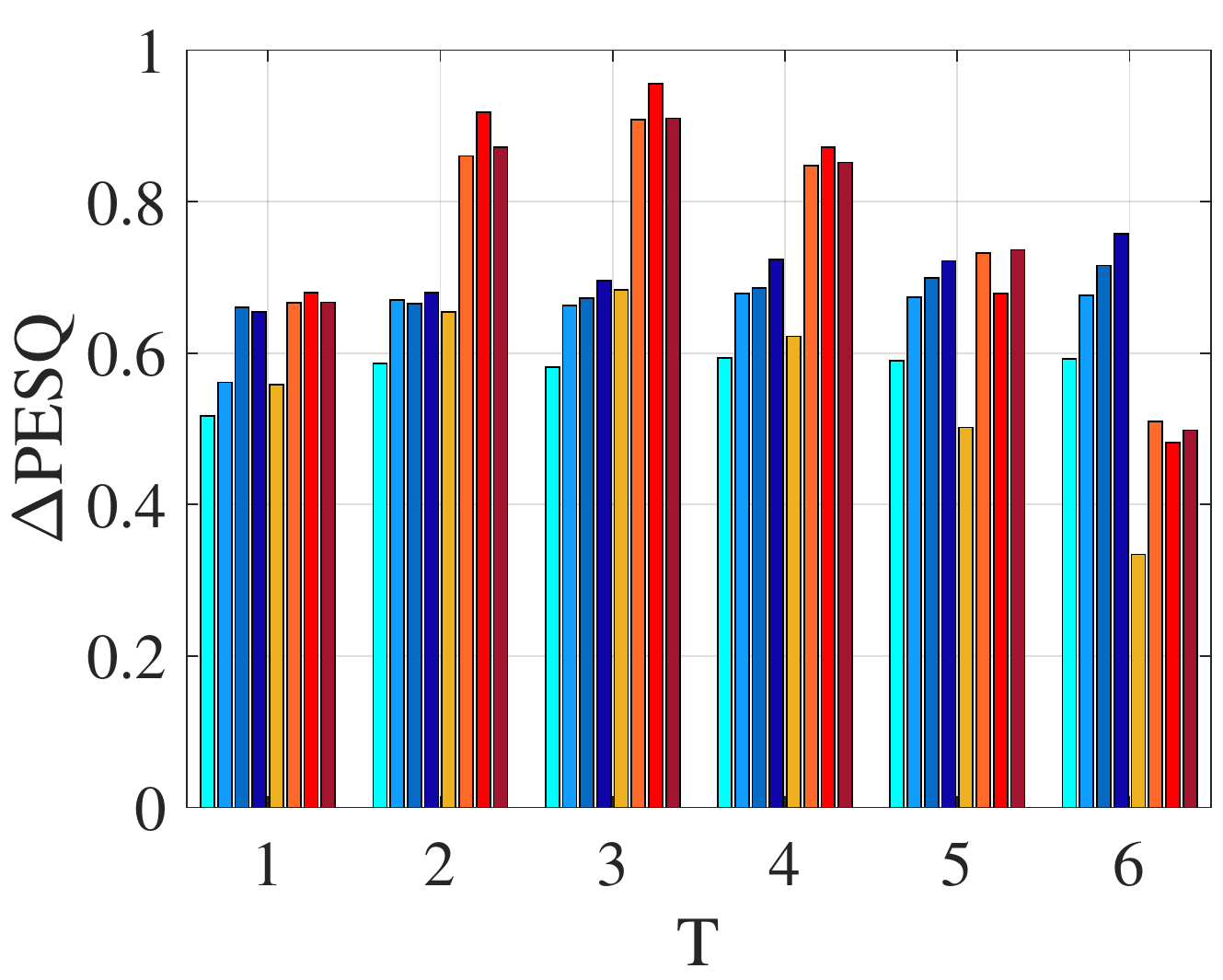}
\label{fig:DeltaPESQ}
}
\subfigure[\hspace{-0.6cm}]{
\includegraphics[scale=0.4]{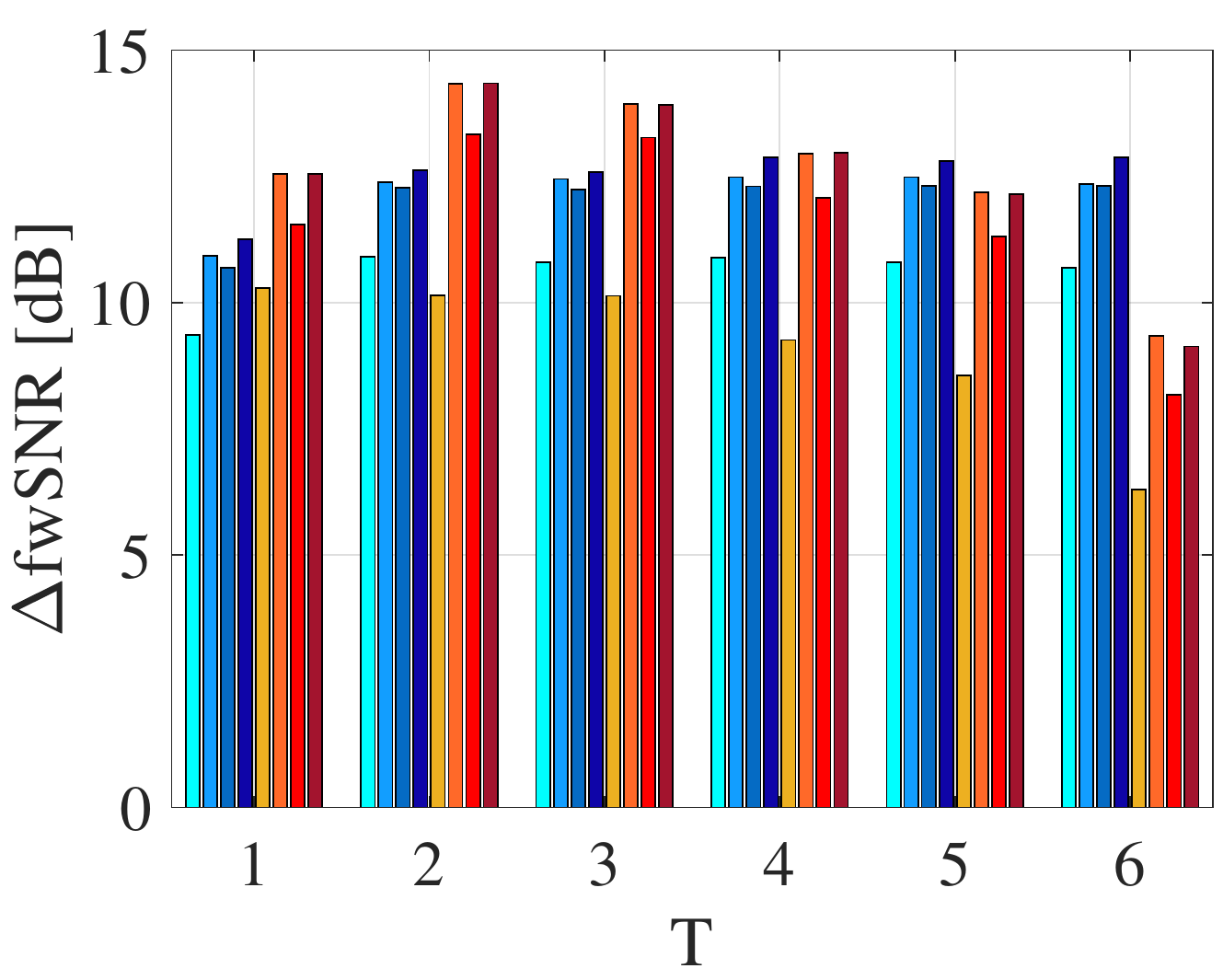}
\label{fig:DeltafwSNR}
}
\subfigure[\hspace{-0.6cm}]{
\includegraphics[scale=0.4]{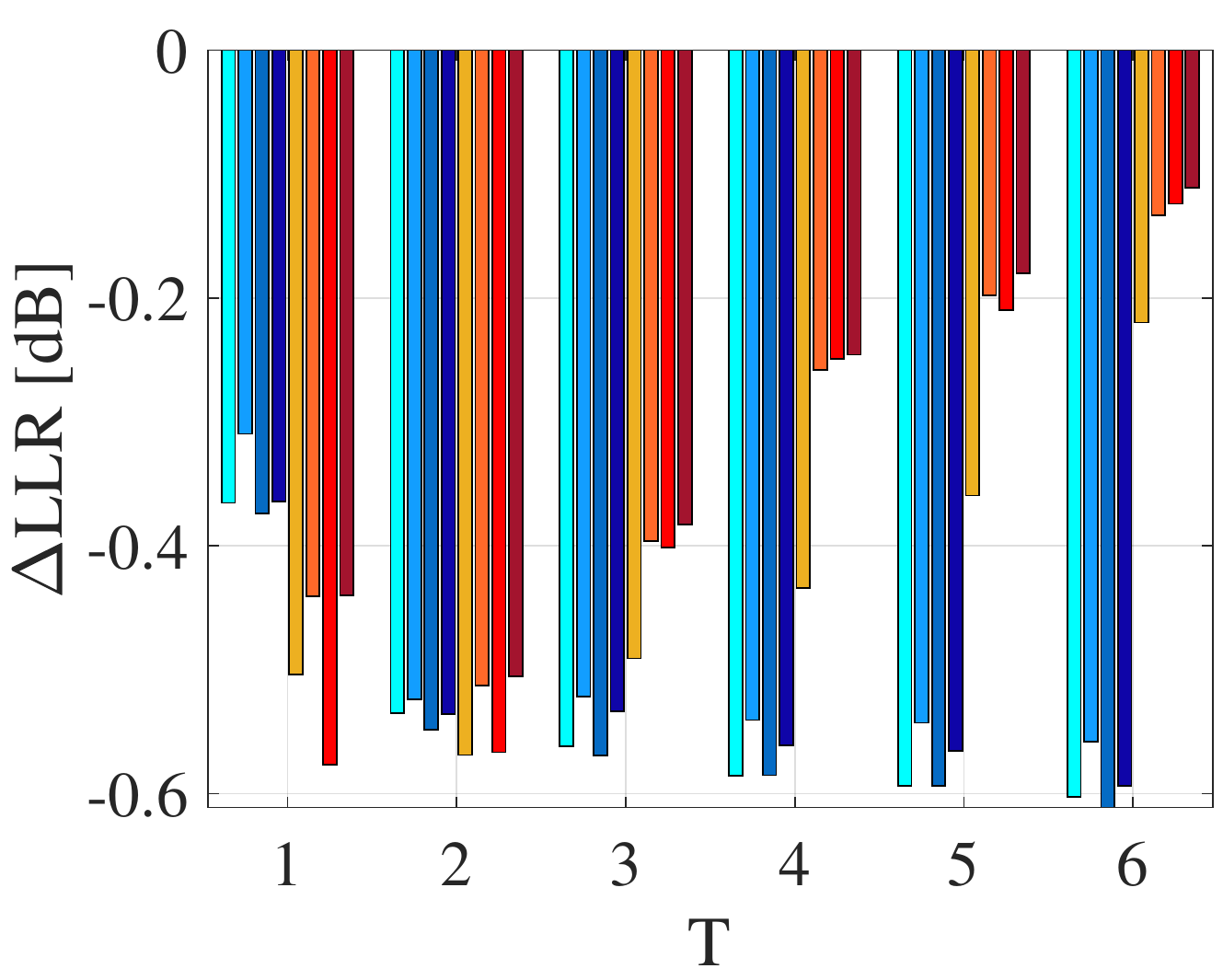}
\label{fig:DeltaLLR}
}
\caption{Results versus $T$, for $T_{60}=610$~msec, $N=8$, $\text{RSNR}=5\text{dB}$:
(a) $\Delta\text{PESQ}$,
(b) $\Delta\text{fwSNR}$,
and
(c) $\Delta\text{LLR}$.
}
\label{fig:Perf_vs_T_610ms}
\end{figure*}

%%%%%%%%%%%%%%%%%%%%%%%%%%%%%%%%%%%%%%%%%%%%%%%%%%%%%%%%%%%%%%%%%%%%%%%%%%%
\subsubsection{Experimental Results}
\label{sec:Results}

First, we examine the performance obtained with various \ac{RSNR} levels.
Tables~\ref{tab:table_T=2_TV2_360ms} and \ref{tab:table_T=2_TV2_610ms}
summarize the results for both reverberation levels.
Note that negative $\Delta\text{LLR}$ indicates a performance improvement.
The best results are highlighted in boldface. 
%%%%%%%%%%%%%%%%%%%%%%%
It is evident that the proposed ML methods outperform the baseline methods of \cite{kodrasi18joint}
for all the considered scenarios.
%%%%
The advantage of the proposed method can be attributed to the fact
that the full $T \times T$ noise PSD matrix is estimated in each frame,
and thus the non-stationarity of all the noise sources can be tracked, whereas the baseline model of a time-invariant
spatial coherence matrix with a single controllable gain parameter cannot track this dynamics.

%%%
In order to analyze the behaviour of the various \ac{ML} implementations,
recall that the non-blocking-based and the blocking-based noise and reverberation \acp{MLE}
were proved to be identical in Section~\ref{sec:comparison}.
%%%%
Since both the non-blocking \ac{ML} DD and the blocking \ac{ML} DD
use also the same speech \ac{PSD} estimate in \eqref{eq:phi_s_inst},
it follows that both methods yield the same performance.
%%%
However, the non-blocking \ac{ML} Dir and the blocking \ac{ML} Dir implementations differ,
due to the fact that they use different speech \ac{PSD} estimators
in \eqref{eq:phi_S_closed-form} and \eqref{eq:phi_S_BM}, respectively.

%%%%%%%
From Tables~III and IV, it is evident that the non-blocking-based \ac{ML} Dir method
is preferable for its high and stable scores, as it obtains the best $\Delta\text{PESQ}$ and $\Delta\text{fwSNRseg}$ 
scores in most cases, and also yields high $\Delta\text{LLR}$ results. 
The advantage of the non-blocking \ac{ML} Dir over the blocking \ac{ML} Dir can be attributed to the fact that the non-blocking \ac{ML} Dir
uses the speech PSD estimate in \eqref{eq:phi_S_closed-form}, which is independent of the estimated noise PSD.
In contrast, the blocking-based \ac{ML} Dir estimates the speech \ac{PSD} with \eqref{eq:phi_S_BM}, which depends on the estimated noise \ac{PSD}, and thus can be affected by estimation errors of the noise \ac{PSD}.
Indeed, there is a slight disadvantage for the non-blocking \ac{ML} Dir in terms of $\Delta\text{LLR}$, which mainly quantifies the performance due to speech distortion. This can be related to the fact that the speech PSD estimate in \eqref{eq:phi_S_closed-form}
generates nulls towards the noise sources, which in some cases can affect speech distortion, for example 
when the speech and noise sources are not fully spatially separated.

%%%
Comparing the direct and the DD implementations of the non-blocking \ac{ML} approach,
it can be observed that there is a clear advantage for the direct implementation in most cases. 
This can be explained by the fact that the direct method optimally estimates the speech \ac{PSD} by maximizing the \ac{ML} criterion,
while the decision-directed approach estimates the speech PSD using \eqref{eq:decision}--\eqref{eq:phi_s_inst},
which is derived in a more heuristic manner. However, in adverse conditions of high noise and reverberation levels,
the DD has some advantage 
due to its use of a smoother estimator compared to the instantaneous estimator of the direct form, 
thus improving the robustness of the speech PSD estimate.
%%%%
To conclude, the non-blocking \ac{ML} Dir is the preferable method among the proposed estimators.

%%%%%%%%%%%%%%%%%%%%%%%%%%%%%%%%%%
Next, the performance is investigated as a function of the number of noise sources.
A representative scenario with $\text{RSNR}=5$~dB and $N=8$ was inspected.
Since our model assumes that $T\leq N-2$, the following values of $T$ were examined:
$T \in \lbrace 1,\ldots,6 \rbrace$.
Figs.~\ref{fig:Perf_vs_T_360ms} and \ref{fig:Perf_vs_T_610ms} depict the results 
for both reverberation levels, where NBB denotes the non-blocking-based method,
and BB refers to the blocking-based method.
%%%
It is shown that the proposed methods outperform the baseline methods \cite{kodrasi18joint} in most cases,
which is inline with the results obtained in Tables~\ref{tab:table_T=2_TV2_360ms} and \ref{tab:table_T=2_TV2_610ms},
showing the advantage of the proposed rank-deficient noise model.
The baseline methods have an advantage only for values of $T$ close to $N-2$.
%%%
This can be attributed to the fact that when the number of non-stationary noise sources increases, 
their sum tends to be stationary, especially in high reverberation conditions.
%%%
In this case, it might be preferred to use the simpler approximation of the noise \ac{PSD} matrix,
namely a time-invariant spatial coherence matrix multiplied by a scalar time-varying \ac{PSD},
rather than the proposed model that tracks the non-stationarity of each of the noises, and requires the estimation of a larger amount of parameters.

\begin{figure*}[t]
\centering
\includegraphics[scale=0.51]{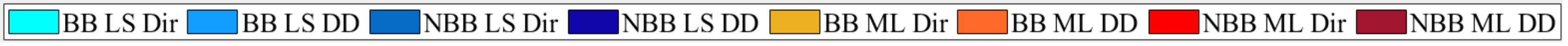}\\

\centering
\subfigure[\hspace{-0.55cm}]{
\includegraphics[scale=0.4]{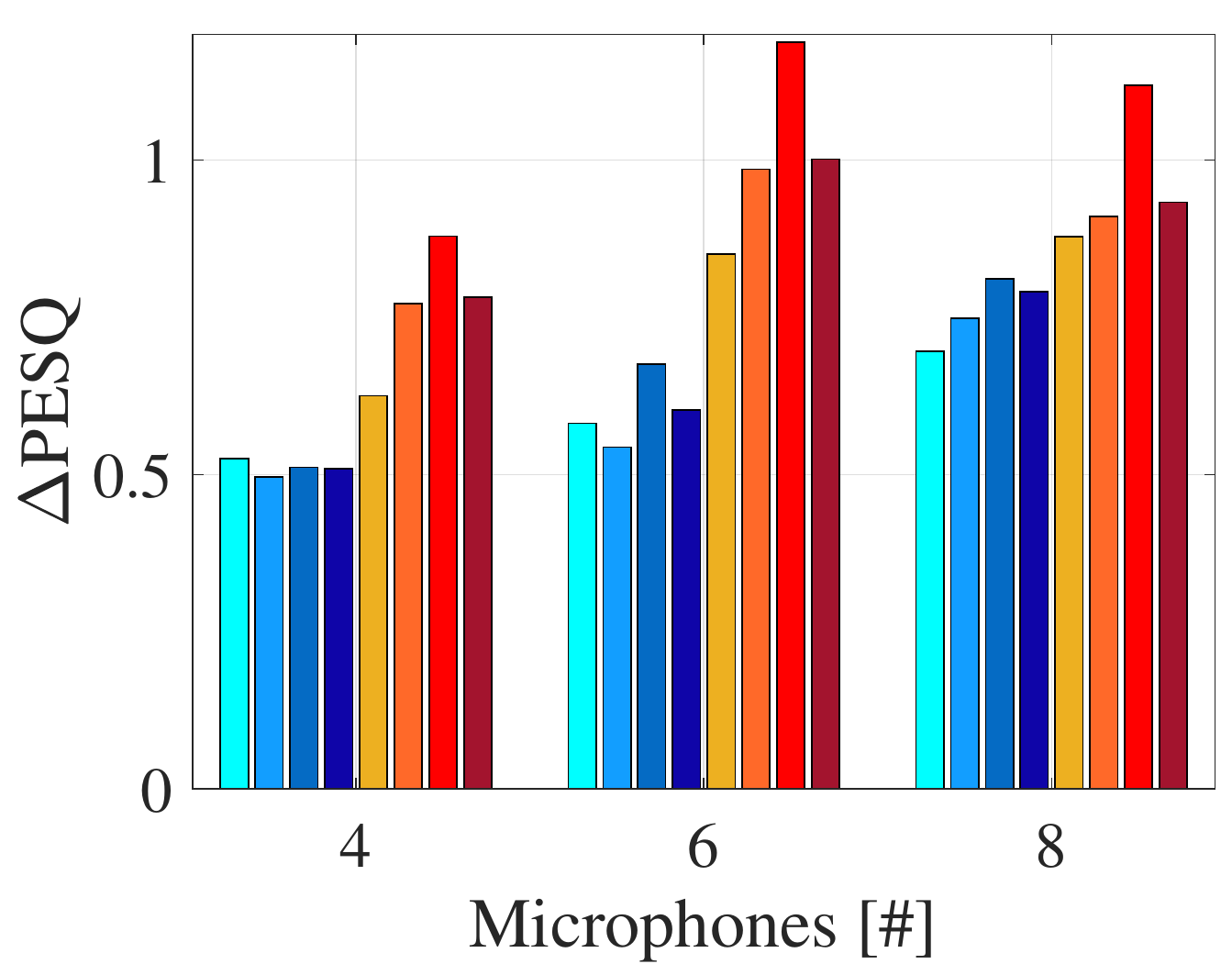}
\label{fig:DeltaPESQ}
}
\subfigure[\hspace{-0.55cm}]{
\includegraphics[scale=0.4]{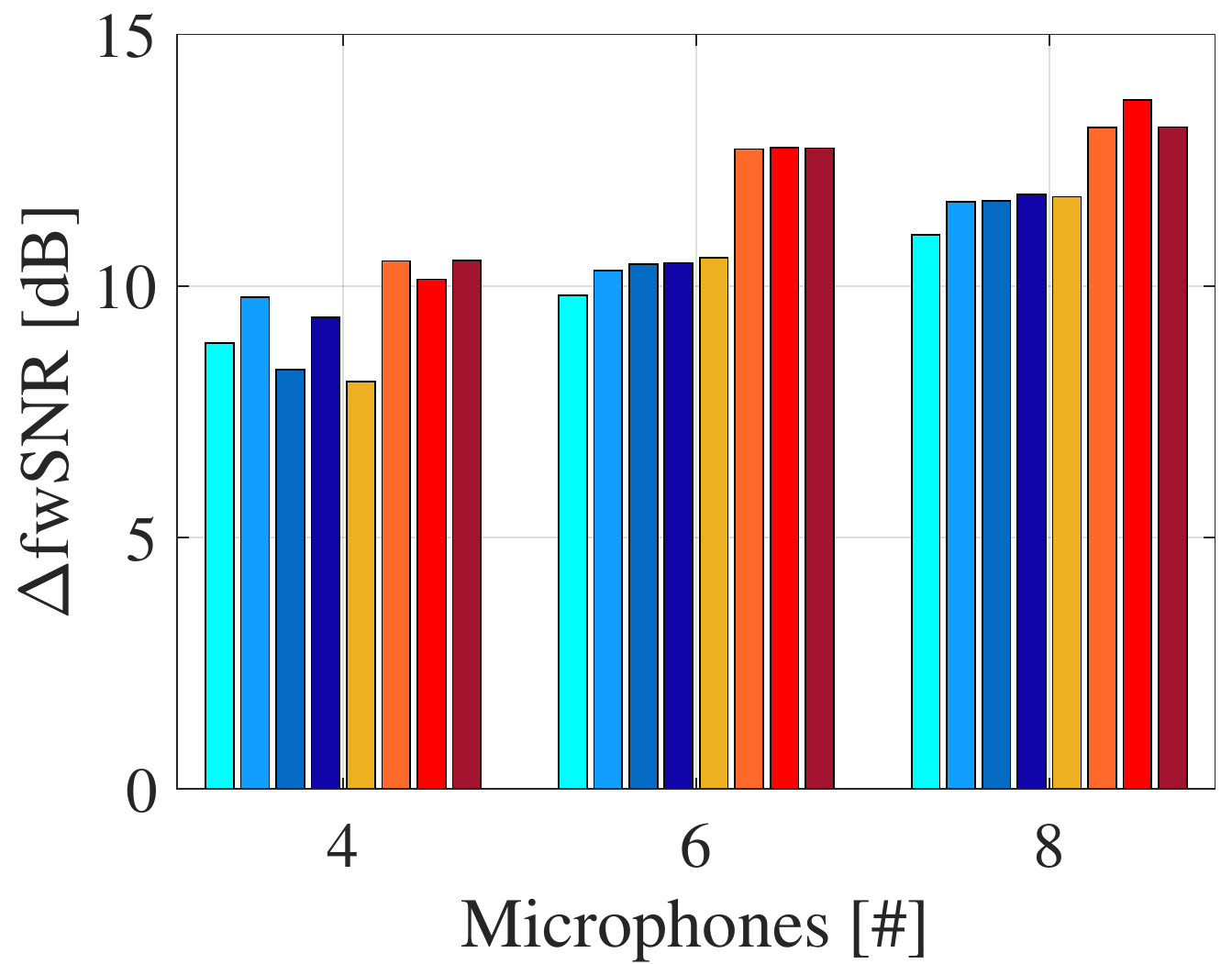}
\label{fig:DeltafwSNR}
}
\subfigure[\hspace{-0.55cm}]{
\includegraphics[scale=0.4]{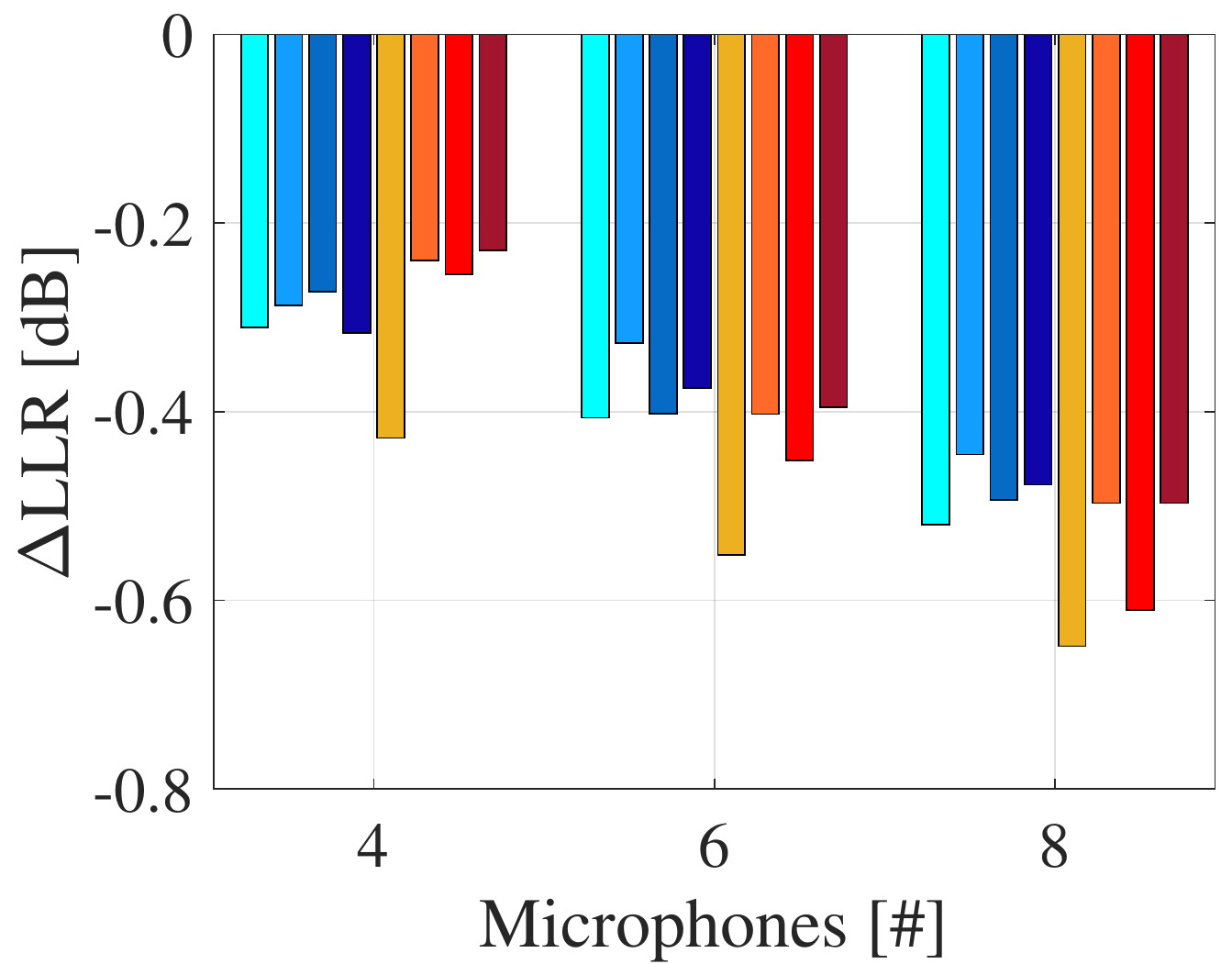}
\label{fig:DeltaLLR}
}
\caption{Results versus $N$, for $T_{60}=360$~msec, $T=2$, $\text{RSNR}=5\text{dB}$:
(a) $\Delta\text{PESQ}$,
(b) $\Delta\text{fwSNR}$,
and
(c) $\Delta\text{LLR}$.
}
\label{fig:Perf_vs_N_360ms}
\end{figure*}

\begin{figure*}[t]
\centering
\includegraphics[scale=0.51]{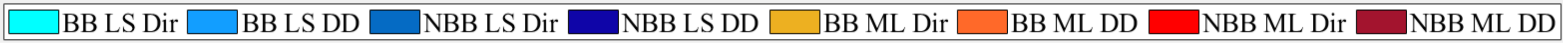}\\

\centering
\subfigure[\hspace{-0.55cm}]{
\includegraphics[scale=0.4]{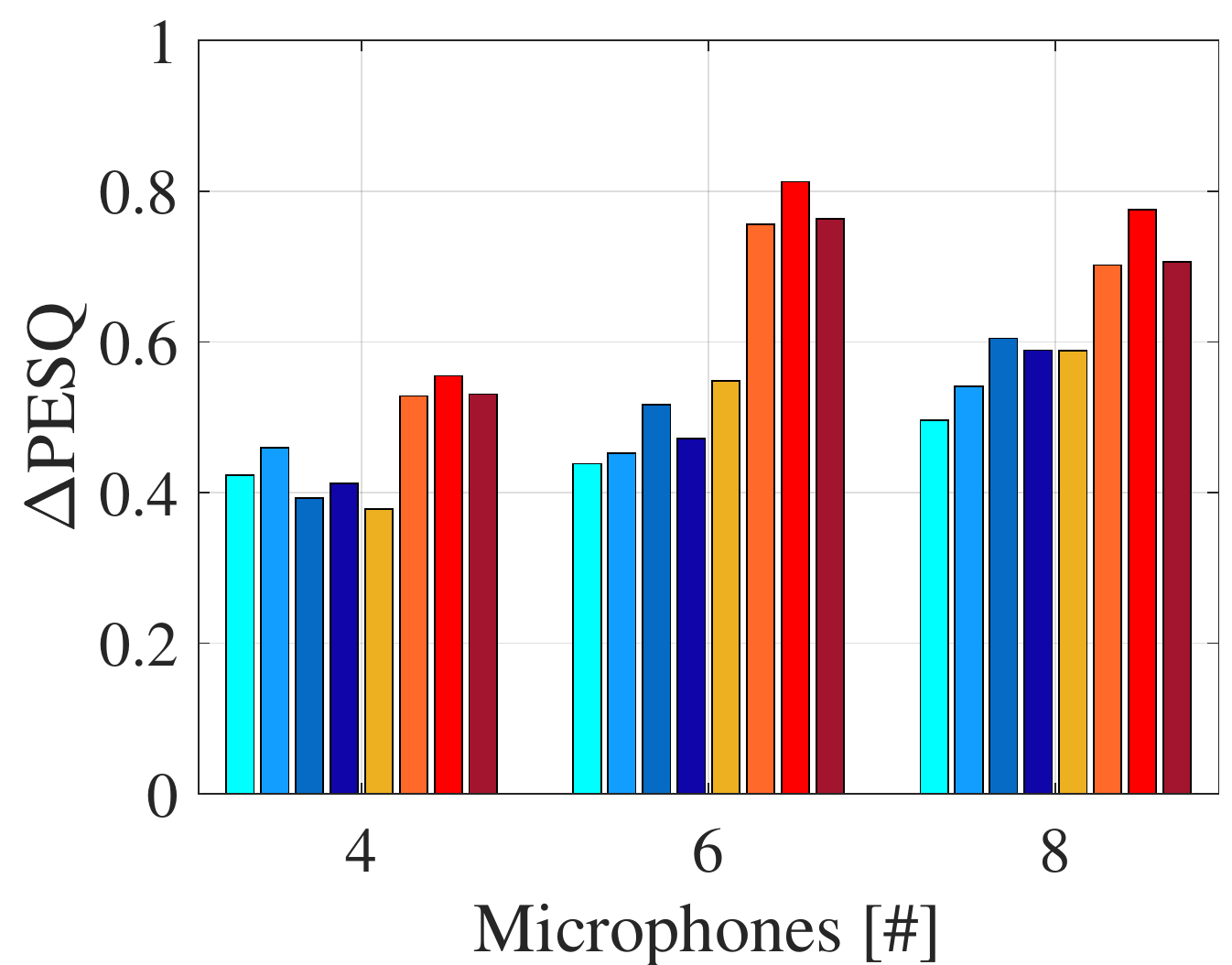}
\label{fig:DeltaPESQ}
}
\subfigure[\hspace{-0.55cm}]{
\includegraphics[scale=0.4]{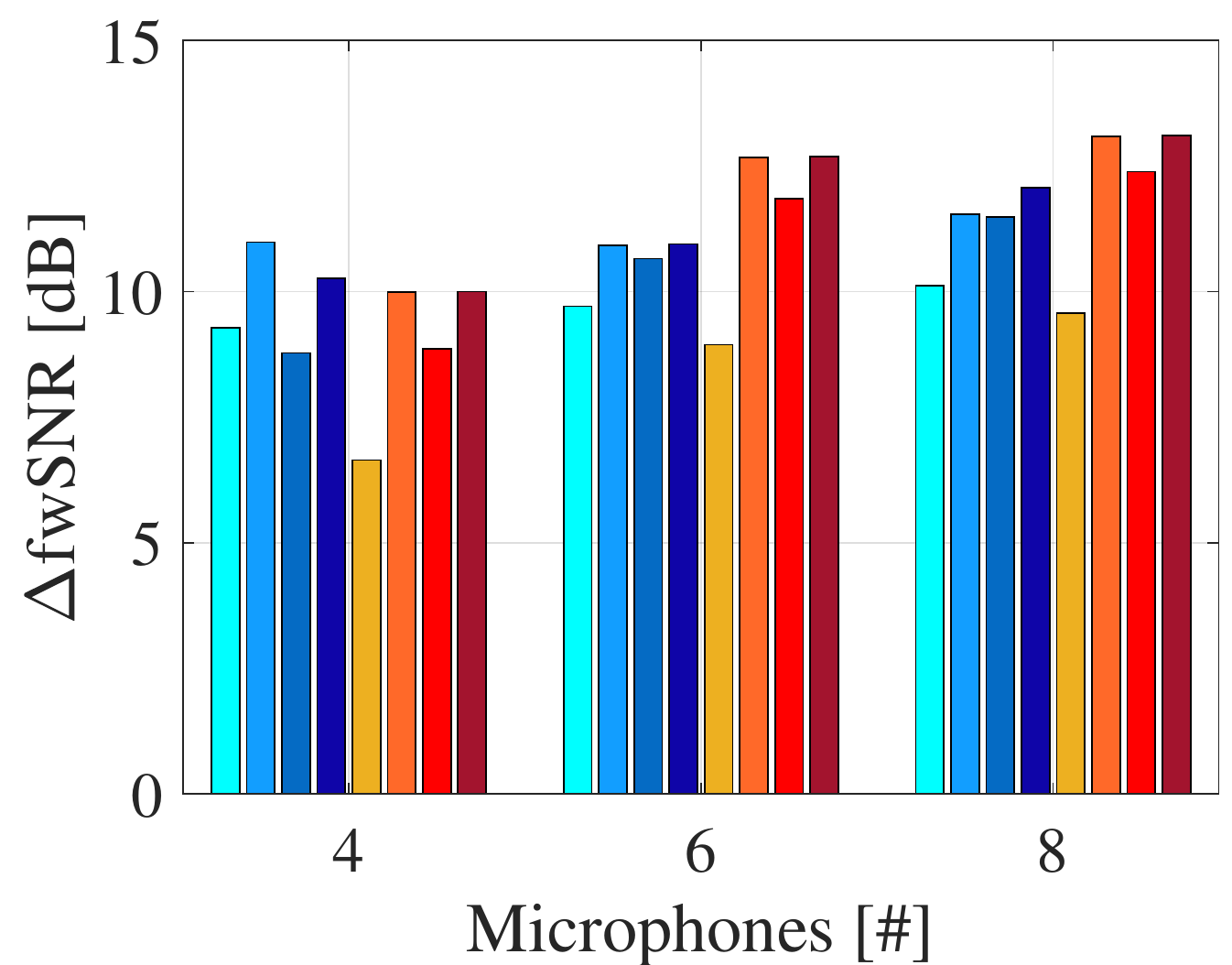}
\label{fig:DeltafwSNR}
}
\subfigure[\hspace{-0.55cm}]{
\includegraphics[scale=0.4]{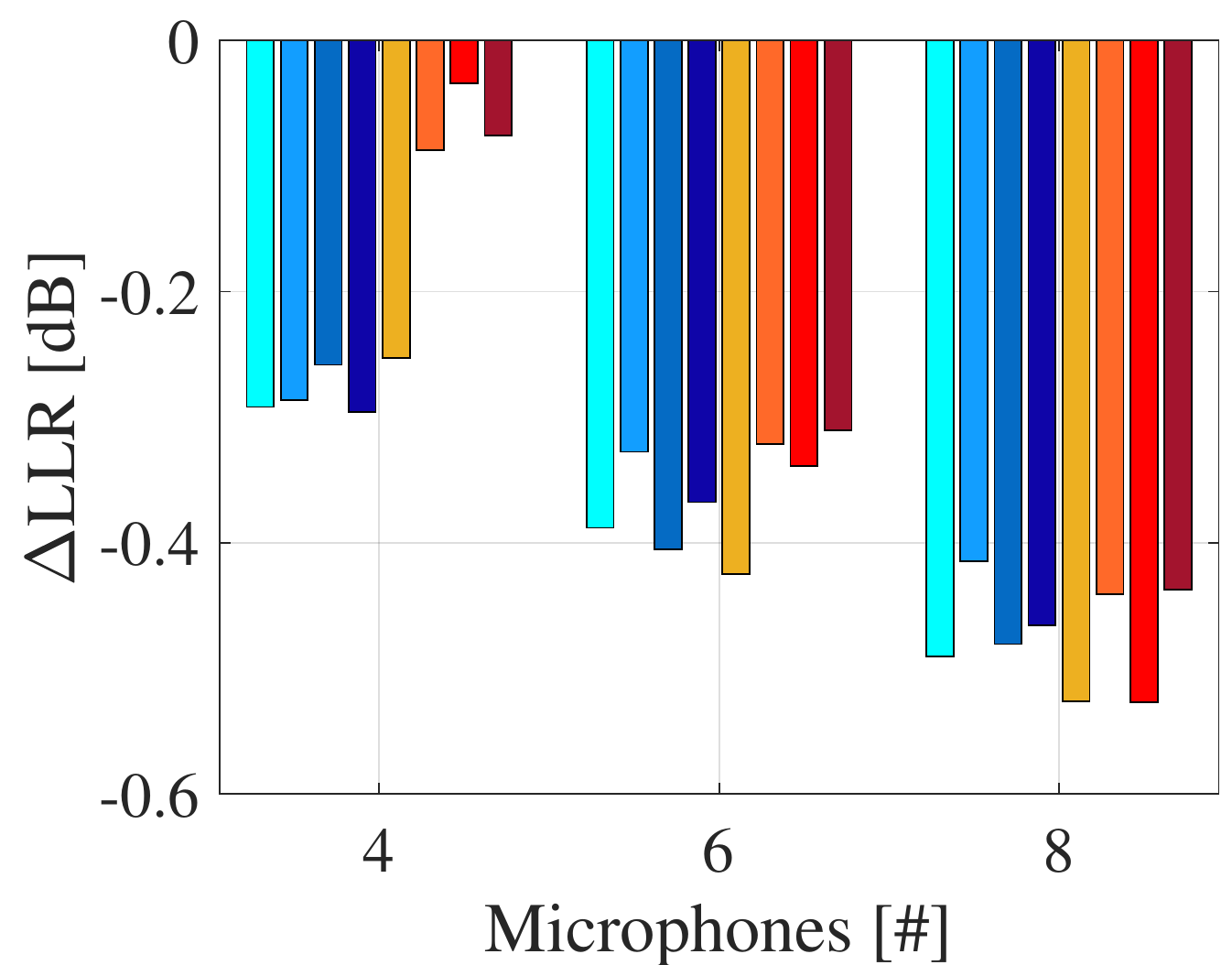}
\label{fig:DeltaLLR}
}
\caption{Results versus $N$, for $T_{60}=610$~msec, $T=2$, $\text{RSNR}=5\text{dB}$:
(a) $\Delta\text{PESQ}$,
(b) $\Delta\text{fwSNR}$,
and
(c) $\Delta\text{LLR}$.
}
\label{fig:Perf_vs_N_610ms}
\end{figure*}

%%%%%%%%%%%%%%%%%%
Finally, we test the influence of the number of microphones on the performance.
Figs.~\ref{fig:Perf_vs_N_360ms} and \ref{fig:Perf_vs_N_610ms} depict 
the measures obtained with different number of microphones, i.e.~$N \in \lbrace 4,6,8 \rbrace$,
for both reverberation times, where $T=2$ and $\text{RSNR}=5$~dB.
It is evident that the proposed methods outperform the baseline methods in almost all cases.

%ver 68

%%%%%%%%%%%%%%%%%%%
%ver 68

\begin{figure*}[t]
\centering
\subfigure[Clean direct speech at microphone \#$1$.]{
\includegraphics[width=.35\textwidth]{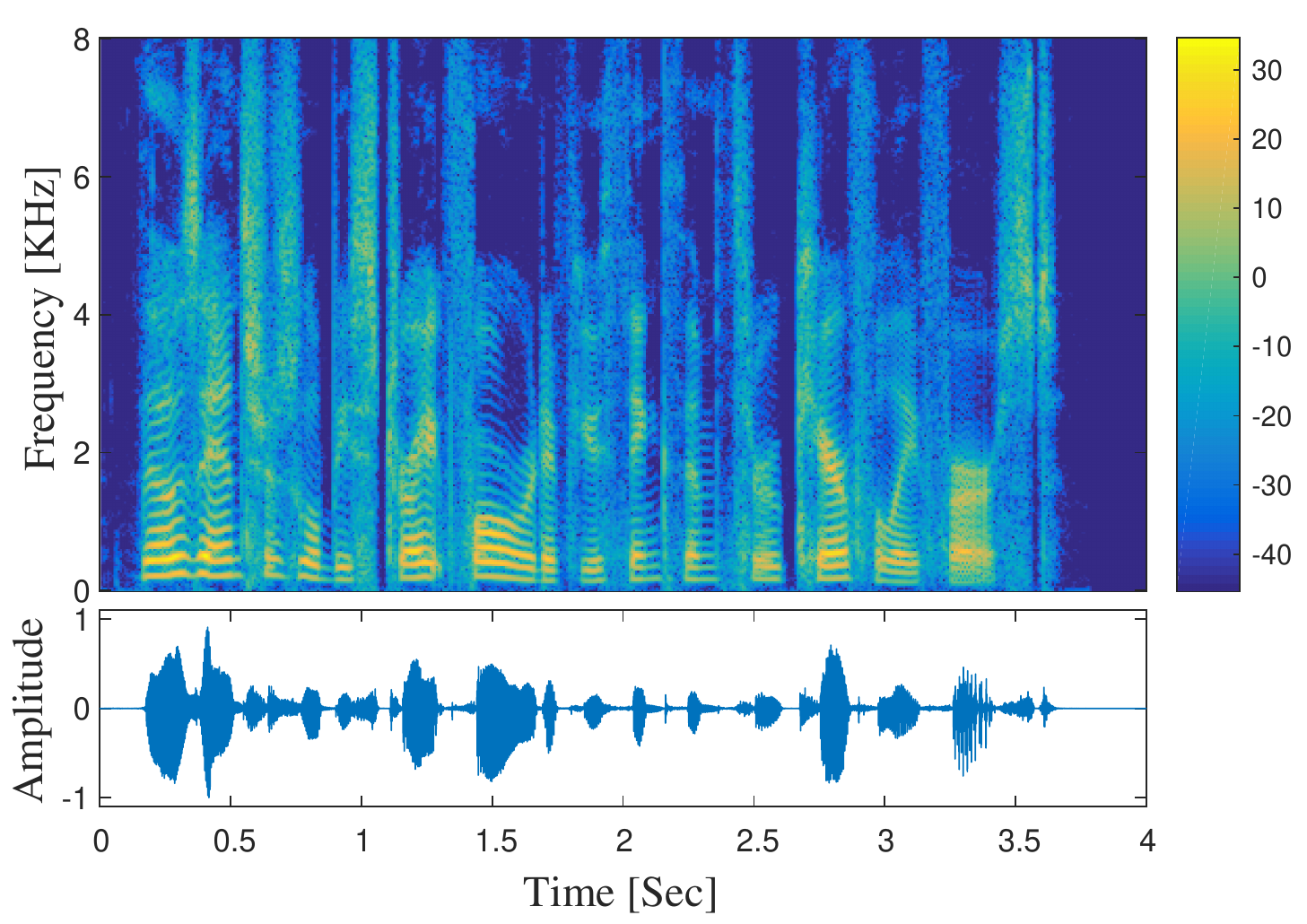}
\label{fig:spectrogram1}
}
\hspace{0.3cm}
\subfigure[Noisy and Reverberant signal at microphone \#$1$.]{
\includegraphics[width=.35\textwidth]{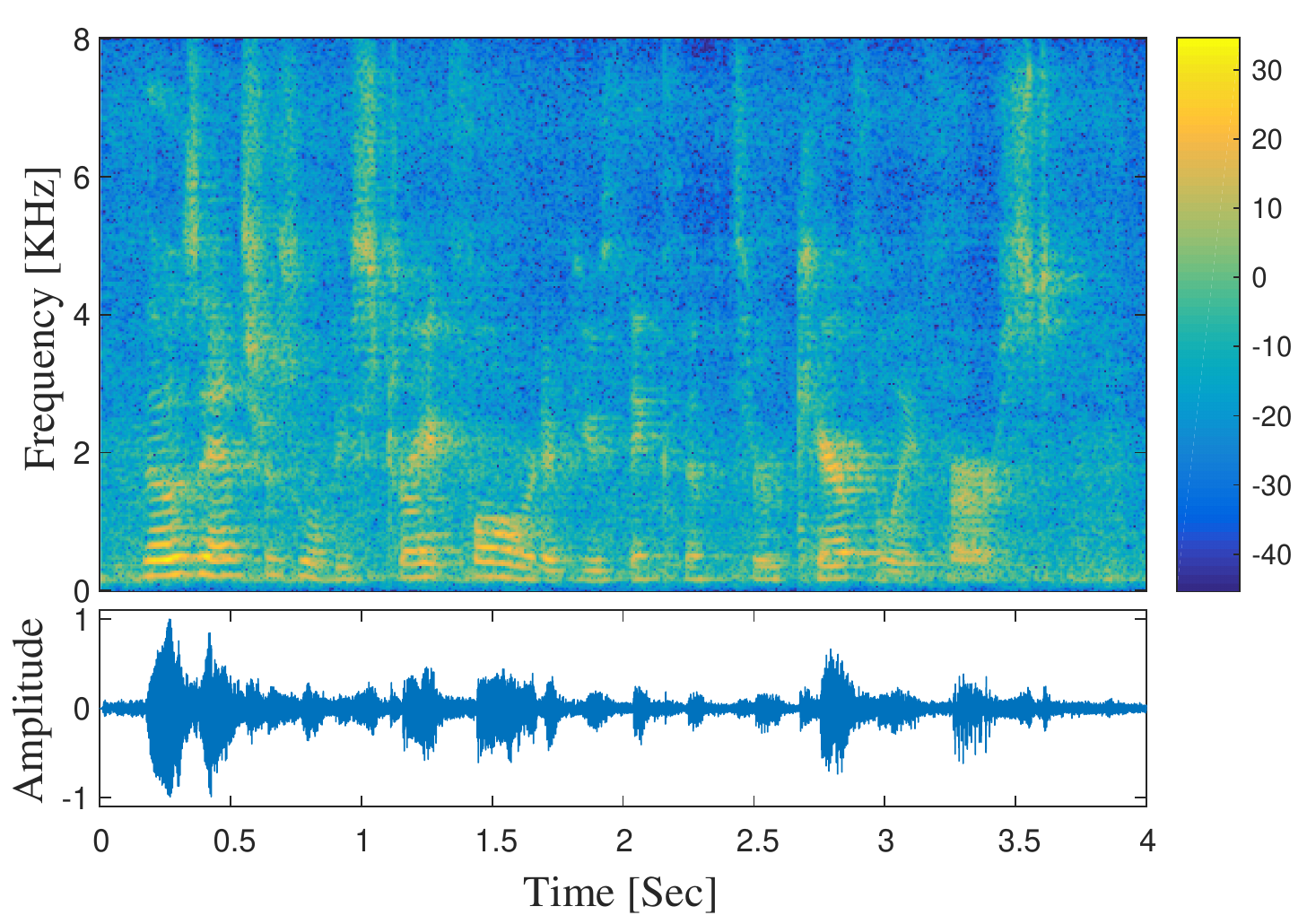}
\label{fig:spectrogram2}
}
\subfigure[Output of the MCWF (direct implementation) using the proposed blocking-based MLEs.]{
\includegraphics[width=.35\textwidth]{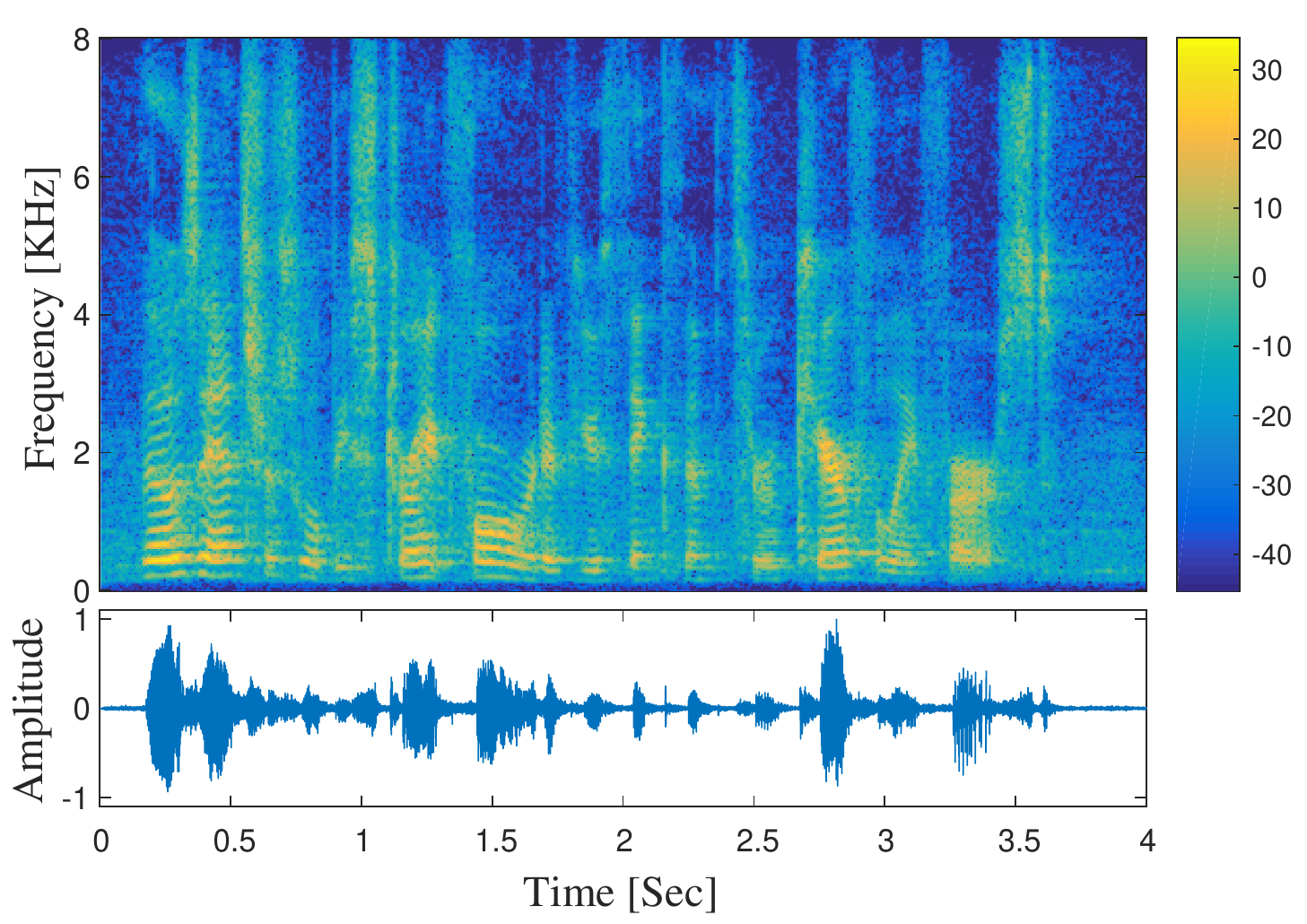}
\label{fig:spectrogram3}
}
\hspace{0.3cm}
\subfigure[Output of the MCWF (direct implementation) using the proposed non-blocking-based MLEs.]{
\includegraphics[width=.35\textwidth]{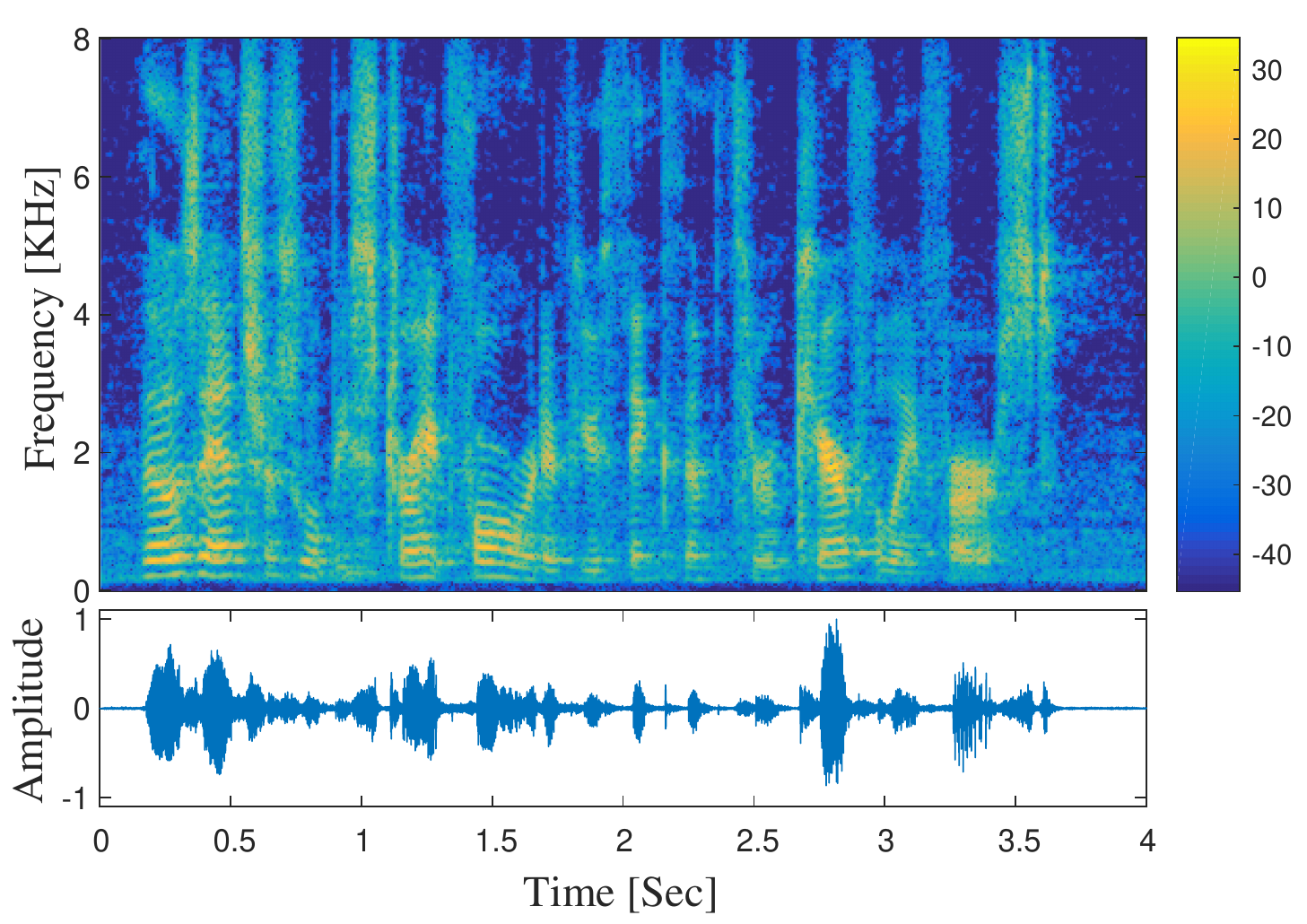}
\label{fig:spectrogram4}
}
\caption{An example of audio sonograms, with $T_{60}=610$~msec and $\text{RSNR}=15\text{dB}$.}
\label{fig:spectrograms}
\end{figure*}

%%%%%
Fig.~\ref{fig:spectrograms} depicts several sonogram examples of the various signals at 
$T_{60}=610$~msec and \ac{RSNR} of $15$~dB.
%%%%
Fig.~\ref{fig:spectrogram1} shows $s$, the direct speech signal as received by the first microphone.
Fig.~\ref{fig:spectrogram2} depicts $y_1$, the noisy and reverberant signal at the first microphone.
Figs.~\ref{fig:spectrogram3} and \ref{fig:spectrogram4} show the \ac{MCWF} output computed with \eqref{eq:gamma_dir},
using the proposed blocking-based and non-blocking-based \acp{MLE}, respectively.
%%%%%%
We conclude that the application of the \ac{MCWF}, implemented based on the proposed \acp{MLE}, 
reduces significantly noise and reverberation, while maintaining low speech distortion.

%%%%%%%%%%%%%%%%%%%%%%%%%%%%%%%%%%%%%%%%%%%%%%%%%%%%%%%%%%%%%%%%%%%%%%%%%%%
\section{Conclusions}
\label{sec:conclusions}

In this contribution, we discussed the problem of joint dereverberation and noise reduction,
in the presence of directional noise sources, forming a rank-deficient noise \ac{PSD} matrix.
As opposed to state-of-the-art methods which assume the knowledge of the noise \ac{PSD} matrix,
we propose to estimate also the time-varying noise \ac{PSD} matrix,
assuming that a basis that spans the noise \ac{ATF} subspace is known.
\acp{MLE} of the reverberation, speech and noise \acp{PSD} are derived for both
the non-blocking-based and the blocking-based methods.
The resulting \acp{MLE} are of closed-form and thus have low computational complexity.
The proposed estimators are theoretically analyzed and compared.
For both the reverberation and the noise \ac{PSD} estimators,
it is shown that the non-blocking-based \ac{MLE} and the blocking-based \ac{MLE} are identical.
The estimators were shown to be unbiased, and the corresponding \acp{MSE} were calculated.
Moreover, \acp{CRB} on the various \acp{PSD} were derived,
and were shown to be identical to the \acp{MSE} of the proposed estimators.
%%%%%%%%%
The discussion is supported by an experimental study based on both simulated data
and real-life audio signals.
It is shown that using the proposed estimators yields
a large performance improvement with respect to a competing method.

%%%%%%%%%%%%%%%%%%%%%%%%%%%%%%%%%%%%%%%%%%%%%%%%%%%%%%%%%%%%%%%%%%%%%%%%%%%
\appendices
\section{} \label{app:Phi_r_No_BM}
The proof follows the lines of \cite{ye1995maximum}.
%%%%%
First, we use \eqref{eq:phi_u_V}
and rewrite the \ac{PSD} matrix of the microphone signals in \eqref{eq:Phi_y_rank_deff} as
\begin{equation}\label{eq:Phi_y2}
 \bm{\Phi}_{\mathbf{y}}(m)  = \vect{A} \bm{\Phi}_{sv}(m) \vect{A}^\hermitian
  + \phi_{R}(m) \, \bm{\Gamma}_R,
\end{equation}
where $\vect{A}$ is defined in \eqref{eq:A_def}
and $\bm{\Phi}_{sv}(m)\in \mathbb{C}^{(T+1)  \times (T+1)}$ is given by
\begin{equation}\label{eq:Phi_S_partition}
\bm{\Phi}_{sv}(m)=
%%%%
\left[
\begin{array}{c|c}
\phi_S(m) & \mathbf{0}^\top\\
\hline
\bm{0}& \bm{\Psi}_{\mathbf{v}}(m)
\end{array}
\right].
\end{equation}
%%%%%
Then, we define $\vect{P} \in \mathbb{C}^{N  \times N}$
as an orthogonal projection matrix onto the speech-plus-noise subspace
\begin{align}\label{eq:P}
\vect{P}=\vect{A}\left(\vect{A}^\hermitian\bm{\Gamma}_R^{-1}\vect{A} \right)^{-1}\vect{A}^\hermitian\bm{\Gamma}_R^{-1},
\end{align}
and $\vect{Q}=\vect{I}_{N}-\vect{P}$ is the orthogonal complement.
We will make use of the following properties, which can be easily verified:
\begin{subequations}
\begin{flalign}
\vect{P}&=\vect{P} \vect{P}, \label{eq:PP_prop}
&&\\
\vect{Q}&=\vect{Q} \vect{Q},  \label{eq:QQ_prop} 
\\
\vect{Q}\bm{\Phi}_{\mathbf{y}}(m)&=\phi_{R}(m)\vect{Q}\bm{\Gamma}_{R}, \label{eq:QPhi_prop} 
&&\\
\bm{\Phi}^{-1}_{\mathbf{y}}(m)\vect{Q}&=\phi_{R}^{-1}(m)\bm{\Gamma}_{R}^{-1}\vect{Q}, \label{eq:PhiQ_prop} 
&&\\
\bm{\Phi}^{-1}_{\mathbf{y}}(m)\vect{A}&=\bm{\Gamma}_{R}^{-1}\vect{A}
\left(\bm{\Phi}_{sv}(m)\vect{A}^\hermitian \bm{\Gamma}_{R}^{-1} \vect{A}+ \phi_{R}(m)\vect{I}\right)^{-1}. &&
\label{eq:PhiA_prop} 
\raisetag{1\normalbaselineskip}
\end{flalign}
\end{subequations}
%%%%%%%%%%%%%%
Using \cite[Eq. (15.47-15.48)]{Kay1993}, 
the derivative of the log of the likelihood function in \eqref{eq:likelihood_y}
 w.r.t.~$\phi_R(m)$ is given by
\begin{flalign}\label{eq:ML_phi_R}
{\cal{L}}\big(\phi_R(m)\big)
&\triangleq
\frac  {\partial  \log p \big(\bar{\mathbf{y}}(m); \bm{\phi}(m) \big)}{  \partial \phi_R(m)  }
\nonumber &&\\
& 
\hspace{-0.2cm}
=L\; \Tr \left[  \left( \bm{\Phi}^{-1}_{\mathbf{y}}(m)  \vect{R}_{\mathbf{y}}(m) - \vect{I}      \right)     \bm{\Phi}^{-1}_{\mathbf{y}}(m)  \bm{\Gamma}_R \right].&&
\raisetag{1\normalbaselineskip}
\end{flalign}
Using \eqref{eq:PP_prop} and \eqref{eq:QQ_prop} 
it follows that
\begin{flalign}\label{eq:ML_phi_R_2}
{\cal{L}}\big(\phi_R(m)\big) 
&= L\; \Tr \Big[\vect{Q}\bm{\Gamma}_R   \left( \bm{\Phi}^{-1}_{\mathbf{y}}(m)  \vect{R}_{\mathbf{y}}(m) - \vect{I}\right)\bm{\Phi}^{-1}_{\mathbf{y}}(m)\vect{Q}  
\nonumber &&\\
&\hspace{0.15cm} +
\vect{P}\bm{\Gamma}_R   \left( \bm{\Phi}^{-1}_{\mathbf{y}}(m)  \vect{R}_{\mathbf{y}}(m) - \vect{I}\right)\bm{\Phi}^{-1}_{\mathbf{y}}(m)\vect{P}  \Big].&&
\raisetag{1.3\normalbaselineskip}
\end{flalign}
However, by substituting \eqref{eq:P} into \eqref{eq:ML_phi_R_2}, it can be shown that the second term vanishes
(follows from setting the derivative of the log likelihood w.r.t. $\bm{\Phi}_{sv}$ in \eqref{eq:ML_phi_S_V} to zero).
%%%%%%%%%%%%%%%%%%%%%
Hence,
\begin{align}\label{eq:ML_phi_R_3}
{\cal{L}}\big(\phi_R(m)\big)
&=
L\;\Tr \Big[\vect{Q}\bm{\Gamma}_R \bm{\Phi}^{-1}_{\mathbf{y}}(m)   
\big(   \vect{R}_{\mathbf{y}}(m) - \bm{\Phi}_{\mathbf{y}}(m)\big)
\nonumber\\
&\hspace{1cm}\times \bm{\Phi}^{-1}_{\mathbf{y}}(m)\vect{Q}  \Big]
\nonumber\\
&\hspace{-1cm}\stackrel{(a)}{=}
L\phi_{R}^{-2}(m)\Tr \left[\vect{Q}   
\big(   \vect{R}_{\mathbf{y}}(m) - \bm{\Phi}_{\mathbf{y}}(m)\big)
\bm{\Gamma}_{R}^{-1}\vect{Q} \right]
\nonumber\\
&\hspace{-1cm}\stackrel{(b)}{=}
L\phi_{R}^{-2}(m)
\Tr \left[\vect{Q}\vect{R}_{\mathbf{y}}(m)\bm{\Gamma}_{R}^{-1}
-  \phi_{R}(m)\vect{Q}
\right],
\end{align}
where (a) follows from \eqref{eq:QPhi_prop} and \eqref{eq:PhiQ_prop},
and (b) follows from \eqref{eq:QQ_prop} and \eqref{eq:QPhi_prop}.
Finally, using \eqref{eq:Q1} we have $\Tr \left[\vect{Q} \right ]= N-(T+1)$.
Thus, setting \eqref{eq:ML_phi_R_3} to zero yields \eqref{eq:phi_R_closed-form}.

%%%%%%%%%%%%%%%%%%%%%%%%%%%
\section{} \label{app:Phi_S_No_BM}
Using \cite[Eq. (12)]{ye1995maximum}, the derivative of the log likelihood function 
$\log p\big(\bar{\mathbf{y}}(m);\bm{\phi}(m) \big)$ w.r.t.~$\bm{\Phi}_{sv}$ is given by
\begin{flalign}\label{eq:ML_phi_S_V}
{\cal{L}}\big(\bm{\Phi}_{sv}(m)\big) 
&\triangleq
\frac  {\partial  \log p \big(\bar{\mathbf{y}}(m);\bm{\phi}(m) \big)}{  \partial \bm{\Phi}_{sv}(m)  }
\nonumber &&\\
&\hspace{-0.5cm} = L \left(\vect{A}^\hermitian 
 \left( \bm{\Phi}^{-1}_{\mathbf{y}}(m)  \vect{R}_{\mathbf{y}}(m) - \vect{I}      \right)     \bm{\Phi}^{-1}_{\mathbf{y}}(m)
\vect{A} 
\right)^\top. &&
\raisetag{1.05\normalbaselineskip}
\end{flalign}
Using \eqref{eq:PhiA_prop} and setting the result to zero yields
\begin{align}\label{eq:Phi_S_closed-form}
\bm{\Phi}^{\mathrm{ML}}_{sv}(m)=
%%%%%%%%%%%%%%
&\left(\vect{A}^\hermitian\bm{\Gamma}_R^{-1}\vect{A} \right)^{-1}\vect{A}^\hermitian\bm{\Gamma}_R^{-1}
\left(\vect{R}_{\mathbf{y}}(m)-\phi^{\mathrm{ML},\bar{\mathbf{y}}}_{R}(m)\bm{\Gamma}_R  \right)
\nonumber \\
&\times
\bm{\Gamma}_R^{-1}\vect{A}\left(\vect{A}^\hermitian\bm{\Gamma}_R^{-1}\vect{A} \right)^{-1}.
\end{align}
%%%%%%%%%%%
In order to simplify the expression of \eqref{eq:Phi_S_closed-form}, we define a partitioned matrix
\begin{equation}\label{eq:partition}
\vect{A}^\hermitian\bm{\Gamma}_R^{-1}\vect{A}=
\left[
\begin{array}{c|c}
\mathbf{g}_{d}^\hermitian\bm{\Gamma}_R^{-1}\mathbf{g}_{d} & \mathbf{g}_{d}^\hermitian\bm{\Gamma}_R^{-1}\vect{V} \\
\hline
\vect{V}^\hermitian\bm{\Gamma}_R^{-1}\mathbf{g}_{d} & \vect{V}^\hermitian\bm{\Gamma}_R^{-1}\vect{V} 
\end{array}
\right].
\end{equation}
Using the formula of the inverse of a partitioned matrix
and then taking the $(1,1)$ entry of $\bm{\Phi}^{\mathrm{ML}}_{sv}$,
the \ac{MLE} of $\phi_{S}$ in \eqref{eq:phi_S_closed-form} is obtained.

%%%%%%%%%%%%%%%%%%%%%%%%%%%
\section{} \label{app:Phi_v_No_BM}

We note that $\bm{\Psi}_{\mathbf{v}}(m)$ is the $T \times T$ lower-right block of the full $(T+1) \times (T+1)$
matrix $\bm{\Phi}_{sv}(m)$ in \eqref{eq:Phi_S_partition}.
Using again the formula for the inverse of the partitioned matrix in \eqref{eq:partition}
and taking the corresponding entries, yields \eqref{eq:phi_v1}.

%%%%%%%%%%%%%%%%%%%%%%%%%%%%%%%%%%
\section{}\label{app:Q_simplified}
The proof follows by substituting the inverse of the partitioned matrix \eqref{eq:partition}
into \eqref{eq:Q1}, and then using the definitions in \eqref{eq:MVDR_S} and \eqref{eq:LCMV_V}.

%%%%%%%%%%%%%%%%%%%%%%%%%%%
\section{} \label{app:Phi_r_BM}

The \ac{PSD} matrix of the \ac{BM} output is given in \eqref{eq:Phi_u2} as
\begin{equation}\label{eq:Phi_u3}
 \bm{\Phi}_{\mathbf{z}}(m)  
   = \widetilde{\vect{V}}\bm{\Psi}_{\mathbf{v}}(m)\widetilde{\vect{V}}^{\hermitian}
  + \phi_{R}(m) \,\vect{B}^\hermitian  \bm{\Gamma}_R \vect{B},
\end{equation}
where $\widetilde{\vect{V}}$ is defined in \eqref{eq:V_til_BV}.
The proof is now similar to that of Appendix~\ref{app:Phi_r_No_BM}, 
with the following changes:
$\vect{A}$, $\bm{\Phi}_{sv}(m)$, $\bm{\Gamma}_R$ and $\vect{R}_{\mathbf{y}}(m)$ are replaced with
$\widetilde{\vect{V}}$, $\bm{\Psi}_{\mathbf{v}}(m)$, $\vect{B}^\hermitian \bm{\Gamma}_R\vect{B}$
and $\vect{R}_{\mathbf{z}}(m)$, respectively.

%%%%%%%%%%%%%%%%%%%%%%%%%%%
\section{} \label{app:Phi_v_BM}

Using \eqref{eq:Phi_u2}, the \ac{MLE} of $\bm{\Psi}_{\mathbf{v}}(m)$ can be calculated 
in a similar manner to \eqref{eq:Phi_S_closed-form}, with
\begin{align}\label{eq:phi_tilde_ML}
\bm{\Psi}_{\mathbf{v}}^{\mathrm{ML},\bar{\mathbf{z}}}(m)
=
\widetilde{\vect{W}}_{\mathbf{u}}^\hermitian
\left(\vect{R}_{\mathbf{z}}(m)-\phi^{\mathrm{ML},\bar{\mathbf{z}}}_{R}(m)
\left(\vect{B}^\hermitian\bm{\Gamma}_R \vect{B}\right) \right)
\widetilde{\vect{W}}_{\mathbf{u}},
\end{align}
where $\widetilde{\vect{W}}_{\mathbf{u}}$ is given in \eqref{eq:LCMV_BM}.

%%%%%%%%%%%%%%%%%%%%%%%%%%%
\section{} \label{app:w_B_relation}
Substituting \eqref{eq:V_til_BV} into \eqref{eq:LCMV_BM}, and then using \eqref{eq:Identity_Phi_R2},
leads to
\begin{equation}\label{eq:LCMV_BM2}
\widetilde{\vect{W}}_{\mathbf{u}}^\hermitian= 
\left(\vect{V}^\hermitian
\bm{P}_{g}^{\perp}\bm{\Gamma}_R^{-1}
\vect{V} \right)^{-1}
\vect{V}^\hermitian \vect{B}
\left(\vect{B}^\hermitian\bm{\Gamma}_R \vect{B}\right)^{-1}.
\end{equation}
Right multiplying \eqref{eq:LCMV_BM2} by $\vect{B}^\hermitian$,
using again \eqref{eq:Identity_Phi_R2} and then comparing to \eqref{eq:LCMV_V},
yields \eqref{eq:w_relation}.

%%%%%%%%%%%%%%%%%%%%%%%%%%%
\section{} \label{app:lambda_v_variance}

Using \eqref{eq:smooth0}, \eqref{eq:phi_R_closed-form} and the i.i.d. assumption, 
the variance of \eqref{eq:lambda_v1} can be recast as
\begin{align}
&\text{var}\left(\psi^{\mathrm{ML},\bar{\mathbf{y}}}_{i}(m)\right)
\nonumber \\
&=\frac{1}{L}
\text{var}
\Bigg[
\mathbf{y}^\hermitian (m) 
\Big(
\vect{w}_{i}\vect{w}_{i}^\hermitian 
-\frac{\vect{w}_{i}^\hermitian \bm{\Gamma}_R\vect{w}_{i}}{N-1-T}\bm{\Gamma}_R^{-1}\bm{Q} 
\Big)
\mathbf{y}(m)
\Bigg].
\end{align}
To proceed, we use \eqref{eq:Kay_identity_15.30}, \eqref{eq:Gamma_r_Q_identity},
and \eqref{eq:w_mvdr_s_Q_w_lcmv_v_Q} to obtain
\begin{align}
&\text{var}\left(\psi^{\mathrm{ML},\bar{\mathbf{y}}}_{i}(m)\right)
\nonumber \\
&=\frac{1}{L}
\Bigg[
\left(\vect{w}_{i}^\hermitian   \bm{\Phi}_{\mathbf{y}}(m) \vect{w}_{i}\right)^2
+
\frac{\left(\vect{w}_{i}^\hermitian
\phi_R(m)\bm{\Gamma}_R\vect{w}_{i}\right)^2}
{N-1-T}
\Bigg].
\end{align}
Finally, using \eqref{eq:Phi_y_rank_deff} and \eqref{eq:w_lcmv_dot_gd,V}
yields \eqref{eq:lambda_V_variance}.
%%%%
The right hand side of \eqref{eq:xi} is obtained by substituting \eqref{eq:wt} into the definition 
of $\xi_i$, and noting that 
$\left(\bm{P}_{\bar{\vect{V}}_{i}}^{\perp}\bm{P}_{g}^{\perp}\right)^\hermitian
=\bm{\Gamma}_R\bm{P}_{g}^{\perp}\bm{P}_{\bar{\vect{V}}_{i}}^{\perp}   \bm{P}_{g}^{\perp}\bm{\Gamma}_{R}^{-1}$.

%%%%%%%%%%%%%%%%%%%%%%%%%%%%%%%%%%%%%%%%%%%%%%%%%%%%%%%%%%%%%%%
\section{} \label{app:CRB_phi_R}

In this Appendix, the \ac{CRB} on the reverberation \ac{PSD} is derived.
We will make use of the following identities\cite{graham2018kronecker}:
\begin{align}
\Tr \left[ \vect{X}\vect{Y} \right]&=\left(\text{vec}\left(\vect{X}^\hermitian \right)\right)^\hermitian \text{vec}\left(\vect{Y}\right),
\label{eq:kronecker_1}\\
\text{vec}\left(\vect{X}\vect{Y}\vect{Z}\right)&=\left(\vect{Z}^\top \otimes \vect{X} \right)\text{vec}\left(\vect{Y}\right),
\label{eq:kronecker_2}\\
\left(\vect{W} \otimes \vect{X} \right)\left(\vect{Y} \otimes \vect{Z} \right)&=
\left(\vect{W}\vect{Y} \right)\otimes\left(\vect{X} \vect{Z} \right)\label{eq:kronecker_3}.
\end{align}

%%%%%%%%%%%%
Using the definitions of $\bm{\Phi}_{\mathbf{y}}$ and $\bm{\Phi}_{sv}$
in \eqref{eq:Phi_y2} and \eqref{eq:Phi_S_partition},
we denote the set of unknown parameters by
$\bm{\alpha}\triangleq \left[\phi_R,\bm{\phi}_{SV}^\top  \right]^\top
\in \mathbb{C}^{\left((T+1)^2+1\right)}$,
where $\bm{\phi}_{SV}\triangleq\text{vec}\left(\bm{\Phi}_{sv}\right)\in \mathbb{C}^{(T+1)^2}$.
%%%%%%
As $\bm{\Phi}_{\mathbf{y}}$ is a \ac{PSD} matrix of a Gaussian vector,
the Fisher information of each pair of parameters is given by \cite{bangs1972array}:
\begin{equation}\label{eq:fisher1}
  \mathrm{I}^{\bar{\mathbf{y}}}_{ij} = L \; \Tr \left[  \bm{\Phi}^{-1}_{\mathbf{y}}  \frac{\partial \bm{\Phi}_{\mathbf{y}} }{ \partial \alpha_{i}  }  \bm{\Phi}^{-1}_{\mathbf{y}}   \frac{\partial \bm{\Phi}_{\mathbf{y}} }{ \partial \alpha_{j}  }       \right],
\end{equation}
where $\mathrm{I}^{\bar{\mathbf{y}}}_{ij}$ is the Fisher information of $\alpha_{i}$ and $\alpha_{j}$ 
and $i,j = 1,\ldots,(T+1)^2+1$.
%%%%%
In order to facilitate the derivation, we use \eqref{eq:kronecker_2} and
vectorize \eqref{eq:Phi_y2} to obtain an $N^2 \times 1$ vector:
\begin{equation}\label{eq:Phi_y_vectorize}
 \bm{\phi}_{y}
 \triangleq 
 \text{vec}\left(\bm{\Phi}_{\mathbf{y}}\right)  =
 \left(\vect{A}^{*} \otimes  \vect{A} \right)
 \text{vec}\left(\bm{\Phi}_{sv} \right)
 +\phi_{R}\, \text{vec}\left(\bm{\Gamma}_R\right).
\end{equation}
%%%%%
Using \eqref{eq:kronecker_1}, \eqref{eq:kronecker_2}, \eqref{eq:fisher1}
and \eqref{eq:Phi_y_vectorize}, the full 
\ac{FIM} writes \cite{stoica2001stochastic,gershman2002stochastic}
\begin{equation}\label{eq:fisher2}
\frac{1}{L}  \mathrm{I}^{\bar{\mathbf{y}}} = 
  \left(\frac{\partial  \bm{\phi}_{y}}{ \partial \bm{\alpha}^\top } \right)^\hermitian
  \left(\bm{\Phi}^{-\top}_{\mathbf{y}} \otimes \bm{\Phi}^{-1}_{\mathbf{y}} \right)
  \left(\frac{\partial  \bm{\phi}_{y}}{\partial \bm{\alpha}^\top} \right).
\end{equation}
%%%
Next, we define the following partitioned matrix:
\begin{equation}\label{eq:partitioned_CRB}
\left[
\mathbf{g} \middle| \bm{\Delta} 
\right]
\triangleq
\left(\bm{\Phi}^{-\top /2}_{\mathbf{y}} \otimes \bm{\Phi}^{-1/2}_{\mathbf{y}} \right)
\left[
\frac{\partial  \bm{\phi}_{y}}{\partial \phi_R} \middle| \frac{\partial  \bm{\phi}_{y}}{\partial \bm{\phi}_{SV}^\top }
\right].
\end{equation}
%%%%%%%%%%%%%%%%
Using \eqref{eq:partitioned_CRB} and \eqref{eq:kronecker_3}, 
the \ac{FIM} in \eqref{eq:fisher2} writes
\begin{align}\label{eq:fisher3}
\frac{1}{L}  \mathrm{I}^{\bar{\mathbf{y}}} = 
  \left[
\begin{array}{c}
\mathbf{g}^\hermitian\\
\hline
\bm{\Delta}^\hermitian
\end{array}
\right]
\left[
\mathbf{g} \middle| \bm{\Delta} 
\right]
= \left[
\begin{array}{c|c}
\mathbf{g}^\hermitian\mathbf{g}& \mathbf{g}^\hermitian\bm{\Delta}\\
\hline
\bm{\Delta}^\hermitian\mathbf{g}& \bm{\Delta}^\hermitian \bm{\Delta}
\end{array}
\right].
\end{align}
%%%%%%%%%%%%%%%%%%%%%
The \ac{CRB} for $\phi_R$ is given by
$\text{CRB}\left(\phi_R \right)=\left[\left(\mathrm{I}^{\bar{\mathbf{y}}}\right)^{-1}\right]_{1,1}$.
Using the formula of the inverse of a partitioned matrix, the \ac{CRB} writes
\begin{align}\label{eq:CRB1}
\text{CRB}\left(\phi_R \right)
&=\frac{1}{L} 
\left(\mathbf{g}^\hermitian \mathbf{g}-
\mathbf{g}^\hermitian \bm{\Delta} \left(\bm{\Delta}^\hermitian \bm{\Delta}\right)^{-1} \bm{\Delta}^\hermitian  \mathbf{g}\right)^{-1}
\nonumber \\
&=\frac{1}{L} \left(\mathbf{g}^\hermitian \bm{\Pi}_{\bm{\Delta}}^{\perp}\mathbf{g} \right)^{-1},
\end{align}
where $\bm{\Pi}_{\bm{\Delta}}^{\perp}\in \mathbb{C}^{N^2 \times N^2}$ is a projection matrix onto the subspace orthogonal to $\bm{\Delta}$:
\begin{equation}\label{eq:Q_Delta}
\bm{\Pi}_{\bm{\Delta}}\triangleq
\bm{\Delta} \left(\bm{\Delta}^\hermitian \bm{\Delta}\right)^{-1} \bm{\Delta}^\hermitian \hspace{0.15cm} , \hspace{0.15cm}
\bm{\Pi}_{\bm{\Delta}}^{\perp}=
\vect{I}_{N^2}-\bm{\Pi}_{\bm{\Delta}}.
\end{equation}

%%%%%%%%%%%%%%%%%%%%%%%%%%%%%%%%%%%%%%%%%%%%%%%%%%%%%%
We now simplify \eqref{eq:CRB1}.
First, we use \eqref{eq:Phi_y_vectorize} along with \eqref{eq:kronecker_2} to write $\mathbf{g}$ as
\begin{align}\label{eq:g}
\mathbf{g}
&=\left(\bm{\Phi}^{-\top /2}_{\mathbf{y}} \otimes \bm{\Phi}^{-1/2}_{\mathbf{y}} \right)
 \text{vec}\left(\bm{\Gamma}_R\right)
 \nonumber \\
&
=\text{vec}\left(\bm{\Phi}^{-1/2}_{\mathbf{y}} \bm{\Gamma}_R \bm{\Phi}^{-1/2}_{\mathbf{y}}\right).
\end{align}
%%%
Similarly, $\bm{\Delta}$ is simplified by using \eqref{eq:Phi_y_vectorize} along with \eqref{eq:kronecker_3},
\begin{align}\label{eq:Delta}
\bm{\Delta}
&=\left(\bm{\Phi}^{-\top /2}_{\mathbf{y}} \otimes \bm{\Phi}^{-1/2}_{\mathbf{y}} \right)
 \left(\vect{A}^{*} \otimes  \vect{A} \right)
 \nonumber \\
 &=\left(\bm{\Phi}^{-\top /2}_{\mathbf{y}}\vect{A}^{*}  \right)
 \otimes \left(\bm{\Phi}^{-1/2}_{\mathbf{y}} \vect{A} \right).
\end{align}
%%%%%%%%%%%%%%%%%%%%%%%%%%%%%%%%%%%%%%%%%%%
In order to calculate $\bm{\Pi}_{\bm{\Delta}}^{\perp}\mathbf{g}$, the following identity is used \cite{gershman2002stochastic}:
\begin{equation}\label{eq:Q_xy_identity}
\bm{\Pi}_{\vect{X} \otimes \vect{Y}}^{\perp}
=\vect{I}\otimes \bm{\Pi}_{\vect{Y}}^{\perp}+\bm{\Pi}_{\vect{X}}^{\perp}\otimes \vect{I}
-\bm{\Pi}_{\vect{X}}^{\perp} \otimes  \bm{\Pi}_{\vect{Y}}^{\perp}
%\vect{Q}_{\vect{X} \otimes \vect{Y}}
%=\vect{I}\otimes \vect{Q}_{\vect{Y}}+\vect{Q}_{\vect{X}}\otimes \vect{I}
%-\vect{Q}_{\vect{X}} \otimes  \vect{Q}_{\vect{Y}}.
\end{equation}
Hence,
\begin{align}\label{eq:Qdelta_g}
\bm{\Pi}_{\bm{\Delta}}^{\perp}\mathbf{g}
&\stackrel{(a)}{=}
\bm{\Pi}^{\perp}_{\left(\bm{\Phi}^{-1/2}_{\mathbf{y}}\vect{A}  \right)^{*}
 \otimes \left(\bm{\Phi}^{-1/2}_{\mathbf{y}} \vect{A} \right)}
 \text{vec}\left(\bm{\Phi}^{-1/2}_{\mathbf{y}} \bm{\Gamma}_R \bm{\Phi}^{-1/2}_{\mathbf{y}}\right)
 \nonumber \\
&\hspace{-0.5cm}\stackrel{(b)}{=}
\Big(
\vect{I}\otimes \bm{\Pi}_{\bm{\Phi}^{-1/2}_{\mathbf{y}}\vect{A}}^{\perp}
+\bm{\Pi}_{\left(\bm{\Phi}^{-1/2}_{\mathbf{y}}\vect{A} \right)^{*}}^{\perp}
\otimes \vect{I}
\nonumber \\
&\hspace{-0.5cm}
\quad -\bm{\Pi}_{\left(\bm{\Phi}^{-1/2}_{\mathbf{y}}\vect{A} \right)^{*}}^{\perp}\otimes  \bm{\Pi}_{\bm{\Phi}^{-1/2}_{\mathbf{y}}\vect{A}}^{\perp} 
  \Big) 
\times
 \text{vec}\left(\bm{\Phi}^{-1/2}_{\mathbf{y}} \bm{\Gamma}_R \bm{\Phi}^{-1/2}_{\mathbf{y}}\right)
\nonumber\\
& \hspace{-0.5cm}\stackrel{(c)}{=}
\text{vec}\Big(\bm{\Pi}_{\bm{\Phi}^{-1/2}_{\mathbf{y}}\vect{A}}^{\perp}\bm{\Phi}^{-1/2}_{\mathbf{y}} \bm{\Gamma}_R \bm{\Phi}^{-1/2}_{\mathbf{y}} 
\nonumber \\  
& \hspace{-0.5cm}\hspace{1cm}
 +\bm{\Phi}^{-1/2}_{\mathbf{y}} \bm{\Gamma}_R \bm{\Phi}^{-1/2}_{\mathbf{y}}
  \bm{\Pi}_{\bm{\Phi}^{-1/2}_{\mathbf{y}}\vect{A}}^{\perp}
\nonumber \\  
& \hspace{-0.5cm}\hspace{1cm} 
-\bm{\Pi}_{\bm{\Phi}^{-1/2}_{\mathbf{y}}\vect{A}}^{\perp} 
\left(\bm{\Phi}^{-1/2}_{\mathbf{y}} \bm{\Gamma}_R \bm{\Phi}^{-1/2}_{\mathbf{y}}\right)
\bm{\Pi}_{\bm{\Phi}^{-1/2}_{\mathbf{y}}\vect{A}}^{\perp}  \Big),
%\nonumber\\*
\end{align}
where (a) follows by \eqref{eq:g} and \eqref{eq:Delta},
(b) follows by \eqref{eq:Q_xy_identity}
and (c) follows by \eqref{eq:kronecker_2}.
%%%%%%%%%%%%%%
Left multiplying \eqref{eq:Qdelta_g} by $\mathbf{g}^\hermitian$
yields the reciprocal of the \ac{CRB} in \eqref{eq:CRB1}:
\begin{align}\label{CRB_inv}
\mathbf{g}^\hermitian \bm{\Pi}_{\bm{\Delta}}^{\perp}\mathbf{g} 
&\stackrel{(a)}{=}
\left(\text{vec}\left(\bm{\Phi}^{-1/2}_{\mathbf{y}} \bm{\Gamma}_R \bm{\Phi}^{-1/2}_{\mathbf{y}}\right)
\right)^\hermitian
\nonumber \\
&
\hspace{-1.5cm}
\times
\text{vec}\Big(\bm{\Pi}_{\bm{\Phi}^{-1/2}_{\mathbf{y}}\vect{A}}^{\perp}\bm{\Phi}^{-1/2}_{\mathbf{y}} \bm{\Gamma}_R \bm{\Phi}^{-1/2}_{\mathbf{y}}
 +\bm{\Phi}^{-1/2}_{\mathbf{y}} \bm{\Gamma}_R \bm{\Phi}^{-1/2}_{\mathbf{y}}
  \bm{\Pi}_{\bm{\Phi}^{-1/2}_{\mathbf{y}}\vect{A}}^{\perp} 
\nonumber \\  
&
-\bm{\Pi}_{\bm{\Phi}^{-1/2}_{\mathbf{y}}\vect{A}}^{\perp} 
\left(\bm{\Phi}^{-1/2}_{\mathbf{y}} \bm{\Gamma}_R \bm{\Phi}^{-1/2}_{\mathbf{y}}\right)
\bm{\Pi}_{\bm{\Phi}^{-1/2}_{\mathbf{y}}\vect{A}}^{\perp}  \Big)
\nonumber \\
& \hspace{-1.5cm} \stackrel{(b)}{=}
\Tr \Bigg[
 2 \bm{\Phi}^{-1/2}_{\mathbf{y}} \bm{\Gamma}_R \bm{\Phi}^{-1/2}_{\mathbf{y}}
 \bm{\Pi}_{\bm{\Phi}^{-1/2}_{\mathbf{y}}\vect{A}}^{\perp}
 \bm{\Phi}^{-1/2}_{\mathbf{y}} \bm{\Gamma}_R \bm{\Phi}^{-1/2}_{\mathbf{y}}
  \nonumber \\  
& \hspace{-1.5cm}
 - \bm{\Phi}^{-1/2}_{\mathbf{y}}
\bm{\Pi}_{\bm{\Phi}^{-1/2}_{\mathbf{y}}\vect{A}}^{\perp}
\bm{\Phi}^{-1/2}_{\mathbf{y}} \bm{\Gamma}_R \bm{\Phi}^{-1/2}_{\mathbf{y}}
\bm{\Pi}_{\bm{\Phi}^{-1/2}_{\mathbf{y}}\vect{A}}^{\perp}\bm{\Phi}^{-1/2}_{\mathbf{y}}\bm{\Gamma}_R
\Bigg],
\raisetag{1.6\normalbaselineskip}
\end{align}
where (a) follows by \eqref{eq:g} and (b) follows by \eqref{eq:kronecker_1}.
%%%%%%%%
To proceed, let us define $\vect{E}\in \mathbb{C}^{N \times (N-T-1)}$ as 
a matrix that spans the nullspace of the matrix $\vect{A}$
s.t.~$\vect{E}^\hermitian \vect{A}=\vect{0}$, i.e.~a blocking matrix of the speech-plus-noise signals.
Since $\left(\bm{\Phi}^{1/2}_{\mathbf{y}}\vect{E}\right)^\hermitian
\bm{\Phi}^{-1/2}_{\mathbf{y}}\vect{A}=\vect{0}$,
it follows that \cite{gershman2002stochastic}
\begin{equation}\label{eq:Pi_D}
\bm{\Pi}_{\bm{\Phi}^{-1/2}_{\mathbf{y}}\vect{A}}^{\perp}
=\bm{\Pi}_{\bm{\Phi}^{1/2}_{\mathbf{y}}\vect{E}},
\end{equation}
and thus 
\begin{align}\label{eq:Phiy_Q_Phiy}
&\bm{\Phi}^{-1/2}_{\mathbf{y}} \bm{\Pi}_{\bm{\Phi}^{-1/2}_{\mathbf{y}}\vect{A}}^{\perp}
 \bm{\Phi}^{-1/2}_{\mathbf{y}}
=
 \bm{\Phi}^{-1/2}_{\mathbf{y}} \bm{\Pi}_{\bm{\Phi}^{1/2}_{\mathbf{y}}\vect{E}}
 \bm{\Phi}^{-1/2}_{\mathbf{y}}
  \nonumber \\
  &\hspace{1cm}\stackrel{(a)}{=}
  \vect{E} \left(\phi_R\vect{E}^\hermitian \bm{\Gamma}_R  \vect{E}\right)^{-1}\vect{E}^\hermitian
  \nonumber \\
      &\hspace{1cm}=
\phi_R^{-1}\bm{\Gamma}_R^{-1}
\left(
\bm{\Gamma}_R \vect{E}
\left(\vect{E}^\hermitian \bm{\Gamma}_R  \vect{E}\right)^{-1}
\vect{E}^\hermitian \bm{\Gamma}_R  \right)
\bm{\Gamma}_R^{-1}
    \nonumber \\
      &\hspace{1cm}\stackrel{(b)}{=}
\phi_R^{-2}\bm{\Gamma}_R^{-1}
\left(
\bm{\Phi}_{\mathbf{y}} \vect{E}
\left(\vect{E}^\hermitian \bm{\Phi}_{\mathbf{y}} \vect{E}\right)^{-1}
\vect{E}^\hermitian \bm{\Phi}_{\mathbf{y}}  \right)
\bm{\Gamma}_R^{-1}
    \nonumber \\
      &\hspace{1cm}\stackrel{(c)}{=}
\phi_R^{-2}\bm{\Gamma}_R^{-1}
\left(
\bm{\Phi}_{\mathbf{y}}^{1/2} 
\bm{\Pi}_{\bm{\Phi}_{\mathbf{y}}^{-1/2} \vect{A}}^{\perp}
 \bm{\Phi}_{\mathbf{y}}^{1/2}  \right)
\bm{\Gamma}_R^{-1},
\end{align}
where (a) and (b) follows since $\vect{E}^\hermitian \bm{\Phi}_{\mathbf{y}}= \phi_R\vect{E}^\hermitian \bm{\Gamma}_R$
and (c) follows by \eqref{eq:Pi_D}.
%%%%%%%%%%%%%%%%%%%%%%%%%%
Substituting \eqref{eq:Phiy_Q_Phiy} into \eqref{CRB_inv}
and using the property 
$\bm{\Pi}_{\vect{X}}^{\perp}\bm{\Pi}_{\vect{X}}^{\perp}=\bm{\Pi}_{\vect{X}}^{\perp}$,
yields
\begin{align}\label{CRB_inv3}
\mathbf{g}^\hermitian \bm{\Pi}_{\bm{\Delta}}^{\perp}\mathbf{g} 
&=\phi_R^{-2} \Tr \left[
\bm{\Pi}_{\bm{\Phi}^{-1/2}_{\mathbf{y}}\vect{A}}^{\perp}
\right]
= \phi_R^{-2}\big(N-(T+1) \big).
\end{align}
Substituting \eqref{CRB_inv3} into \eqref{eq:CRB1} yields \eqref{eq:CRB1_2}.

%%%%%%%%%%%%%%%%%%%%%%%%%%%%%%%
\section{} \label{app:CRB_phi_S}

We define the following partitioned matrix:
\begin{equation}\label{eq:partitioned_CRB_phi_s}
%\left[
%\begin{array}{c|c}
%\mathbf{s}& \bm{\Sigma}\\
%\end{array}
%\right]
%\triangleq
%\left(\bm{\Phi}^{-\top /2}_{\mathbf{y}} \otimes \bm{\Phi}^{-1/2}_{\mathbf{y}} \right)
%\left[
%\begin{array}{c|c}
%\frac{\partial  \bm{\phi}_{y}}{\partial \phi_S} & \frac{\partial  \bm{\phi}_{y}}{\partial \bm{\sigma}^\top } \\
%\end{array}
%\right],
%%%%%%%%%%%%%
\left[
\mathbf{s}\vert \vect{W}
\right]
\triangleq
\left(\bm{\Phi}^{-\top /2}_{\mathbf{y}} \otimes \bm{\Phi}^{-1/2}_{\mathbf{y}} \right)
\left[
\frac{\partial  \bm{\phi}_{y}}{\partial \phi_S} \middle| \frac{\partial  \bm{\phi}_{y}}{\partial \bm{\sigma}^\top } ,
\frac{\partial  \bm{\phi}_{y}}{\partial \phi_R}
\right],
\end{equation}
where $\bm{\sigma}\triangleq \text{vec}\left(\lbrace \bm{\phi}_{SV}\rbrace \setminus \lbrace\phi_S \rbrace \right)
%\in \mathbb{C}^{(T+1)^2}
$.
Similarly to \eqref{eq:fisher3}--\eqref{eq:CRB1},
we use \eqref{eq:partitioned_CRB_phi_s} to construct the \ac{FIM},
and then compute its inverse and take the corresponding component,
\begin{align}\label{eq:CRB_phi_S}
\text{CRB}\left(\phi_{S}  \right)
=\frac{1}{L} \left(\mathbf{s}^\hermitian \bm{\Pi}_{\vect{W}}^{\perp}\mathbf{s}\right)^{-1}.
\end{align}
We define a partition of $\vect{W}$ as
\begin{equation}\label{eq:partitioned_CRB_phi_s2}
%\bm{\Sigma}
%=
%\left[
%\begin{array}{c|c}
%\mathbf{s}& \bm{\Sigma}\\
%\end{array}
%\right]
%\triangleq
%\left(\bm{\Phi}^{-\top /2}_{\mathbf{y}} \otimes \bm{\Phi}^{-1/2}_{\mathbf{y}} \right)
%\left[
%\begin{array}{c|c}
%\frac{\partial  \bm{\phi}_{y}}{\partial \bm{\sigma}^\top } & \frac{\partial  \bm{\phi}_{y}}{\partial \phi_R} \\
%\end{array}
%\right],
%%%%%%%%%%%%%%%%%%%%%%
\vect{W}
=
\left(\bm{\Phi}^{-\top /2}_{\mathbf{y}} \otimes \bm{\Phi}^{-1/2}_{\mathbf{y}} \right)
\left[
\frac{\partial  \bm{\phi}_{y}}{\partial \bm{\sigma}^\top }\middle| \frac{\partial  \bm{\phi}_{y}}{\partial \phi_R}
\right]
\triangleq
\left[\bm{\Sigma} \vert \mathbf{g}\right]
.
\end{equation}
%%%%%%
Using the blockwise formula for projection matrices \cite{radhakrishna1999linear},
\begin{equation}\label{eq:Projection_W_blockwise}
\bm{\Pi}_{\vect{W}}^{\perp}
=\bm{\Pi}_{\bm{\Sigma}}^{\perp}
-\bm{\Pi}_{\bm{\Sigma}}^{\perp}\mathbf{g}
\left(\mathbf{g}^\hermitian\bm{\Pi}_{\bm{\Sigma}}^{\perp}\mathbf{g} \right)^{-1}
\mathbf{g}^\hermitian\bm{\Pi}_{\bm{\Sigma}}^{\perp}.
\end{equation}
We note that $\bm{\Sigma}$ can be further partitioned as
\begin{align}\label{eq:Projection_Sigma_partition}
\bm{\Sigma}
&=
\left(\bm{\Phi}^{-\top /2}_{\mathbf{y}} \otimes \bm{\Phi}^{-1/2}_{\mathbf{y}} \right)
\nonumber \\
& \quad \times
\left[
\vect{V}^{*} \otimes  \vect{V} \big| \mathbf{g}_{d}^{*} \otimes  \vect{V},
\vect{V}^{*} \otimes  \mathbf{g}_{d}
\right]
\triangleq
\left[\vect{C} \vert \vect{D}\right].
\end{align}
In a similar manner to \eqref{eq:Projection_W_blockwise}, we obtain
\begin{equation}\label{eq:Projection_Sigma_blockwise}
\bm{\Pi}_{\bm{\Sigma}}^{\perp}=\bm{\Pi}_{\vect{C}}^{\perp}-\bm{\Pi}_{\vect{C}}^{\perp}\vect{D}
\left(\vect{D}^\hermitian\bm{\Pi}_{\vect{C}}^{\perp}\vect{D} \right)^{-1}
\vect{D}^\hermitian\bm{\Pi}_{\vect{C}}^{\perp}.
\end{equation}
Substituting \eqref{eq:Projection_Sigma_blockwise} 
into \eqref{eq:Projection_W_blockwise} and then into \eqref{eq:CRB_phi_S}, yields
\begin{align}\label{eq:CRB_phi_S2}
\text{CRB}\left(\phi_{S}  \right)
&=\frac{1}{L} 
\left(
\alpha_s-\bm{\alpha}_{ds}^\hermitian \bm{\Theta}_d^{-1}\bm{\alpha}_{ds}
-\beta_{sg}^2
\beta_{g}^{-1}
%-(\alpha_{sg}-\bm{\alpha}_{ds}^\hermitian \bm{\Theta}_d^{-1}\bm{\alpha}_{dg})
%\left(\alpha_g-\bm{\alpha}_{dg}^{\hermitian}\bm{\Theta}_d^{-1} \bm{\alpha}_{dg} \right)^{-1}
%(\alpha_{sg}-\bm{\alpha}_{ds}^\hermitian \bm{\Theta}_d^{-1}\bm{\alpha}_{dg})^\hermitian 
\right)^{-1},
\end{align}
where
$\alpha_s=\mathbf{s}^\hermitian \bm{\Pi}_{\vect{C}}^{\perp}\mathbf{s}$,
$\bm{\alpha}_{ds}=\vect{D}^\hermitian \bm{\Pi}_{\vect{C}}^{\perp}\mathbf{s}$,
$\bm{\Theta}_d=\vect{D}^\hermitian \bm{\Pi}_{\vect{C}}^{\perp}\vect{D}$,
$\beta_{sg}=\alpha_{sg}-\bm{\alpha}_{ds}^\hermitian \bm{\Theta}_d^{-1}\bm{\alpha}_{dg}$,
$\beta_{g}=\alpha_g-\bm{\alpha}_{dg}^\hermitian \bm{\Theta}_d^{-1}\bm{\alpha}_{dg}$,
and
$\alpha_{sg}=\mathbf{s}^\hermitian\bm{\Pi}_{\vect{C}}^{\perp}\mathbf{g}$,
$\bm{\alpha}_{dg}=\vect{D}^\hermitian \bm{\Pi}_{\vect{C}}^{\perp}\mathbf{g}$,
$\alpha_g=\mathbf{g}^\hermitian \bm{\Pi}_{\vect{C}}^{\perp}\mathbf{g}$.
%%%%%%%%%%%
These quantities can be computed using similar techniques to those used in the derivation of
\eqref{eq:g}--\eqref{CRB_inv3}.
%%%
Due to space constraints, the detailed derivation is omitted.
%%%
Collecting all the terms and substituting into \eqref{eq:CRB_phi_S2},
yields the same expression as the \ac{MSE} in \eqref{eq:phi_S_variance}.

%%%%%%%%%%%%%%%%%%%%%%%%%%%%%%%%%%%%%%%%%%%%%%%%%%%%%%%%%%%%%%%%%%%%%
\balance
\bibliographystyle{IEEEtran}

\bibliography{papers}

%%%%%%%%%%%%%%%%%%%%%%%%%%%%%%%%%%%%%%%%%%%%%%%%%%%%%%%%%%%%%%%%%%%%%%

\end{document}